\documentclass[prd,english,preprintnumbers,amsmath,amssymb,nofootinbib,twocolumn,superscriptaddress,aps,10pt]{revtex4-1}

\pdfoutput=1

\usepackage[utf8]{inputenc}
\usepackage{graphicx}
\usepackage{bbm}
\usepackage{amssymb}
\usepackage{amsmath}
\usepackage{tabularx}
\usepackage{slashed}

\usepackage{dsfont}
\usepackage{babel}

\def\0#1#2{\frac{#1}{#2}}

\def\s0#1#2{\mbox{\small{$ \frac{#1}{#2} $}}}

\def\CC{{\mathcal C}}

\newcommand{\E}{\mathrm{e}}
\newcommand{\I}{\mathrm{i}}
\newcommand{\be}{\begin{eqnarray}}
\newcommand{\ee}{\end{eqnarray}}
\newcommand{\del}{\partial}

\newcommand{\nn}{\nonumber }
\newcommand{\fslash}{\hspace*{-0.2cm}\slash }

\newcommand{\beq}{\begin{equation}}
\newcommand{\eeq}{\end{equation}}
\newcommand{\bea}{\begin{eqnarray}}
\newcommand{\eea}{\end{eqnarray}}
\newcommand{\psib}{\bar{\psi}}
\newcommand{\Nc}{N_{\rm c}}
\newcommand{\Nf}{N_{\rm f}}

\newcommand\numberthis{\addtocounter{equation}{1}\tag{\theequation}}

\newcommand{\lFOrthogonalP}{l^\text{(F)}_{\perp +}(\tau, 0, -\I \tilde \mu _\tau)}
\newcommand{\lFOrthogonalPM}{l^\text{(F)}_{\perp \pm}(\tau, 0, -\I \tilde \mu _\tau)}
\newcommand{\lFParallelP}{l^\text{(F)}_{\parallel +}(\tau, 0, -\I \tilde \mu _\tau)}
\newcommand{\lFParallelPM}{l^\text{(F)}_{\parallel \pm}(\tau, 0, -\I \tilde \mu _\tau)}

\newcommand{\SpPm}{{(S+P)_{-}}}

\newcommand{\SpPmAdj}{{(S+P)_{-}^\mathrm{adj}}}
\newcommand{\Csc}{{\mathrm{csc}}}
\newcommand{\SigmaPion}{\text{($\sigma $-$\pi $)}}
\newcommand{\VmAPar}{{(V-A)_{\parallel}}}
\newcommand{\VmAPer}{{(V-A)_{\perp}}}
\newcommand{\VpAPar}{{(V+A)_{\parallel}}}
\newcommand{\VpAPer}{{(V+A)_{\perp}}}

\newcommand{\VmAPerAdj}{{(V-A)_{\perp}^{\mathrm{adj}}}}
\newcommand{\VpAParAdj}{{(V+A)_{\parallel}^{\mathrm{adj}}}}

\newcommand{\dif}{\ensuremath{\mathrm{d}}}

\newcommand{\br}{\nn \\ 
						    && \hspace*{.2cm}}

\makeatother
\begin{document}

\title{Fierz-complete NJL model study II:\\ towards the fixed-point and phase structure of hot and dense two-flavor QCD}

\author{Jens Braun}
\affiliation{Institut f\"ur Kernphysik (Theoriezentrum), Technische Universit\"at Darmstadt, 
D-64289 Darmstadt, Germany}
\affiliation{ExtreMe Matter Institute EMMI, GSI, Planckstra{\ss}e 1, D-64291 Darmstadt, Germany}
\author{Marc Leonhardt} 
\affiliation{Institut f\"ur Kernphysik (Theoriezentrum), Technische Universit\"at Darmstadt, 
D-64289 Darmstadt, Germany}
\author{Martin Pospiech}
\affiliation{Institut f\"ur Kernphysik (Theoriezentrum), Technische Universit\"at Darmstadt, 
D-64289 Darmstadt, Germany}

\begin{abstract}
{\it Nambu}--{\it Jona}-{\it Lasinio}-type models are often employed as low-energy models for the theory of the strong interaction 
to analyze its phase structure at finite temperature and quark chemical potential. In particular at low temperature and large
chemical potential, where the application of fully first-principles approaches is 
{currently difficult at best,
this class of models still plays a prominent} role to guide our understanding of the dynamics of dense strong-interaction matter. 
In this work, we consider a {\it Fierz}-complete version of the {\it Nambu}--{\it Jona}-{\it Lasinio} model with two massless
quark flavors and study its renormalization group flow and fixed-point structure at 
leading order of the derivative expansion of the effective action. {Sum rules for the various
four-quark couplings then allow} us to monitor the strength of the breaking of {the axial $U_{\rm A}(1)$ symmetry}
close to and above the phase boundary. 
We find that the dynamics in the ten-dimensional {\it Fierz}-complete space of four-quark couplings can only be reduced to a one-dimensional 
space associated with {the scalar-pseudoscalar coupling} in the strict large-$\Nc$ limit. Still, the interacting fixed point 
associated with this one-dimensional subspace appears to govern the dynamics at small quark chemical potential even beyond the large-$\Nc$ limit.
{At large chemical potential, corrections} beyond the large-$\Nc$ limit become important and
{the dynamics is dominated by diquarks, favoring the formation of a chirally symmetric diquark condensate. 
In this regime, our study suggests that 
the phase boundary is shifted to higher temperatures when} 
a {\it Fierz}-complete set of four-quark interactions is considered.
\end{abstract}

\maketitle

%
\section{Introduction}
Despite the tremendous progress that has been made in the development of fully first-principles approaches to the theory of the strong
interaction in recent years, low-energy models of the theory of the strong interaction (Quantumchromodynamics, QCD) 
are still considered very valuable 
for a variety of reasons.
In particular in the high-density regime, which is 
at least difficult to access with lattice Monte Carlo techniques, the {\it Nambu--Jona-Lasinio} (NJL) model
and its various variations and relatives, such as quark-meson (QM) models, provide us with an insight into
the rich symmetry breaking patterns that may potentially be at work in this regime.
The interest in strong-interaction matter at densities beyond the nuclear saturation density is by no means academic but
rather essential for our understanding of the dynamics of astrophysical objects. For example, 
studies of neutron stars rely on the equation of state of strong-interaction matter as input which, however, is currently
plagued by significant uncertainties at least at high density, see, e.g., Ref.~\cite{Watts} 
for a recent review.

From its early days on, the NJL model 
-- originally introduced as an effective theory to describe spontaneous symmetry 
breaking in particle physics based on an analogy with 
superconducting materials~\cite{Nambu:1961tp, *Nambu:1961fr} --
has guided our
understanding of the dynamics underlying the QCD phase structure, see 
Refs.~\cite{Klevansky:1992qe,Buballa:2003qv,Fukushima:2011jc,Andersen:2014xxa} for reviews.

The great relevance of NJL-type model studies for our understanding of dense strong-interaction matter 
is undisputed. Still, it is also part of the truth that the corresponding predictions 
suffer from generic features of the NJL model as well as from approximations {underlying these studies.
First of all, it should} be mentioned that NJL-type models 
in {\it four} space-time dimensions are non-renormalizable, both on the perturbative as well as on the 
non-perturbative level (see, e.g., Refs.~\cite{Braun:2011pp,Braun:2012zq}). Therefore, the 
ultraviolet~(UV) cutoff scale becomes a parameter of the model and, as an immediate consequence, the
regularization scheme belongs to the definition of the model. In particular, this implies that a given value of the UV cutoff scale
has always to be viewed against the background of the chosen regularization scheme. 
From a renormalization group (RG) standpoint, this scale should anyhow {\it not} be considered as 
an actual UV extent of the model but rather as the scale where the couplings of the model
are fixed.
In addition, we note that 
the use of so-called three-dimensional/spatial regularization schemes for studies of hot and dense matter 
may not be unproblematic. This issue originates from the fact that 
this class of schemes explicitly breaks {\it Poincar\'{e}} invariance, 
even at zero temperature where the model parameters are usually fixed. This 
may then eventually lead to spuriously emerging 
symmetry breaking patterns~\cite{Braun:2017srn}.

Let us now turn to more specific aspects of NJL-type model studies. The so-called classical action underlying the latter 
is usually given by a kinetic term for the quarks and a   
set of four-quark interactions which are usually selected by a phenomenological reasoning. These interactions may be viewed as
dynamically generated by  
quark-gluon interactions at high energy scales. 
In fact, two-gluon exchange diagrams
do not only generate immediately a scalar-pseudoscalar four-quark self-interaction channel as mostly considered 
in NJL-type models, but all four-quark self-interaction channels compatible with the fundamental symmetries of~QCD. 
From the standpoint of an RG evolution of QCD from high to low energies, 
the gluon-induced four-quark interactions may then become strong enough to trigger spontaneous symmetry 
breaking at some intermediate energy scale, 
depending on, e.g., the temperature and the quark chemical potential. This scale may be associated
with the UV cutoff scale of NJL-type models. 
In this spirit, we may therefore consider the general form of NJL-type models to be rooted in QCD.
In practice, however, the four-quark couplings in NJL-type models 
are usually not fixed in this way but rather by tuning them 
such that the correct values of a given set of low-energy observables is reproduced at, e.g., 
vanishing temperature and quark chemical potential. Unfortunately, the values of the chosen
set of low-energy observables may in general be reproduced by various different set of parameters. Moreover,
the parameters may depend on, e.g., the temperature and the quark chemical potential as suggested 
by the before-mentioned RG evolution of gluon-induced four-quark interactions~\cite{Springer:2016cji}.

From a phenomenological point of view, the four-quark interaction channels may be recast into 
effective bosonic degrees of freedom.
{In particular at large chemical potential, 
the effective degrees of freedom 
associated with the scalar-pseudoscalar channel, namely the $\sigma$-meson and the pions, are no longer expected to dominate the 
low-energy physics.
Here, other degrees of freedom, such as diquarks, play a dominant role,
see, e.g., Refs.~\cite{Bailin:1983bm,Buballa:2003qv,Alford:2007xm,Anglani:2013gfu} for reviews.
Apart from this phenomenologically guided point of view, the inclusion of more than one 
four-quark channel is of field-theoretical relevance as a given point-like four-quark interaction channel is reducible by
means of so-called {\it Fierz} transformations. This naturally causes 
ambiguities in studies where the considered set of four-quark interaction channels is 
{incomplete and also implies the necessity to use very general ans\"atze for the quark propagator 
as employed in, e.g., {\it Dyson}-{\it Schwinger}-type studies~\cite{Rusnak:1995ex}.} 
QCD low-energy model studies are in general treated in approximations which are indeed incomplete with
respect to {\it Fierz} transformations. Note also that, even in studies of the conventional NJL/QM model {defined
with only a scalar-pseudoscalar four-quark} interaction channel other four-quark channels
are dynamically generated due to quantum fluctuations.

By {considering an NJL model}
with a single fermion species
in Ref.~\cite{Braun:2017srn}, we have demonstrated   
that {\it Fierz}-completeness as associated with the inclusion of more ``exotic" four-fermion channels does not 
only play a prominent role at large chemical potential but also affect the dynamics at small chemical
potential. The latter is illustrated by a significant dependence of the curvature of the finite-temperature phase boundary at small chemical potential
on the number of four-fermion channels included in the calculations.
In the present work, which is a sequel of Ref.~\cite{Braun:2017srn}, we now extend our analysis to 
{an NJL model with} 
{\it massless} quark flavors coming in~$\Nc$ colors and two flavors
 to gain a better understanding of how {\it Fierz}-incomplete approximations of 
QCD low-energy models affect the 
predictions for the phase structure at finite temperature and density. In particular, 
we take into account the explicit symmetry breaking arising from the
presence of a heat bath and the chemical potential in our present study anchored at the leading order 
of the derivative expansion of the effective action. 
Within our {\it Fierz}-complete framework including ten four-quark channels,
we observe that channels associated with an explicit breaking of 
{\it Poincar\'{e}} invariance tend to increase significantly the critical temperature at large chemical potential. In 
 {accordance with
many conventional model studies (see, e.g., Refs.~\cite{Buballa:2003qv,Alford:2007xm,Fukushima:2011jc,Anglani:2013gfu} for reviews),
diquarks} are nevertheless found to be the most dominant degrees of freedom in this regime.  

In Sec.~\ref{sec:model}, we discuss our model and aspects of symmetries which are relevant for our present analysis. 
{Whereas the number of colors~$\Nc$ is a free parameter in this more general discussion, we 
emphasize that we exclusively consider~$\Nc=3$ in all numerical studies in this work.
Our RG framework is introduced} in Sec.~\ref{sec:fwall}.
However, since this work is a sequel of Ref.~\cite{Braun:2017srn}, where also
the theoretical setting underlying our present analysis has {been thoroughly discussed, 
we refrain here from a presentation of details regarding our RG framework.
The RG fixed-point and phase structure of our  
model at finite temperature and density 
at leading order of the derivative expansion of the effective action is analyzed in Sec.~\ref{sec:fpps} where
we begin with a discussion of the relation of the mean-field approximation and an RG study of
four-quark interactions in a one-channel approximation in Subsec.~\ref{sec:1c} 
The actual analysis of the fixed-point and phase structure of our 
{\it Fierz}-complete NJL-type model can be found in Subsecs.~\ref{sec:sbfc} - \ref{sec:2c}. 
The symmetry breaking patterns are discussed in Subsec.~\ref{sec:sbfc}.
The effect of~$U_{\rm A}(1)$ breaking and its fate at high temperature 
is analyzed in Subsec.~\ref{sec:ua1}. In order to gain a better understanding of the mechanisms underlying 
symmetry breaking at finite temperature and density in our {\it Fierz}-complete study involving ten four-quark
interaction channels, we then discuss 
the RG flow of the {\it Fierz}-complete system in the large-$\Nc$ limit in Subsec.~\ref{sec:ln}.   
In Subsec.~\ref{sec:2c}, we finally analyze the dynamics of the {\it Fierz}-complete system with the aid of 
a two-channel approximation which is composed of the before-mentioned scalar-pseudoscalar channel and a diquark channel
often employed in conventional NJL model studies. There, we also comment on the effect of {\it Fierz}-incomplete approximations 
on the curvature of the finite-temperature phase boundary at small chemical potential.}
Our conclusions can be found in Sec.~\ref{sec:conc}.

\section{Model}\label{sec:model}

A frequently employed approximation in terms of NJL-type models is to consider an action which only consists of a kinetic 
term for the quarks and a scalar-pseudoscalar four-quark interaction channel. By means of a {\it Hubbard}-{\it Stratonovich} transformation,
auxiliary fields can be introduced and the four-quark interaction channel is converted into a (screening) mass term for the
auxiliary fields and a Yukawa interaction channel
between the latter and the quarks. Conventionally, the auxiliary fields are chosen to carry the quantum numbers of the $\sigma$-meson
and the pions in case of the scalar-pseudoscalar four-quark interaction channel. The interactions between the quarks are then said
to be mediated by an exchange of the before-mentioned mesons.
For a study of chiral symmetry breaking this choice for the auxiliary fields is particularly convenient because it allows for a straightforward
projection on the chiral order parameter. In this work, however, we shall consider a purely fermionic formulation of the NJL model
with massless quarks coming in two flavors and~$\Nc$ colors, aiming at an analysis of the effect
of {\it Fierz}-incomplete approximations on the phase structure at finite temperature and density in QCD low-energy model studies.
In order to relate our work to the latter studies, we start our discussion by considering a 
so-called classical action~$S$ which essentially consists of a kinetic term for the quarks and a scalar-pseudoscalar four-quark 
interaction channel in four {\it Euclidean} space-time dimensions:
\be
 &S[\bar\psi,\psi]=&\int_0^\beta\dif \tau \int\dif^3 x\ \Big\{\bar{\psi}\left( \I \slashed \partial - \I \mu \gamma_0  \right) \psi\nn  \\
&&\hspace{.5cm} + \frac{1}{2}\bar\lambda _{\text{($\sigma $-$\pi $)}} \left[\left(\bar\psi \psi\right)^2-\left(\bar\psi \gamma_5 \tau_i \psi\right)^2\right]\Big\}\,.
\label{eq:SNJL}
\ee
Here, $\beta=1/T$ is the inverse temperature,~$\mu$ is the quark chemical potential, and~$\bar\lambda _{\text{($\sigma $-$\pi $)}}$
is the coupling associated {with the scalar-pseudoscalar channel.}
The $\tau_i$'s represent the {\it Pauli} matrices and couple the spinors in flavor 
space. Note that the quark fields~$\psi$ {carry {\it Dirac}, color, and} flavor indices. In the 
{following, they are}
assumed to be contracted pairwise, 
e.g.~$(\bar{\psi}{\mathcal O}\psi)\equiv \bar{\psi}_{\chi}{\mathcal O}_{\chi\xi}\psi_{\xi}$, where~$\xi$ and~$\chi$ represent  
collective indices for the {\it Dirac}, flavor and color indices and~$\mathcal O$ represents an operator specifying the properties of the interaction channel.
Suitable insertions of~$\mathds{1}$-operators {in {\it Dirac}, color, and} flavor space are tacitly assumed. 

{Let us begin our symmetry analysis by noting that the action~\eqref{eq:SNJL} is invariant under 
(global) $SU(\Nc)$ color rotations of the quark fields.
As we do not allow for an explicit quark mass term, we also have an invariance under}
(independent) global flavor rotations of the left- and right-handed quark fields, $\psi_{\text{L},\text{R}}=\frac{1}{2}(1\pm \gamma_5)\psi$,
i.e. the action is 
invariant under~$SU_\text{L}(2)\otimes SU_\text{R}(2)$ transformations. 
The spontaneous breakdown of this so-called
chiral symmetry is associated with the formation of a corresponding chiral {condensate~$\langle\bar{\psi}\psi\rangle$
rendering the quarks massive.}

The action~\eqref{eq:SNJL} is also invariant
under simple global phase transformations,
\be
U_\text V (1) &:& \psib \mapsto  \psib \E^{-\I \alpha}, \enspace \psi\mapsto \E^{ \I \alpha} \psi\,,
\ee
but is {\it not} invariant under axial 
phase transformations:
\be
U_\text A (1) &:&  \psib \mapsto \psib  \E^{ \I \gamma_5 \alpha}, \enspace  \psi \mapsto \E^{ \I \gamma_5 \alpha} \psi\,.
\ee
In both cases,~$\alpha$ denotes the ``rotation" angle. Note that, in contrast to the case of a single fermion species, 
i.e. the case of one color and one {flavor, a broken~$U_{\text{A}}(1)$ symmetry} does not necessarily entail
the existence of a finite expectation value~$\langle\bar{\psi}\psi\rangle$ as associated with spontaneous
chiral symmetry breaking. However, the spontaneous breakdown of the chiral symmetry also entails the breakdown of
{the~$U_{\text{A}}(1)$ symmetry~\cite{Pokorski:1987ed}. In nature, it turns out that the~$U_{\text{A}}(1)$ symmetry is not realized  
but anomalously broken by topologically non-trivial gauge configurations~\cite{tHooft:1976rip,tHooft:1976snw},  
even if the chiral~$SU_\text{L}(2)\otimes SU_\text{R}(2)$ is restored. 
This absence of $U_{\rm A}(1)$ symmetry 
may be deduced from the fact that we do not observe parity doubling of hadronic states~\cite{Cheng:1985bj}, at least 
at low energies. 
In} any case, in the action~\eqref{eq:SNJL}, we can artificially restore the~$U_{\text{A}}(1)$ symmetry by adding
an additional four-quark channel,
\be
\sim \det\left(\bar{\psi}(1+\gamma_5)\psi\right) + \det\left(\bar{\psi}(1-\gamma_5)\psi\right)
\,,
\label{eq:det}
\ee
provided that the coupling associated with this channel is adjusted suitably relative to the coupling~$\bar\lambda _{\text{($\sigma $-$\pi $)}}$
of the scalar-pseudoscalar channel, see also Ref.~\cite{Klevansky:1992qe}.\footnote{Note that the determinant 
in Eq.~\eqref{eq:det} is taken in flavor space.}
Indeed, the topologically non-trivial gauge configurations violating the~$U_{\text{A}}(1)$ symmetry can be 
recast into a four-quark interaction channel of the form~\eqref{eq:det} in the case of 
two-flavor QCD~\cite{tHooft:1976snw,Shifman:1979uw,Shuryak:1981ff,Schafer:1996wv,Pawlowski:1996ch}.
We shall come back to the issue of~$U_{\text{A}}(1)$ symmetry breaking below.

Apart from the chiral symmetry and the~$U_{\text{A}}(1)$ symmetry, the~$U_\text{V} (1)$ symmetry associated with baryon number 
conservation may also be spontaneously broken. In contrast to chiral symmetry breaking, 
this is indicated by the {formation of, e.g., a diquark condensate~$\delta^{a}$ carrying a net baryon 
and net color} {charge:
\be
\delta^{a} \sim\langle \I \bar \psi^C \gamma_5 \epsilon_{f}\varepsilon_{c}^l\psi \rangle\,.
\label{eq:diq}
\ee
Here,~$\epsilon_{f}\equiv \epsilon_f^{(\alpha,\beta)}$ and $\varepsilon_{c}^{l}\equiv\varepsilon_c^{l(m,n)}$ are antisymmetric} tensors in flavor and color space, respectively. 
Moreover, we have introduced charge conjugated fields~$\psi^C = \CC \bar\psi ^T$ and $\bar \psi^C = \psi^T \CC $ 
with $\mathcal{C}=\I \gamma_2 \gamma_0$ being related to the charge conjugation operator.

The diquark condensate~$\delta^{a}$ is a state with $J^{P}=0^{+}$ and has been found to be 
most dominantly generated by one-gluon exchange~\cite{Alford:1997zt} 
and topologically non-trivial gauge configurations~\cite{Alford:1997zt,Rapp:1997zu}, see 
also Refs.~\cite{Schafer:1996wv,Alford:2007xm,Buballa:2008zza} for reviews.
The flavor antisymmetric structure {of this color-superconducting condensate corresponds} 
to a singlet representation of the global chiral group 
which implies that the formation of such a condensate does not violate the chiral symmetry of the theory~\cite{Shovkovy:2004me}.
Note that this is different in QED-like theories where the formation of a superconducting ground state also requires the 
chiral symmetry to be broken, see, e.g., Ref.~\cite{Braun:2017srn}. 
Instead, the {formation of the before-mentioned 
color-superconducting condensate~$\delta^a$} in QCD comes at the price of a broken $SU(\Nc)$ color symmetry.

In addition to the breaking of the before-mentioned symmetries, we have to deal with the 
explicit breaking of {\it Poincar\'{e}} invariance in our calculations 
because of the presence of a heat bath and a finite
quark chemical potential. This implies that the {\it Euclidean} time direction is distinguished and
we are only left with rotational invariance among the spatial components.
With respect to the fundamental symmetries associated with charge conjugation, time reversal,
and parity, we add that only invariance under parity transformations and time reversal transformations
remain intact in the presence of a finite quark chemical potential as the latter also breaks
explicitly the charge conjugation symmetry. 

Let us now discuss the general form of the quantum effective 
action~$\Gamma$ of our model at leading order of the derivative expansion.
Our symmetry considerations constrain the classical action of the model underlying our studies.
The latter action may be considered as the zeroth-order approximation 
of the corresponding quantum effective action~$\Gamma$.
The full quantum effective action is obtained from the 
path integral associated with the classical action by means of a Legendre transformation. By computing
the quantum corrections to the classical action~\eqref{eq:SNJL}, we immediately observe that four-quark channels other than 
the scalar-pseudoscalar channel are generated. In fact, any operator respecting the symmetries of 
our model can in principle be generated. 
Structuring our calculations by means of a derivative expansion, we find
that,  at leading order (LO), the most general ansatz for the effective average action compatible
with the symmetries of the theory reads\footnote{Quark self-interactions 
of higher order {are also generated 
dynamically but, at leading order of the derivative expansion, they do
not contribute to the RG flow of the four-quark self-interactions  
and are} therefore not
included in our present study.}
\be
&& \Gamma_{\text{LO}}[\bar\psi,\psi]\nn\\
&&\quad =\int_0^\beta\dif \tau \int\dif^3 x\ \Big\{ 
\bar{\psi}\left(Z^\parallel \I \gamma_0 \partial_{0} + Z^\perp \I \gamma_i \partial_{i}- Z_\mu \I \mu \gamma_0 \right) \psi \nn \\
&& \qquad\qquad\qquad\qquad\qquad\qquad\quad +\frac{1}{2}\sum_{j \in\, \mathcal{B}}\ Z_j{\bar\lambda_{j}}\,\mathcal{L}_{j} \Big\}\,,
\label{eq:GLO}
\ee
where the {elements $\mathcal{L}_{j}$ form a}
{ten-component {\it Fierz}-complete basis~${\mathcal B}$ of pointlike} four-quark interactions 
accompanied by the associated bare couplings~$\bar{\lambda}_i$ and the corresponding vertex renormalizations~$Z_j$.\footnote{The leading
order of the derivative expansion implies that the four-quark self-interactions are treated in the pointlike limit.}
Any other pointlike four-quark interaction invariant under the symmetries of our model is indeed reducible by means of 
{\it Fierz} transformations.\footnote{The couplings~$\bar{\lambda}_{j}$
appearing in the effective action~\eqref{eq:GLO} should not be confused with the couplings~$\bar{\lambda}_j$ appearing in the classical action~$S$, see, e.g., Eq.~\eqref{eq:SNJL}.
The couplings appearing in the effective action include quantum corrections whereas, from an RG standpoint, the couplings appearing in the classical
action only determine the values of the RG flows of the four-quark couplings at the initial scale~$\Lambda$.}
Recall that
we assume here that the $U_{\text{A}}(1)$ symmetry is broken explicitly, see below for a detailed discussion of this issue.
The renormalization factors associated with the kinetic term are given by~$Z^{\parallel}$ and~$Z^{\perp}$, respectively. In 
the following we set~$Z^{\parallel}=Z^{\perp}\equiv 1$ as the RG flow of these quantities vanishes 
identically at leading order 
of the derivative expansion, see Ref.~\cite{Braun:2011pp}.
In general, the chemical potential is also accompanied by a renormalization factor~$Z_{\mu}$. 
As a consequence of the so-called {\it Silver}-{\it Blaze} property of general {quantum field theories~\cite{Cohen:2003kd,Marko:2014hea,Khan:2015puu,*Fu:2015naa}, however, 
we have~$Z_{\mu}^{-1}=Z^{\parallel}=Z^{\perp}$ at~$T=0$, provided that the renormalized chemical potential~$\mu_{\rm r}\equiv Z_{\mu}\mu$ is smaller 
than} the dynamically generated renormalized (pole) mass~$\bar{m}_{\rm q}=\bar{m}_{\rm q}/Z^{\perp}$ 
{of the quarks~\cite{Braun:2017srn}.}

{Now we specify} the {\it Fierz}-complete basis of pointlike four-quark interactions which we use to parametrize the effective
action~\eqref{eq:GLO}. As indicated above, we find that 
{this basis is composed of ten four-quark channels. We choose six of them to be invariant 
under $SU(\Nc)\otimes SU_\text{L}(2)\otimes SU_\text{R}(2)\otimes U_\text{V}(1) \otimes U_\text{A}(1)$ transformations:}
\be
\mathcal{L}_\VpAPar&=&\left(\bar\psi\gamma_0\psi\right)^2+\left(\bar\psi\I\gamma_0\gamma_5\psi\right)^2\,,\label{eq:firstchan}
\\
\mathcal{L}_\VpAPer&=&\left(\bar\psi\gamma_i\psi\right)^2+\left(\bar\psi\I\gamma_i\gamma_5\psi\right)^2\,,\\
\mathcal{L}_\VmAPar&=&\left(\bar\psi\gamma_0\psi\right)^2-\left(\bar\psi\I\gamma_0\gamma_5\psi\right)^2\,,\\
\mathcal{L}_\VmAPer&=&\left(\bar\psi\gamma_i\psi\right)^2-\left(\bar\psi\I\gamma_i\gamma_5\psi\right)^2\,,\\
\mathcal{L}_\VpAParAdj&=&\left(\bar\psi\gamma_0 T^a\psi\right)^2+\left(\bar\psi\I\gamma_0\gamma_5 T^a\psi\right)^2\,,
\label{Eq:VpAParAdj}\\
\mathcal{L}_\VmAPerAdj&=&\left(\bar\psi\gamma_iT^a\psi\right)^2-\left(\bar\psi\I\gamma_i\gamma_5 T^a\psi\right)^2\,.
\ee
The remaining four channels are then chosen to be invariant {under~$SU(\Nc)\otimes SU_\text{L}(2)\otimes SU_\text{R}(2)\otimes U_\text{V}(1)$ 
transformations}
but break the $U_\text{A}(1)$ symmetry explicitly:
\be
\mathcal{L}_{\text{($\sigma $-$\pi $)}}&=&\left(\bar\psi \psi\right)^2\!-\! \left(\bar\psi \gamma_5 \tau_i \psi\right)^2\,,\\
\mathcal{L}_{(S+P)_{-}}&=&\left(\bar\psi \psi\right)^2\!-\!\left(\bar\psi \gamma_5 \tau_i \psi\right)^2 \nn\\
&&\hspace{1cm} \!+\!\left(\bar\psi \gamma_5 \psi\right)^2\!-\!\left(\bar\psi \tau_i \psi\right)^2\,,\\
\mathcal{L}_\Csc &=& 4 \left( \I \bar \psi \gamma_5 \tau_2\, T^{A} \psi^C \right) \left( \I \bar \psi^C \gamma_5 \tau_2\, T^{A} \psi \right)\,,
\label{eq:cscdef}\\
\mathcal{L}_{(S+P)_{-}^\mathrm{adj}} &=&\left(\bar\psi T^a\psi\right)^2\!-\!\left(\bar\psi \gamma_5 \tau_i T^a\psi\right)^2 \nn\\
&&\hspace{1cm}\!+\!\left(\bar\psi \gamma_5 T^a\psi\right)^2\!-\!\left(\bar\psi \tau_i T^a\psi\right)^2\,,
\label{eq:detcp}
\ee
where, e.g., $\left(\bar\psi \gamma_5 \tau_i \psi\right)^2\equiv \left(\bar\psi \gamma_5 \tau_i \psi\right)\left(\bar\psi \gamma_5 \tau_i \psi\right)$ 
and the $T^a$'s denote the generators of $SU(\Nc)$. Note that this basis is not unique. In principle, we can combine elements of the 
basis to perform a basis transformation. Our present choice is motivated 
by the four-quark channels conventionally employed in QCD low-energy models. Apparently, 
the scalar-pseudoscalar channel appearing in Eq.~\eqref{eq:SNJL} is given by the channel~$\mathcal{L}_{\text{($\sigma $-$\pi $)}}$. 
A channel of the form of Eq.~\eqref{eq:det} is associated 
with the presence of topologically non-trivial gauge configurations and is given by the channel~$\mathcal{L}_{(S+P)_{-}}$ up to 
a numerical constant.
There is also a channel associated with the formation of a diquark condensate of the type~\eqref{eq:diq} in our basis. 
In fact, taking into account that such a condensate leaves the chiral symmetry intact, the corresponding four-quark channel~$\mathcal{L}_\Csc$
can be constructed from the 
tensor structure of the condensate~\eqref{eq:diq}. 
Our conventions in Eq.~\eqref{eq:cscdef} are such 
that we only sum over the antisymmetric~($A$) generators {of the $SU(\Nc)$ color} group. {The normalization of this channel is 
chosen as in the standard literature (see, e.g., Ref.~\cite{Buballa:2003qv}).}
Note that the channel~$\mathcal{L}_\Csc$ 
is invariant {under~$SU(\Nc)\otimes SU_\text{L}(2)\otimes SU_\text{R}(2)\otimes U_\text{V}(1)$ transformations. The} 
formation of a diquark 
condensate then goes along with the breakdown 
{of the~$U_\text{V}(1)$ symmetry as well as the~$SU(\Nc)$ color symmetry.}
Finally, we add that 
the channel~\eqref{eq:detcp} may be 
viewed as a counterpart of the channel~$\mathcal{L}_{(S+P)_{-}}$ with a non-trivial
color structure.

It is worth pointing out {that our {\it Fierz}-complete set of pointlike four-quark interactions 
allows us to monitor $U_{\text{A}}(1)$ symmetry} breaking.
Indeed, by requiring that the effective action~$\Gamma$ is invariant under~$U_{\text{A}}(1)$ transformations, we find the following 
two sum rules for the four pointlike couplings violating the $U_{\text{A}}(1)$ {symmetry:
\be
\!\!\!\!\!\!\!\!\!\!\mathcal{S}_{U_{\rm A}(1)}^{(1)}&=& \bar{\lambda}_\Csc+ \bar{\lambda}_\SpPmAdj =0\,,\label{eq:SUA1}\\
\!\!\!\!\!\!\!\!\!\!\mathcal{S}_{U_{\rm A}(1)}^{(2)}&=& \bar{\lambda}_\SpPm \!-\! \frac{\Nc \!-\! 1}{2\Nc}\,\bar{\lambda}_\Csc \!+\!\frac{1}{2}\bar{\lambda}_{(\sigma \text{-} \pi)} =0\,.\label{eq:SUA2}
\ee
These} two sum rules are only fulfilled simultaneously 
if the $U_{\text{A}}(1)$ symmetry of the theory is intact. For example, choosing only the scalar-pseudoscalar
coupling~$\bar{\lambda}_{(\sigma\text{-}\pi)}$ to be finite in the classical action~\eqref{eq:SNJL}, we find that 
the $U_{\text{A}}(1)$ symmetry is violated. {This symmetry is 
only found to be approximately restored} on the quantum level at high 
temperatures, see {our discussion in Subsec.~\ref{sec:ua1}.}

{From the sum rules~\eqref{eq:SUA1} and~\eqref{eq:SUA2}, we deduce that 
the four-dimensional space spanned by the~$U_{\text{A}}(1)$-violating 
channels contains 
a~$U_{\text{A}}(1)$-symmetric subspace.
In particular, the two sum rules imply that a {\it Fierz}-complete 
basis of pointlike four-quark interactions in 
case of a theory invariant 
under $SU(\Nc)\otimes SU_\text{L}(2)\otimes SU_\text{R}(2)\otimes U_\text{V}(1) \otimes U_\text{A}(1)$
transformations} is composed of eight \mbox{channels}.\footnote{Albeit possible, we do not use a basis of four-quark channels 
composed of an eight-dimensional subspace invariant 
{under $SU(\Nc)\otimes SU_\text{L}(2)\otimes SU_\text{R}(2)\otimes U_\text{V}(1) \otimes U_\text{A}(1)$ transformations}
and a remaining two-dimensional subspace only invariant 
{under~$SU(\Nc)\otimes SU_\text{L}(2)\otimes SU_\text{R}(2)\otimes U_\text{V}(1)$ transformations
in order to make better contact to conventional QCD model studies.}}

\section{Framework}\label{sec:fwall}
\subsection{Four-fermion interactions and phase transitions: a brief introduction}\label{sec:fw}

Before we actually analyze the fixed-point and
phase structure of our model at finite temperature and quark chemical potential, we briefly discuss
our RG framework. A detailed discussion of the latter
can be found in Ref.~\cite{Braun:2017srn} to which this work is a sequel. 
Moreover, we also give only a brief discussion of how a study of the quantum effective action~\eqref{eq:GLO}
at leading order in the derivative expansion can give us access to the phase structure. 
A detailed discussion of {this issue 
is also given in Refs.~\cite{Braun:2011pp,Braun:2017srn}}. 

We begin our discussion by noting that the four-quark vertex is treated in the so-called pointlike limit at 
leading order of the derivative expansion, i.e.
in the limit of vanishing external momenta. It is then clear that the mass spectrum encoded in the 
momentum structure of, e.g., the general four-quark vertex cannot be accessed at this order. In particular, the dynamics of  
regimes governed by the spontaneous formation
of condensates is not accessible at this order. In fact, the formation
of such condensates is indicated by singularities in the four-quark correlation functions. Nevertheless, 
the effective action~\eqref{eq:GLO}
at leading order in the derivative expansion still allows us to study regimes which are not governed by 
condensate formation, e.g. the high-temperature regime. For a given value of the quark chemical potential, 
we can therefore use the leading order of the 
derivative expansion to determine a critical temperature~$T_{\rm cr}$ below which the pointlike approximation breaks down
and a (fermion) condensate associated with the spontaneous breakdown of one of the symmetries 
of our model is expected to be generated dynamically.
This can also be understood from the standpoint of a {\it Hubbard}-{\it Stratonovich}-transformation~\cite{Hubbard:1959ub,*Stratonovich}. Using such a transformation,
we can reformulate our purely fermionic action in terms of 
quark fields and auxiliary bosonic fields (composites of two fermion fields, e.g. pion-like field or diquark-like field) 
coupled via {\it Yukawa}-type interactions. Such a reformulation 
eventually allows to compute conveniently the {\it Ginzburg}-{\it Landau}-type effective 
potential for the bosonic fields.
The transformation properties of the latter depend 
on the tensor structure of the corresponding four-quark channel. 
Now in case of a second-order phase transition, the sign of at least one of the terms bilinear in the bosonic fields appearing in the {\it Ginzburg}-{\it Landau}-type effective 
potential changes. At a first-order phase transition, the situation is different. Still, taking into account {\it all} quantum fluctuations,
the {\it Ginzburg}-{\it Landau}-type effective 
potential becomes also convex in this case, implying that at least one of the mass-like parameters associated with 
the terms bilinear in the bosonic fields tends to zero in the long-range limit. 
Since the pointlike four-quark couplings are found to be inverse proportional to the mass-like parameters associated with the bosonic fields,
a diverging four-quark coupling 
in the purely fermionic formulation
indicates the onset of spontaneous symmetry breaking. 

With respect to the RG analysis underlying
the present work, these considerations imply that the observation of a divergence of a four-quark coupling at an RG scale~$k_{\rm cr}$ 
can be used as an indicator for the onset of spontaneous symmetry breaking. Such an analysis has indeed been 
successfully applied to compute the phase structure of various systems including gauge theories with many flavors~\cite{Gies:2005as,*Braun:2005uj,*Braun:2006jd,*Braun:2014wja}, 
see Ref.~\cite{Braun:2011pp} for a review.
However, it should also be noted that this type of analysis is limited.\footnote{For a detailed discussion of 
{such an analysis} and its limitations, we refer the reader to Refs.~\cite{Braun:2011pp,Braun:2017srn}.} 
For example, it does not allow us to resolve the order 
of a phase transition. Moreover, the phenomenological meaning of a critical temperature obtained from {such an
analysis} is potentially ambiguous. Different symmetry breaking patterns associated with 
the various four-quark channels exist in our model. Therefore it is at least difficult to 
relate the breakdown of the pointlike approximation to the spontaneous breakdown of a specific symmetry,
even more so since a divergence in a specific four-quark channel entails corresponding divergences in all other
channels. However, a ``dominantly diverging" four-quark channel can in general be identified, i.e. the modulus of the 
coupling of this channel is greater than the ones of the other four-quark couplings. Of course, this does not 
necessarily imply that a condensate associated with this channel is generated. It should only be viewed as 
an indicator for the symmetry breaking scenario at work. {In Sec.~\ref{sec:fpps}, we 
present an analysis of the 
``hierarchy" of the various four-quark interactions in terms of their strength and
show that our ``criterion
of dominance" is at least in accordance with the simplest phenomenological expectation
of the symmetry breaking 
patterns at work at small and large chemical potential~\cite{Buballa:2003qv}. 
We have checked that our results from such an analysis are not altered when 
we rescale the channels~\eqref{eq:firstchan}-\eqref{eq:detcp} with factors of~${\mathcal O}(1)$.}

For the computation of the RG flows of the four-quark couplings, we employ 
the {\it Wetterich} equation~\cite{Wetterich:1992yh} which is an RG equation for the quantum effective action~$\Gamma$.
Within this framework, the effective action~$\Gamma$ depends on the RG scale~$k$ which is 
related to the RG ``time" $t=\ln (k/\Lambda)$. Here, the scale~$k$ defines an infrared (IR) cutoff scale 
and~$\Lambda$ may be chosen to be the scale at which
we fix the initial conditions of the RG flow of the effective action. In this RG approach, the {\it Wilsonian} momentum-shell 
integrations are specified by the regularization sheme which is encoded in form of a so-called regulator function.
Following Ref.~\cite{Braun:2017srn}, we employ here a so-called four-dimensional {\it Fermi}-surface adapted 
regularization scheme which becomes manifestly covariant in the limit~$T\to 0$ and~$\mu\to 0$, see also App.~\ref{app:RG}.

In general, we then find that the RG flow of the four-quark couplings is governed by two classes\footnote{The two classes 
contain diagrams which are associated with contributions longitudinal and transversal to the heat bath.}
 of one-particle irreducible (1PI) diagrams, 
see Fig.~\ref{fig:fd}. These diagrams can be recast into so-called threshold functions which
describe the decoupling of massive modes and
modes in a thermal and/or dense medium. The definitions of these functions 
can be found in Ref.~\cite{Braun:2017srn}.

\subsection{Scale-fixing procedure}\label{sec:sfproc}

{Let us now give a detailed discussion of the scale-fixing procedure underlying our
calculations in this work.}
The values of the ten four-quark couplings 
at the initial RG scale~$k=\Lambda$ can be considered as free parameters of our model.
To pin them down, let us consider RG studies of QCD where the strengths of pointlike gluon-induced four-quark interactions
have been analyzed in detail in the vacuum limit {within a {\it Fierz-complete} 
setting~\cite{Braun:2006zz,Mitter:2014wpa,Cyrol:2017ewj}. There, it} was found that the scalar-pseudoscalar
channel~$\mathcal{L}_{(\sigma\text{-}\pi)}$ is generated predominantly at high momentum 
{scales~$p\sim k$. Moreover, 
it} was found that this channel  
remains to be the most dominant one over a wide range of scales down 
to~$k\gtrsim 1\,\text{GeV}$, i.e. the modulus of any
other four-quark coupling remains smaller than the one of the scalar-pseudoscalar coupling. With respect to 
our present study, it is also reasonable to expect that effects associated with an explicit breaking of {\it Poincar\'{e}} 
and~$\mathcal C$ invariance are subleading 
as long as~$T/k \ll 1$ and~$\mu/k\ll 1$. In the light of these facts, we only
choose the scalar-pseudoscalar coupling~$\bar\lambda _{\text{($\sigma $-$\pi $)}}$ to be finite at the initial RG scale~$\Lambda$ and 
set all other four-quark couplings to zero. Thus, at the initial scale, 
we are left with the action~$S$ given in Eq.~\eqref{eq:SNJL}, $\Gamma_{k=\Lambda}=S$.
This implies that we assume the~$U_{\text{A}}(1)$ symmetry to be broken explicitly at the initial 
RG scale.\footnote{In Subsec.~\ref{sec:ua1}, we discuss the
effect of $U_{\text{A}}(1)$ symmetry breaking in more detail with the aid of the sum rules~\eqref{eq:SUA1} and~\eqref{eq:SUA2}.}
{Clearly, these considerations do not represent a rigorous determination of the initial conditions of our NJL-type model from QCD, which would
require the dynamical inclusion of gauge degrees of freedom in an RG study~\cite{Gies:2005as,*Braun:2005uj,*Braun:2006jd,*Braun:2014wja,
Braun:2006zz,Mitter:2014wpa,Braun:2014ata},
but rather serve as a motivation for our scale-fixing procedure below.}
\begin{figure}[t]
\begin{center}
  \includegraphics[width=0.35\linewidth]{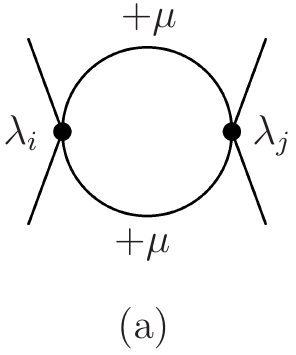}\hspace*{0.7cm}
  \includegraphics[width=0.35\linewidth]{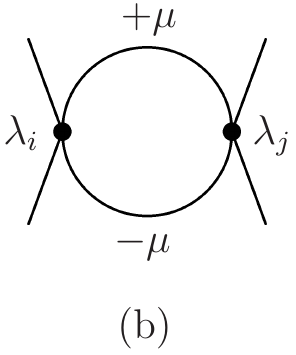}
\end{center}
\caption{The two classes of 1PI diagrams contributing to the RG flow of the four-quark couplings.}
\label{fig:fd}
\end{figure}

The initial condition of the remaining coupling, the scalar-pseudoscalar coupling~$\bar\lambda _{\text{($\sigma $-$\pi $)}}$,
can be fixed in different ways. For example, we may tune it in the vacuum limit 
such that the resulting symmetry breaking scale~$k_{\rm cr}$ leads to a given value for the critical temperature 
at vanishing chemical potential~\cite{Braun:2017srn}. In the following, however, we employ a different procedure which
exploits the mean-field gap equation for the chiral order-parameter field 
in a scalar-pseudoscalar one-channel approximation.
To be more specific, 
performing a {\it Hubbard}-{\it Stratonovich} transformation of the path integral defined
by the action~\eqref{eq:SNJL} and integrating out the quark degrees of freedom subsequently, 
we arrive at a path integral over the auxiliary fields $\sigma \sim (\bar{\psi}\psi)$ 
and~$\vec{\pi}\sim (\bar{\psi}\gamma_5\vec{\tau}\psi)$, respectively. From an evaluation of the latter in the mean-field approximation, we then
obtain the following implicit equation for the 
{constituent quark mass~$\bar{m}_{\rm q}^2=\langle \sigma\rangle^2$}
at~$T=\mu=0$:\footnote{From now on, we identify~$\bar{\lambda}^{(\text{UV})}_{(\sigma\text{-}\pi)}$ 
with the value of the coupling~$\bar{\lambda}_{(\sigma\text{-}\pi)}$ appearing in the classical 
action since the latter determines the value of this coupling at the UV scale~$\Lambda$ in our RG study below.}
\be
\lambda _{\text{($\sigma $-$\pi $)}}^{\ast}{\mathcal J}(0)=\lambda _{\text{($\sigma $-$\pi $)}}^{(\text{UV})}{\mathcal J}(\bar{m}_{\rm q}^2)\,,
\label{eq:gapeq}
\ee
where~$\lambda_{\text{($\sigma $-$\pi $)}}^{\text{UV}}=\Lambda^2\bar{\lambda}_{(\sigma\text{-}\pi)}^{(\text{UV})}$ 
and~$\lambda _{\text{($\sigma $-$\pi $)}}^{\ast}\equiv \lambda _{\text{($\sigma $-$\pi $)}}^{\ast}[r_{\psi}]$ is a dimensionless functional 
of the regularization scheme {since~$\mathcal J$ is} not only a function of~$\bar{m}_{\rm q}$ but also a functional of
the so-called regulator shape function~$r_{\psi}$ specifying the regularization scheme:
\be
\!\!\!\!\! {\mathcal J}(\bar{m}_{\rm q}^2)&=& 8\Nc\int\frac{{\rm d}^4p}{(2\pi)^4}\bigg(\frac{1}{p^2\!+\! \bar{m}_{\rm q}^2} \nn\\
&& \qquad\qquad\quad -\frac{1}{p^2(1+r_{\psi}(\tfrac{p^2}{\Lambda^2}))^2\!+\! \bar{m}_{\rm q}^2}
\bigg)
\,,
\ee
see also Ref.~\cite{Braun:2011pp} for details.
Diagrammatically, this integral 
is associated with a purely fermionic loop integral evaluated at vanishing 
external momenta.\footnote{Loosely speaking, the loop integral
corresponds to the diagram shown in the left panel of Fig.~\ref{fig:fd} with amputated external fermionic legs.}
The parameter~$\Lambda$ may be considered a UV
cutoff scale for the loop-momentum integral. However, from our RG standpoint, 
it should be rather associated with the initial RG scale at which
we fix the initial conditions of the four-quark couplings in our RG study below.

For a given regularization scheme,  the functional~$\lambda _{\text{($\sigma $-$\pi $)}}^{\ast}$ determines the critical value of the four-quark 
coupling above which the ground state is governed by a finite vacuum expectation value~$\langle \sigma\rangle\neq 0$.
{We find
\be
{\lambda _{\text{($\sigma $-$\pi $)}}^{\ast}}=\frac{\Lambda^2}{{\mathcal J}(0)}\,.
\ee
Thus, we} have $\bar{m}_{\rm q} > 0$ for $\lambda_{\text{($\sigma $-$\pi $)}}^{\text{UV}}>\lambda _{\text{($\sigma $-$\pi $)}}^{\ast}$ 
and~$\bar{m}_{\rm q} = 0$ otherwise. 
For example, 
we obtain~${\lambda _{\text{($\sigma $-$\pi $)}}^{\ast}}=2\pi^2/\Nc$ for the four-dimensional sharp cutoff 
often employed in mean-field calculations
and~${\lambda _{\text{($\sigma $-$\pi $)}}^{\ast}}=4\pi^2/\Nc$ for the {\it Litim} regulator~\cite{Litim:2000ci,*Litim:2001up,*Litim:2001fd}.
From here on, however, we shall employ the same scheme as in our studies of the RG flow of  
four-quark couplings to ensure comparability, see App.~\ref{app:RG} for details. 
For this scheme, we find~${\lambda _{\text{($\sigma $-$\pi $)}}^{\ast}}=2\pi^2/\Nc$.
In any case, we deduce from Eq.~\eqref{eq:gapeq} that the actual value of~${\lambda _{\text{($\sigma $-$\pi $)}}^{\ast}}$ 
is of no importance. For a given regularization scheme together with a specific choice for 
the UV scale~$\Lambda$,
the quark mass~$\bar{m}_{\rm q}$ 
only depends on the ``strength"~$\Delta\lambda _{\text{($\sigma $-$\pi $)}}$ of the scalar-pseudoscalar coupling 
relative to its critical value for chiral symmetry breaking:
\be
\Delta\lambda _{\text{($\sigma $-$\pi $)}}=
\frac{\lambda _{\text{($\sigma $-$\pi $)}}^{\text{(UV)}}-\lambda _{\text{($\sigma $-$\pi $)}}^{\ast}}{\lambda _{\text{($\sigma $-$\pi $)}}^{\text{(UV)}}}\,.
\ee
In the following, we shall fix the scale in our studies by setting~$\bar{m}_{\rm q}\approx 0.300\,\text{GeV}$ for the constituent quark 
mass in order to relate our model study to QCD. In terms of the scalar-pseudoscalar coupling, this choice
corresponds to~$\Delta\lambda _{\text{($\sigma $-$\pi $)}}\approx 0.234$ for~$\Lambda/\bar{m}_{\rm q}\approx 10/3$. 

From this discussion it follows immediately that a specific choice for~$\Delta\lambda _{\text{($\sigma $-$\pi $)}}$ also
determines the sign of the curvature~$\bar{m}^2$ of the 
{order-parameter potential~$U$ at} the origin. Indeed, we have
\be
\bar{m}^2:=2\frac{\partial U}{\partial \sigma^2}\Bigg|_{\sigma=0}
=-\Lambda^2\frac{ \Delta \lambda _{\text{($\sigma $-$\pi $)}} }{\lambda_{\text{($\sigma $-$\pi $)}}^{\ast}}\,,
\ee
implying that, at the ``critical point"~$\Delta\lambda _{\text{($\sigma $-$\pi $)}}=0$, the curvature~$\bar{m}^2$ of the order-parameter potential 
changes its sign. From the underlying {\it Hubbard}-{\it Stratonovich} transformation, 
we deduce that 
the renormalized scalar-pseudoscalar coupling~$\bar{\lambda}_{(\sigma\text{-}\pi)}$ is inverse proportional to the curvature~$\bar{m}^2$:
\be
\Lambda^2\bar{\lambda} _{\text{($\sigma $-$\pi $)}}=\frac{\Lambda^2}{\bar{m}^2}
=-\frac{\lambda _{\text{($\sigma $-$\pi $)}}^{\ast}}{ \Delta \lambda _{\text{($\sigma $-$\pi $)}} }\,.
\ee
Thus, the scalar-pseudoscalar four-quark coupling diverges at the ``critical point"~$\Delta\lambda _{\text{($\sigma $-$\pi $)}}=0$.

As discussed on more general grounds in Subsec.~\ref{sec:fw}, these observations regarding the
critical behavior and the formation of a non-trivial ground state can be carried over to studies
of the RG flow of four-quark interactions, even beyond the mean-field limit.
We refer the reader to Ref.~\cite{Braun:2011pp} for a corresponding detailed discussion where
it has also been shown that the value of~$\lambda _{\text{($\sigma $-$\pi $)}}^{\ast}$ in the mean-field 
approximation is indeed nothing but the
value of the non-Gau\ss ian fixed-point of the scalar-pseudoscalar coupling in the large-$\Nc$ limit. 
Thus, this non-Gau\ss ian fixed-point separates a 
regime governed by a trivial ground state from one governed by spontaneous symmetry breaking.

Let us now exploit the relation between the order-parameter potential and the RG flow of  
four-quark couplings to fix the scale in our study of the phase diagram below.
To be specific, we first consider the flow equation for the dimensionless scale-dependent renormalized
scalar-pseudoscalar coupling~$\lambda _{\text{($\sigma $-$\pi $)}}=Z_{(\sigma\text{-}\pi)}k^2\bar\lambda _{\text{($\sigma $-$\pi $)}}/(Z^{\perp})^2$ in
the one-channel approximation at~$T=\mu=0$:
\be
\partial_t  \lambda _{\text{($\sigma $-$\pi $)}} &=& 2 \lambda _{\text{($\sigma $-$\pi $)}} \!-\! 
\frac{\Nc}{\pi^2}\left( 1 + \frac{1}{2\Nc} \right) \lambda _{\text{($\sigma $-$\pi $)}}^2\,.
\label{eq:1cvac}
\ee
Here, we have employed the regularization scheme defined in App.~\ref{app:RG} which is 
identical to the well-known four-dimensional exponential scheme in the {vacuum limit~\cite{Jungnickel:1995fp,Berges:1997eu}, 
see Ref.~\cite{Pawlowski:2005xe} for a detailed discussion of regularization schemes in RG studies.}
The flow equation~\eqref{eq:1cvac} has been extracted from 
the set of equations for the {\it Fierz}-complete basis of four-quark interactions given in App.~\ref{app:fcset} by setting
all couplings but the scalar-pseudoscalar coupling to zero and {also dropping
their flow equations. Note that, because of the {\it Fierz} ambiguity, 
the flow equation~\eqref{eq:1cvac} is ambiguous in the sense that the prefactor of 
the term quadratic in the four-quark
coupling} is not unique. 

The flow equation~\eqref{eq:1cvac} has two fixed points: a Gau\ss ian fixed point and the before-mentioned 
non-Gau\ss ian fixed point~$\lambda _{\text{($\sigma $-$\pi $)}}^{\ast}$,
\be
\lambda _{\text{($\sigma $-$\pi $)}}^{\ast}=\frac{2\pi^2}{\Nc +\frac{1}{2}}\,.
\ee
Thus, the value of the non-Gau\ss ian fixed point indeed agrees with the critical value of the 
scalar-pseudoscalar coupling in the mean-field approximation for~$\Nc\gg 1$. 
Again, with respect to the question of the formation 
of a non-trivial ground state, the actual value of the non-Gau\ss ian fixed point is of no importance, 
only the value of the scalar-pseudoscalar coupling 
at the initial RG scale~$\Lambda$ relative to the value of the non-Gau\ss ian fixed point matters. 
Choosing~$\lambda _{\text{($\sigma $-$\pi $)}}^{(\text{UV})}>\lambda _{\text{($\sigma $-$\pi $)}}^{\ast}$, we find 
that the scalar-pseudoscalar coupling diverges at a finite scale~$k_{\text{cr}}$,
\be
k_{\text{cr}}=\Lambda \left( \Delta\lambda _{\text{($\sigma $-$\pi $)}} \right)^{\frac{1}{2}} 
\theta( \Delta\lambda _{\text{($\sigma $-$\pi $)}} )\,,
\label{eq:kcr}
\ee
indicating the onset of chiral symmetry breaking, i.e. the curvature of the 
order-parameter at the origin changes its sign at this so-called chiral symmetry breaking scale~$k_{\text{cr}}$. 
This scale sets the scale for the (chiral) low-energy observables, such as the constituent quark mass~$\bar{m}_{\rm q}\sim k_{\text{cr}}$. Using
the relation~\eqref{eq:kcr}, we can compute the value of the chiral symmetry breaking scale 
in the mean-field approximation. Using~$\Delta\lambda\approx 0.234$ extracted from the mean-field calculation above
for~$\bar{m}_{\rm q}\approx 0.300\,\text{GeV}$ and~$\Lambda/\bar{m}_{\rm q}\approx 10/3$, we 
obtain~$k_0/\bar{m}_{\rm q}\equiv k_{\text{cr}}/\bar{m}_{\rm q}\approx 1.613$, where~$k_0$ serves as 
a reference scale in the remainder of this work.

At first glance,  
it seems that Eq.~\eqref{eq:kcr} defining~$k_{\text{cr}}$ implies that the low-energy dynamics is independent
of the combinatoric prefactor of the term quadratic in the four-quark coupling in Eq.~\eqref{eq:1cvac}. However, 
this turns out to be too naive. A closer look reveals that the contribution~$\sim 1/\Nc$ is related to quantum corrections to the Yukawa coupling in a 
partially bosonized formulation of our model~\cite{Braun:2008pi,Braun:2011pp}. In a study of the partially bosonized formulation, these corrections
therefore yield $1/\Nc$-corrections to the critical scale~$k_{\text{cr}}$. Moreover, it should be noted that order-parameter fluctuations, 
which are nothing but $1/\Nc$-corrections, tend to restore the chiral 
symmetry in the infrared limit, thereby lowering the value of the 
critical temperature compared to its value in the large-$\Nc$ approximation (see, e.g., Ref.~\cite{Braun:2009si}). 

At the order of the derivative expansion considered in this work, we 
do not have access to low-energy observables such as the constituent quark mass, see  
our discussion in Subsec.~\ref{sec:fw}. Therefore,
we exploit the relation between the chiral order-parameter potential 
in the mean-field approximation and the RG flow of the corresponding scalar-pseudoscalar coupling to 
fix the scale in our calculations. In all our studies of the phase diagram presented below, we shall
set all four-quark couplings to zero at the initial RG scale~$\Lambda$ except for 
the scalar-pseudoscalar coupling~$\lambda _{\text{($\sigma $-$\pi $)}}$. The latter is tuned
at this scale such that, at~$T=\mu=0$, we obtain~$k_{\text{cr}}=k_0$, i.e. the value of the critical scale
is always tuned to agree identically with its value in the mean-field approximation. This ensures comparability 
between the results of our studies from different approximations. Moreover, since~$k_0$ is directly related
{to the constituent} quark mass in the mean-field approximation,~$k_0/\bar{m}_{\rm q}\approx 1.613$, this allows 
at least for
a rough translation of our results for the phase transition temperatures as obtained from, e.g., our {\it Fierz}-complete
set of flow equations into physical units. Of course, such a translation is only approximative. 
We always have to keep in mind that the use of the 
same value for~$k_0$ in different approximations may not necessarily translate into the 
same value for the low-energy observables, such as the constituent quark mass. 
In any case, considering the critical temperature at~$\mu=0$ as an example for an 
low-energy observable being sensitive to the vacuum constituent quark mass and also accessible
within our framework, we find that this quantity does not depend strongly on our approximations
associated with different numbers of four-quark channels. This observation may be 
traced back to the fact that we find the scalar-pseudoscalar channel to be most dominant at~$\mu=0$,
therefore governing the low-energy dynamics in this regime, see our discussion below.

\section{Fixed points and phase structure}\label{sec:fpps}
\subsection{Mean-field and one-channel approximation}\label{sec:1c}

Let us now study the phase diagram in the plane spanned by the temperature and quark chemical potential
for an approximation in which we only take into account the scalar-pseudoscalar four-quark coupling.
The flow equation of the latter reads:
\be
\partial_t  \lambda _{\text{($\sigma $-$\pi $)}} &=& 2 \lambda _{\text{($\sigma $-$\pi $)}} \!-\! 
64 v_4 \left(2 \Nc\!+\! 1\right)\! \lambda _{\text{($\sigma $-$\pi $)}}^2\!\!
   \left(l_{\text{$\parallel $+}}^{\text{(F)}}\left(\tau ,0,-i \tilde\mu _{\tau }\right) \right.\nn\\
    &&\qquad\qquad\qquad\qquad +  \left. l_{\text{$\bot$+}}^{\text{(F)}}\left(\tau ,0,-i \tilde\mu _{\tau }\right)\right)\,,
\label{eq:1cftmu}
\ee
where~$v_4=1/(32\pi^2)$, $\tau=T/k$, and~$\tilde\mu_{\tau}= \mu/(2\pi T)=\mu/(2\pi k\tau)$.
Here and in the following, we do not take into account the renormalization of the 
chemical potential and set~$Z_{\mu} = 1$.
The definitions of the so-called threshold functions also mentioned in Subsec.~\ref{sec:fw}
can be found in Ref.~\cite{Braun:2017srn}. The ones appearing in Eq.~\eqref{eq:1cftmu} 
are associated with the loop integral depicted in the left panel of Fig.~\ref{fig:fd}.
As done for Eq.~\eqref{eq:1cvac}, 
we have extracted the flow equation~\eqref{eq:1cftmu} from 
the set of equations for the {\it Fierz}-complete basis of four-quark interactions given in App.~\ref{app:fcset} by setting
all but the scalar-pseudoscalar coupling to zero and also dropping
their RG flow equations. 
In the limit~$T\to 0$ and~$\mu\to 0$, we therefore recover 
Eq.~\eqref{eq:1cvac} from Eq.~\eqref{eq:1cftmu} since
\be
\left( l_{\text{$\parallel $+}}^{\text{(F)}}\left(\tau ,0,-i \tilde\mu _{\tau }\right)+ l_{\text{$\bot$+}}^{\text{(F)}}\left(\tau ,0,-i \tilde\mu _{\tau }\right)
\right) \to \frac{1}{4}
\ee
for~$T\to 0$ and~$\mu\to 0$, see Ref.~\cite{Braun:2017srn}.

The flow equation~\eqref{eq:1cftmu} can be solved analytically, even at finite temperature and quark
chemical potential~\cite{Braun:2017srn}. The solution can then be employed to compute the critical
temperature~$T_{\rm cr}=T_{\rm cr}(\mu)$ as a function of the quark chemical potential~$\mu$.
The latter is defined as the temperature at which the 
scalar-pseudoscalar four-quark coupling still diverges at~$k\to 0$: 
\be
\lim_{k\to 0} \frac{1}{\lambda _{\text{($\sigma $-$\pi $)}}(T_{\text{cr}},\mu,k)}=0\,,
\label{eq:tcdef}
\ee
i.e. it is defined as the highest temperature for which the four-quark coupling still diverges.
For our studies with more than one channel, this definition can be generalized straightforwardly. The critical
temperature is then defined 
{to be the temperature at} which the four-quark couplings {diverge. Note that}
a divergence in one channel at a scale~$k_{\text{cr}}(T,\mu)$ entails corresponding 
divergences in all the other channels at the same scale. However, the associated
four-quark couplings in general have a different strength relative to each other, see, 
e.g., Subsec.~\ref{sec:sbfc} below and also Ref.~\cite{Braun:2017srn} for a detailed discussion. 

With the definition~\eqref{eq:tcdef}, we obtain the following implicit equation for the critical {temperature $T_{\rm cr}$:
\be
 k_{0} = \Lambda\left(1 + 8\,{\mathcal I}(T_{\rm cr}, \mu , 0)\right)^{\frac{1}{2}}\,.
 \label{eq:Tcana}
\ee
For} convenience, we have introduced the following dimensionless auxiliary function:
\be
{\mathcal I}(T,\mu,k) &=& \frac{1}{\Lambda^2} 
\int_\Lambda^k {\rm d} k^{\prime} k^{\prime}
\left( l_{\text{$\parallel $+}}^{\text{(F)}}\left(\tau^{\prime},0,-i \tilde\mu _{\tau^{\prime} }\right) \right. \nn\\ 
&& \left. \qquad\qquad\quad + l_{\text{$\bot$+}}^{\text{(F)}}\left(\tau^{\prime},0,-i \tilde\mu _{\tau^{\prime} }\right)
\right)\,,
\label{eq:Idef1c}
\ee
where $\tau^{\prime}=T/k^{\prime}$. 

Apparently, the critical temperature~$T_{\text{cr}}$ depends on our choice for the 
UV scale~$\Lambda$ as well as the
scale~$k_0$, i.e. eventually on the constituent quark mass in the vacuum limit.
Recall that the scale~$k_0$ is directly related to the initial condition for the scalar-pseudoscalar
four-quark coupling which we keep fixed to the same value for all 
temperatures and chemical potentials. In the following, we shall therefore measure all physical observables 
in units of~$k_0$.
\begin{figure}[t]
\begin{center}
  \includegraphics[width=1\linewidth]{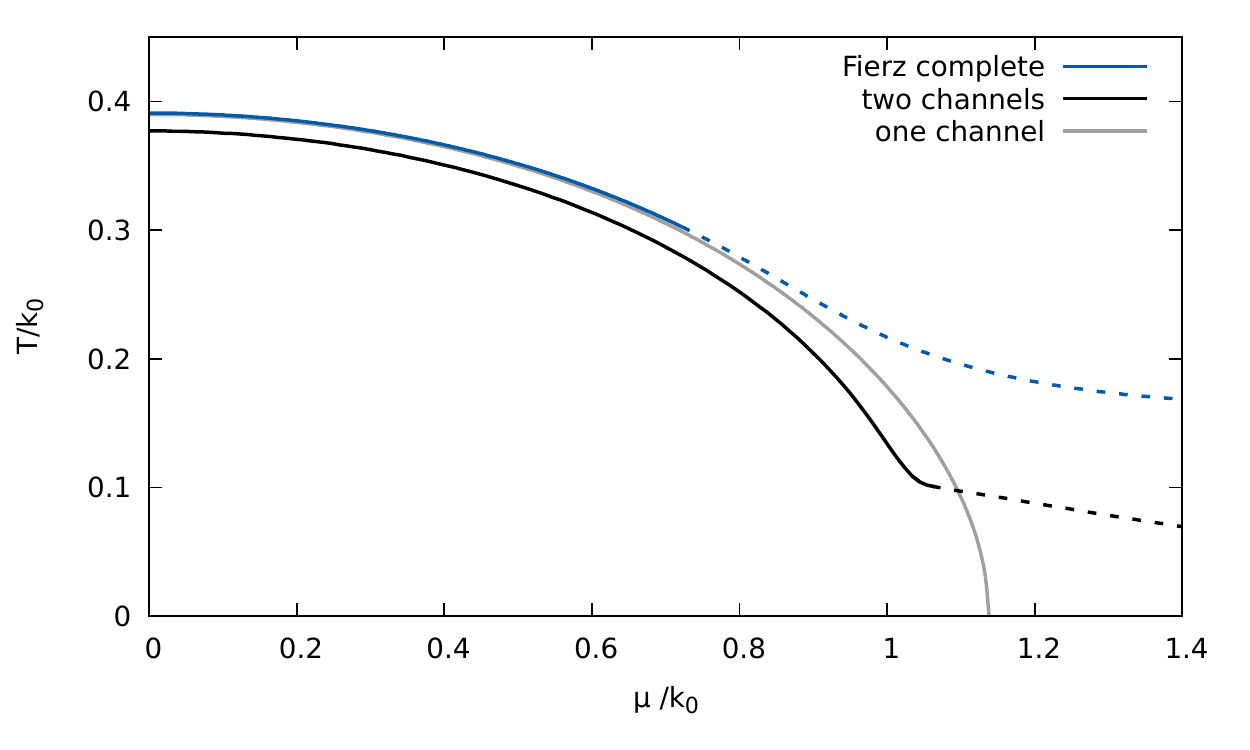}
\end{center}
\caption{(color online) Phase boundary associated with the spontaneous breakdown of at least one of the fundamental symmetries 
of our NJL-type model as obtained from a one-channel, two-channel, and {\it Fierz}-complete study of the ansatz~\eqref{eq:GLO}, 
see main text for details.}
\label{fig:pd}
\end{figure} 

In Fig.~\ref{fig:pd}, we show the critical temperature as a function of the quark chemical potential for
the one-channel approximation (gray line) as obtained
from a solution of Eq.~\eqref{eq:Tcana}. For~$\mu=0$, we obtain~$T_{\text{cr}}/k_0\approx 0.390$ ($T_{\text{cr}}\approx 0.190\,\text{GeV}$). For
increasing~$\mu$, the critical temperature then decreases monotonously and 
eventually vanishes at~$\mu/k_0=\mu_{\text{cr}}/k_0\approx 1.14$ ($\mu_{\text{cr}} \approx 0.552\,\text{GeV}$).

We emphasize again that our definition of the critical temperature is associated with a sign change of the chiral order-parameter potential 
at the origin. In our present approximation, our result for~$T_{\text{cr}}(\mu)$ therefore only describes the phase boundary in case 
of a second-order transition. Our criterion is not sensitive to a first-order transition. However, it {allows 
to detect the line of} metastability separating a regime associated with a negative curvature of the order-parameter potential at the origin (e.g. 
at low temperature and small quark chemical potential) from a regime where the curvature changes its sign but the true ground state is still 
assumed for a finite expectation value of the order-parameter field. Such lines of metastability usually emerge in the vicinity of 
a first-order transition. In particular, for a given temperature, 
the chemical potential associated with the  
emergence of a metastable state at the origin of the potential is less or equal than the chemical potential of the associated first-order transition.

It is instructive to compare the results for the phase boundary from our RG study with those 
obtained from a solution of the mean-field gap equation~\eqref{eq:gapeq}.\footnote{The gap equation for finite~$T$ and~$\mu$ is obtained
from Eq.~\eqref{eq:gapeq} by replacing~$\int {\rm d}^4 p/(2\pi)^4$ with $T\sum_n\int {\rm d}^3 p/(2\pi)^3$. Moreover, we have to 
replace~$p^2$ with $(\nu_n^2 - {\rm i}\mu)^2 +\vec{p}^{\,2}$,
where~$\nu_n=(2n+1)\pi T$, except 
in the argument of the regulator shape function due to our conventions.}
To ensure comparability, we employ of course the same regularization scheme as in our RG study. 
We then find that the phase transition line from our RG study agrees identically with the 
second-order phase 
transition line of the mean-field study up to a first-order endpoint at~$(\mu/k_0,T/k_0)\approx(0.951, 0.207)$. As expected, beyond this point, 
the phase transition line obtained from our RG study agrees identically with the 
line of metastability in the mean-field phase diagram.
The comparatively large extent of the phase boundary in $\mu$-direction 
can be traced back to the comparatively large $\sigma$-meson 
mass~$\bar{m}_{\sigma}/\bar{m}_{\rm q}\approx 2.67$ ($\bar{m}_{\sigma}\approx 0.800\,\text{GeV}$) found
in the vacuum limit of our mean-field calculation 
for the employed set of parameters, i.e.~$\Delta\lambda _{\text{($\sigma $-$\pi $)}}$ 
and~$\Lambda$.\footnote{The 
computation of the $\sigma$-mass requires fixing the Yukawa coupling~$\bar{h}$ 
since $m_q = \langle \sigma \rangle = \bar{h}f_{\pi}$. 
Here, we use~$\bar{h}\approx 3.45$ corresponding to~$f_{\pi}\approx 87.0\,\text{MeV}$ for the pion decay constant.}
In fact, it has already been found in previous mean-field calculations that the critical point separating a first-order phase 
transition line from a second-order phase transition line can be shifted continuously to larger values of the quark chemical potential
by increasing the mass of the~$\sigma$-meson~\cite{Schaefer:2008hk}. Even more, it can even be made disappear, leaving us with
only a second-order transition line.
Note that this highlights the strong scheme dependence 
as the $\sigma$-mass {can be tuned by suitable variations of the} 
constituent quark mass and the UV cutoff~$\Lambda$. Since the actual
value of the latter should always be viewed against the background of the employed regularization scheme, 
the scheme unavoidably belongs to the definition of the model, at least in four {\it Euclidean} space-time dimensions.
Of course, we could also use smaller values
for~$\Lambda$ which would lead to a smaller mass of the~$\sigma$-meson. However, this 
then leads to strong ``cutoff effects" as both the temperatures as well as the quark chemical potentials considered in this work would then 
be of the order of the UV scale~$\Lambda$. In order to at least suppress such unwanted effects, we have 
chosen~$\Lambda/k_0 \approx 2.07$. 
\begin{figure}[t]
\begin{center}
  \includegraphics[width=1\linewidth]{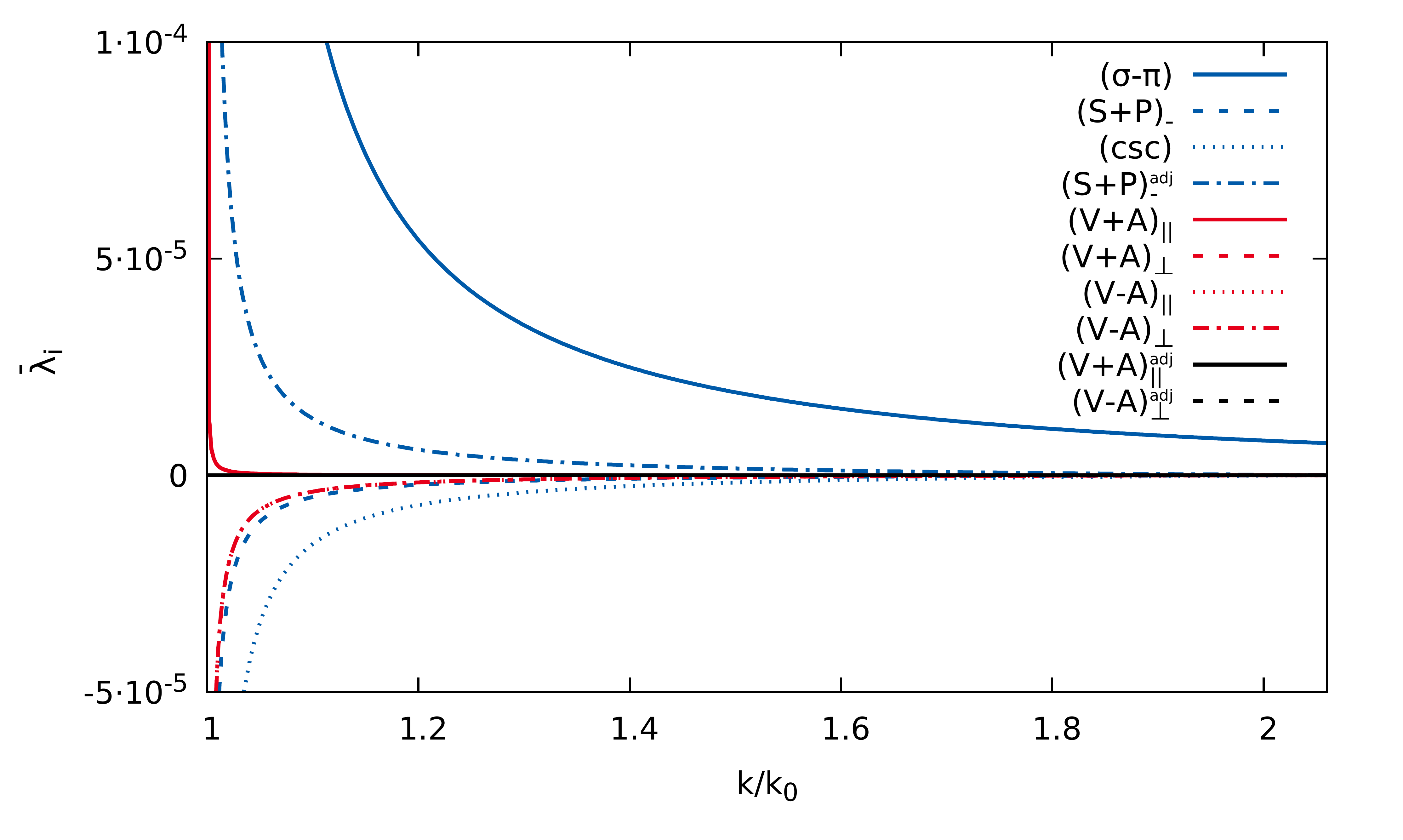}
\end{center}
\caption{(color online) {Renormalized (dimensionful) four-quark} couplings as a function of the RG scale~$k$ at~$T=0$ and~$\mu=0$ as obtained
from our {\it Fierz}-complete study. Note that the {\it Fierz}-complete basis is 
effectively composed of only six channels in the vacuum limit since~$\mathcal C$ invariance 
is intact and  the
Euclidean time direction is not distinguished. In particular, 
we have~$\bar{\lambda}_{\VpAParAdj}\equiv 0\equiv \bar{\lambda}_{\VmAPerAdj}$ in this limit.}
\label{fig:rgflowvac}
\end{figure}

\vfill

\subsection{Symmetry breaking patterns and {\it Fierz} completeness}\label{sec:sbfc}
Let us now analyze {the phase diagram as} obtained from an RG flow study of the {\it Fierz}-complete
set of four-quark interactions, see App.~\ref{app:fcset} for the RG flow equations. Such an analysis
goes well beyond studies in the mean-field limit. Indeed, mean-field studies of NJL-type low-energy 
models have been found to exhibit a residual ambiguity related to the possibility to perform {\it Fierz} transformations, even
if a {\it Fierz}-complete set of four-quark interactions is taken into account~\cite{Jaeckel:2002rm}, see {also Ref.~\cite{Gies:2006wv}
for an introduction.}
Results from 
mean-field calculations therefore potentially depend on an unphysical parameter which reflects the actual choice of the 
mean field in the various channels.

As discussed above, we fix the scale in our {\it Fierz}-complete studies by setting all but the scalar-pseudoscalar coupling 
to zero at the initial RG scale~$\Lambda$. The latter is tuned at~$T=\mu=0$ such that the critical scale~$k_0$ associated with diverging four-quark couplings
agrees identically with its counterpart in the mean-field calculation. For our calculations at finite temperature and/or quark chemical potential, 
we then use the same set of initial conditions as in the vacuum limit, i.e. at~$T=\mu=0$.
The scale dependence of the (dimensionful) renormalized
four-quark couplings at~$T=\mu=0$ is shown in Fig.~\ref{fig:rgflowvac}. We observe that the dynamics of the theory in this case is clearly dominated
by the scalar-pseudoscalar interaction channel; the modulus of all other couplings is at least one order of magnitude smaller than the modulus 
of the scalar-pseudoscalar coupling. The dominance of this channel may indicate that the ground state in the vacuum limit is governed by chiral symmetry breaking.
However, we emphasize again that such an analysis based on the strength of four-quark interactions has to be taken with some care: It neither rules out the possible formation of 
other condensates associated with subdominant channels nor proves the formation of a condensate associated with the most dominant channel.
Such an analysis can only yield indications for the actual structure of the ground state, see Ref.~\cite{Braun:2017srn} for a detailed discussion.
\begin{figure*}[t]
\begin{center}
  \includegraphics[width=0.47\linewidth]{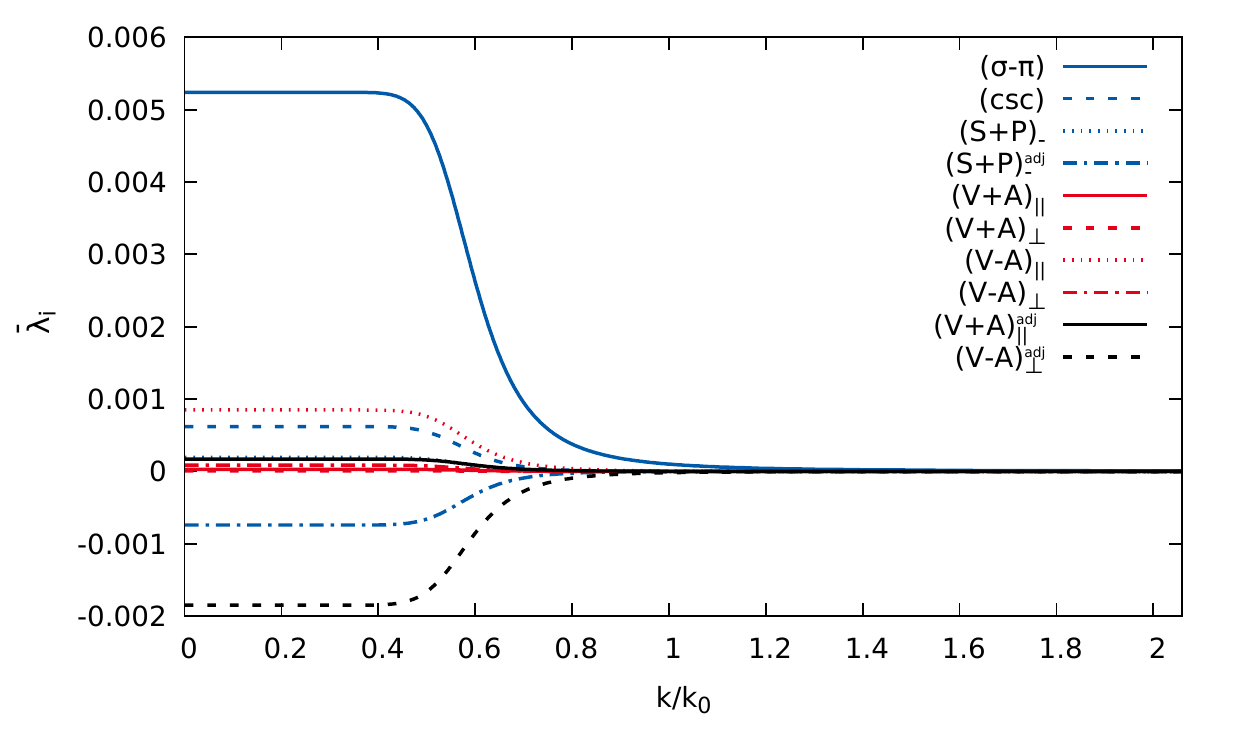}
 \includegraphics[width=0.47\linewidth]{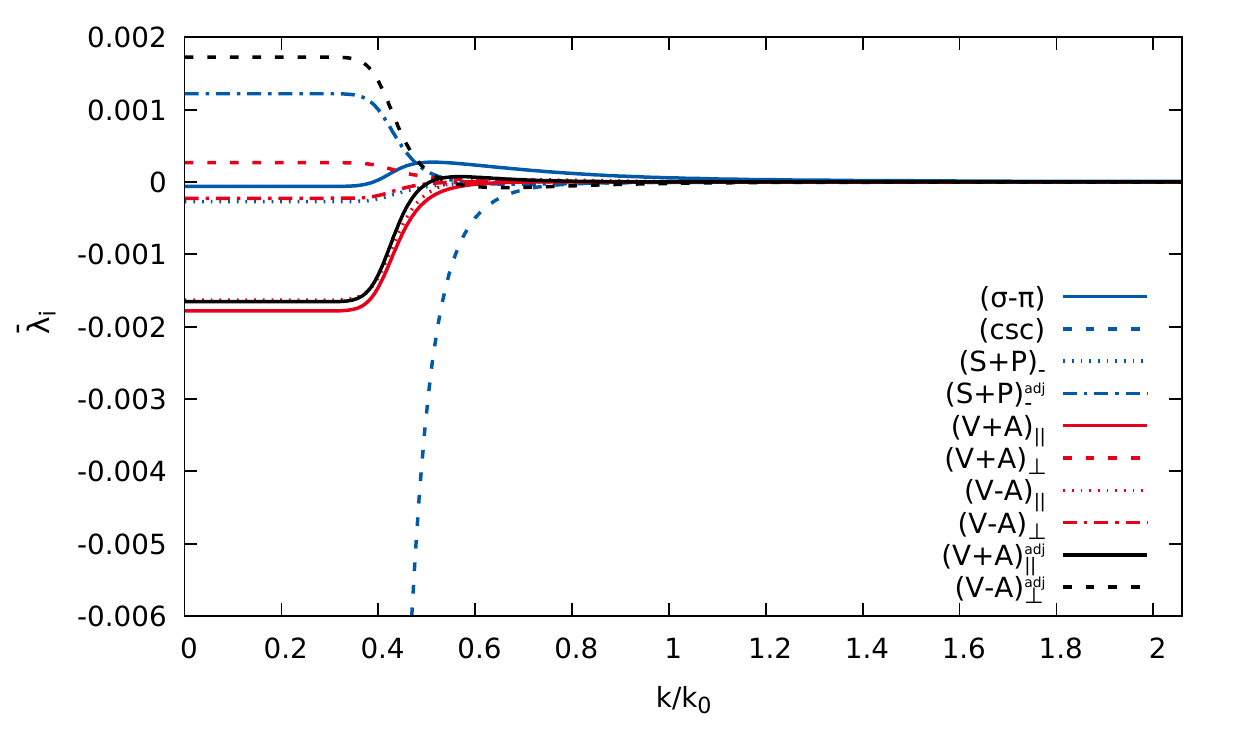}\\
\end{center}
\caption{(Color online) 
Scale dependence of the various {renormalized (dimensionful) four-quark} couplings at~$\mu=0$ 
and~$T/k_0 \simeq T_{\text{cr}}(\mu=0)/k_0 \approx  0.391$ (left panel) as well as at~$\mu/k_0\approx 1.1$ 
and~$T/k_0 \simeq T_{\text{cr}}(\mu)/k_0 \approx  0.196$ (right panel). 
}
\label{fig:rgflowmufq}
\end{figure*}

In the vacuum limit, the observation of the dominance of the scalar-pseudoscalar channel may be considered trivial as it 
may be exclusively triggered by our choice 
for the initial conditions. 
Increasing now the temperature at vanishing quark chemical potential, we still observe a dominance of the scalar-pseudoscalar channel which persists 
even up to high temperatures beyond the critical temperature~$T_{\text{cr}}(\mu=0)/k_0\approx 0.391$. This is illustrated in 
the left panel of Fig.~\ref{fig:rgflowmufq} 
where the scale dependence of the various four-quark couplings is shown for~$T\simeq T_{\text{cr}}(\mu\!=\! 0)$ at~$\mu=0$.
At least in units of~$k_0$, it also appears that the critical temperature at~$\mu=0$ in our {\it Fierz}-complete study agrees very well with the one
from the one-channel approximation. However, we note that 
this could be misleading as choosing the same value for~$k_0$ in our {\it Fierz}-complete study and in our one-channel
approximation may not necessarily lead to the same values of the low-energy observables
(e.g. the constituent quark mass),
although the flow in the vacuum limit is also strongly dominated by the scalar-pseudoscalar channel in the {\it Fierz}-complete analysis.
Thus, direct quantitative comparisons of the results from our various different approximations should be taken with care.
Still, we expect that qualitative comparisons are meaningful.

Following now the critical temperature~$T_{\text{cr}}$ as a function of~$\mu$ starting from~$\mu=0$, we find that the scalar-pseudoscalar channel 
continues to dominate the dynamics {up to~$\mu/k_0=\mu_{\chi}/k_0 \approx 0.734$}, as depicted by the blue solid line in Fig.~\ref{fig:pd}.
In this regime, we also observe that the phase transition temperatures 
from our one-channel approximation agree almost 
identically with those from the {\it Fierz}-complete study, at least in units of the vacuum symmetry breaking 
scale~$k_0$.\footnote{Note that~$\mu_{\chi}/k_0 \approx 0.734$ roughly corresponds to~$\mu_{\chi}/\bar{m}_{\rm q}\approx 1.18$ 
in our mean-field approximation, where~$\bar{m}_{\rm q}\approx 0.3\,\text{GeV}$.} 
At first glance, this may come as a surprise. We shall therefore 
analyze this observation in detail in Subsec.~\ref{sec:ln} below.
Following the phase transition line beyond the point associated with the 
quark chemical potential~$\mu_{\chi}$, we find that the dynamics is now clearly and exclusively 
dominated by the CSC (color superconducting) channel associated 
with the emergence of a diquark condensate~$\delta^{a}$ {and a corresponding gap}
in the quark propagator, see blue dashed line in Fig.~\ref{fig:pd}. Exemplary, this change in the ``hierarchy" of the channels 
is illustrated in the right panel of Fig.~\ref{fig:rgflowmufq} where the scale dependence of the various four-quark couplings is shown for~$\mu/k_0\approx 1.1$ 
and~$T/k_0 \gtrsim T_{\text{cr}}(\mu)/k_0 \approx 0.196$. We emphasize that this change in the ``hierarchy" of the channels is non-trivial as it is fully
triggered by the dynamics of the system when the quark chemical potential is increased. There is no fine-tuning of, e.g., the CSC coupling involved. 
{Recall that we use {identical initial conditions 
in the vacuum limit as well as at finite temperature and/or
quark chemical} potential.}

\subsection{$U_{\rm A}(1)$ symmetry}\label{sec:ua1}
Our choice for the initial conditions of the RG flow equations explicitly breaks the~$U_{\rm A}(1)$ symmetry since
we only choose the coupling~$\lambda _{\text{($\sigma $-$\pi $)}}$ to be finite and set all the other 
four-quark couplings to zero at the initial RG scale~$\Lambda$. 
By using the sum rules~\eqref{eq:SUA1} and~\eqref{eq:SUA2}, we can now
study the fate of the (broken)~$U_{\rm A}(1)$ symmetry at finite temperature and quark chemical potential 
when quantum fluctuations are taken into account. To be specific, in  
case of~$U_{\rm A}(1)$-violating initial conditions, we consider the 
following two dimensionless quantities to ``measure" the 
strength of the explicit~$U_{\rm A}(1)$ symmetry breaking:
\be
R_i &=& \mathcal{N} \left| \mathcal{S}_{U_{\rm A}(1)}^{(i)} \right|\,.
\label{eq:Rdef}
\ee
The normalization~$ \mathcal{N}$ is chosen to be independent of the index~$i$ and
is implicitly determined by
\be
1 =\left( R_1 + R_2 \right)\big|_{k=\Lambda}\,.
\ee
Thus, the auxiliary quantities defined in Eq.~\eqref{eq:Rdef}
essentially measure the strength of~$U_{\rm A}(1)$ symmetry breaking relative to its strength at the initial RG scale~$\Lambda$.
In case of a~$U_{\rm A}(1)$-symmetric theory
defined by a suitable choice for the initial conditions, 
we find that the couplings fulfill the sum rules~\eqref{eq:SUA1} and~\eqref{eq:SUA2} 
at {\it all} scales~$k$ greater than the symmetry breaking scale, as it should be. Therefore, there is no need at all to 
consider the auxiliary quantities defined in Eq.~\eqref{eq:Rdef} in such a scenario.
\begin{figure*}[t]
\begin{center}
  \includegraphics[width=0.47\linewidth]{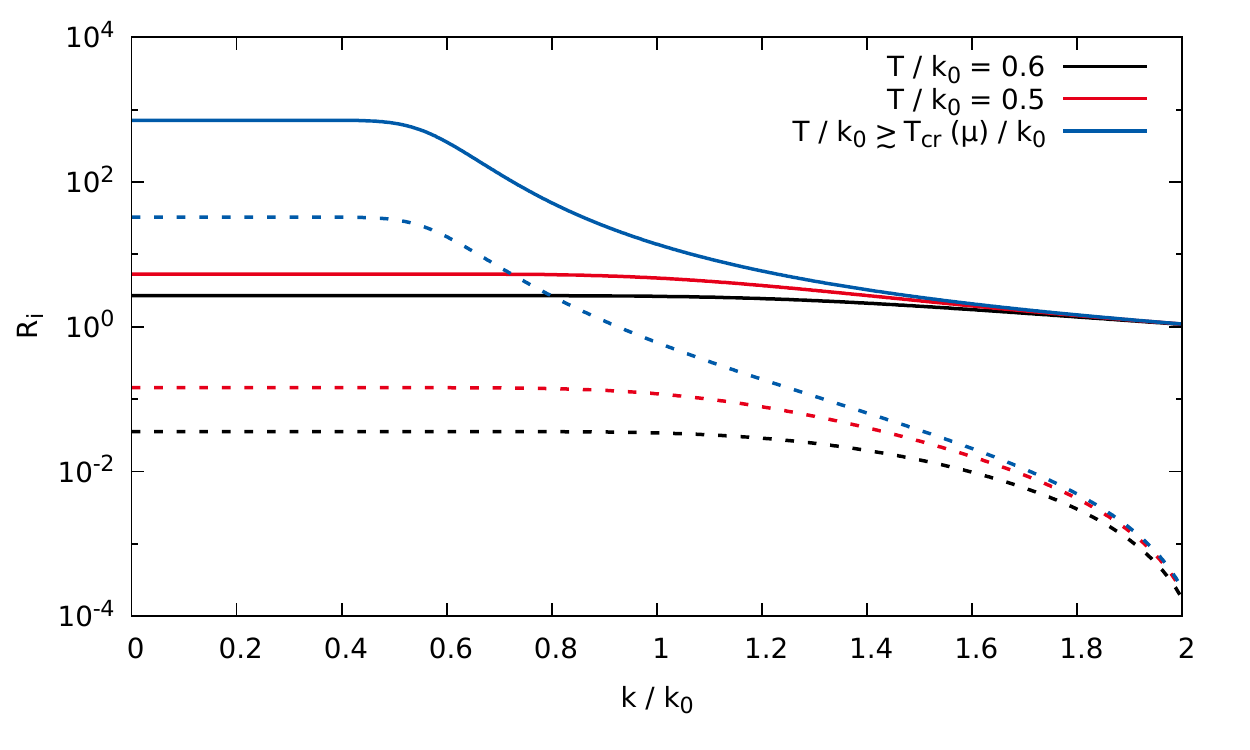}
 \includegraphics[width=0.47\linewidth]{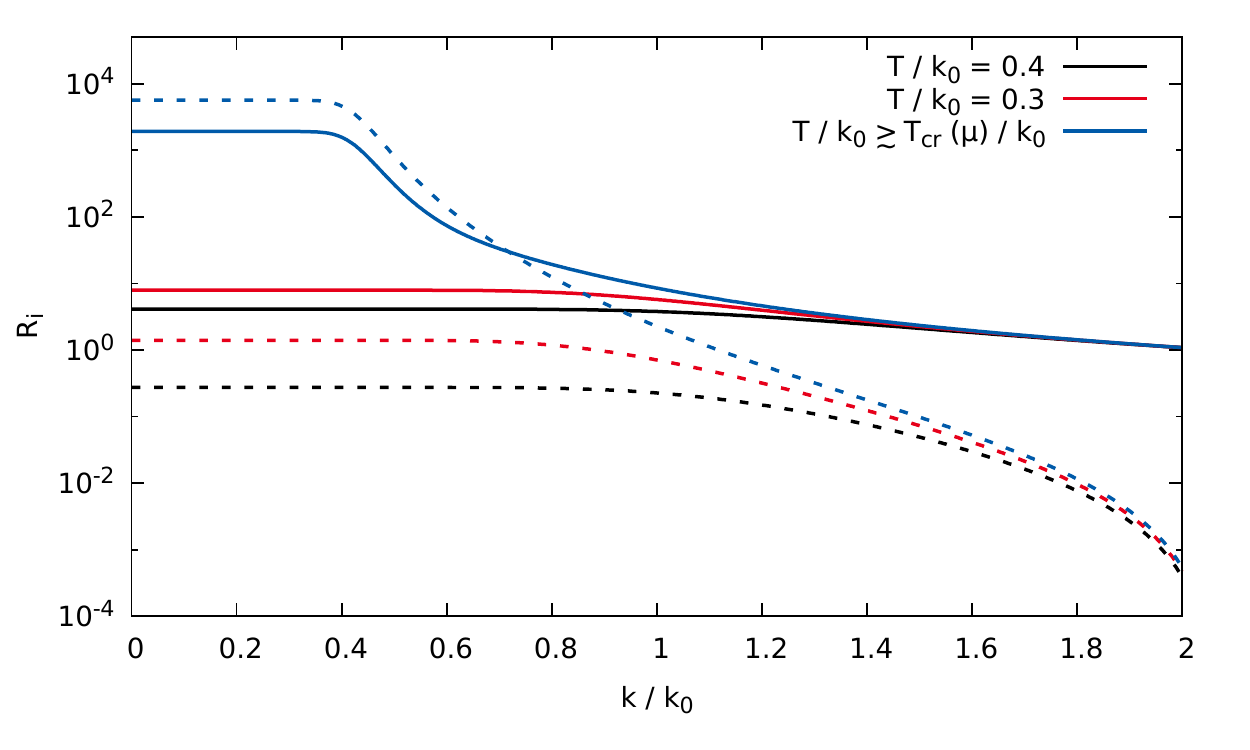}
\end{center}
\caption{(Color online) Scale dependence of the explicit~$U_{\rm A}(1)$ breaking as measured by the 
functions~$R_1$ (dashed lines) and~$R_2$ (solid lines)
at~$\mu=0$ (left panel) and at~$\mu/k_0\approx 1.1$ (right panel)
for three values of the temperature for each of the two cases.}
\label{fig:rgflowmusr}
\end{figure*}

In Fig.~\ref{fig:rgflowmusr}, we show the scale dependence of~$R_1$ and~$R_2$
for two values of the quark chemical potential, $\mu=0$ and~$\mu/k_0\approx 1.1$, 
and three values of the temperature for each of the two cases {as obtained for
our~$U_{\rm A}(1)$-violating initial conditions.}
Let us first note that, for all values of~$\mu$ considered in this work, we observe that~$U_{\rm A}(1)$ breaking
as ``measured" by our sum rules in form of~$R_1$ and~$R_2$ is continuously softened 
when the temperature is increased. More specifically, at~$\mu=0$, for example, we already find that 
the strength of~$U_{\rm A}(1)$ breaking remains on the level of its strength at the initial 
scale~$\Lambda$ for temperatures~$T/T_{\text{c}}\gtrsim 2$, i.e. its strength 
remains on the level as present
in the classical action in this temperature regime. 
A qualitatively similar behavior can also be observed at finite chemical potential
when the temperature is increased,
see Fig.~\ref{fig:rgflowmusr}.
Indeed, the strength of~$U_{\rm A}(1)$ symmetry breaking is controlled by the strength of the 
four-quark couplings. Quantum corrections to the latter are thermally suppressed
at high temperature due to the {presence of a thermal mass of the fermions}. This explains our 
observations at high temperature.

Conversely, approaching the phase transition from above for a given 
value of the chemical potential~$\mu$, 
we find that the violation of the~$U_{\rm A}(1)$ symmetry becomes continuously stronger, in the sense 
that the functions~$R_1$ and~$R_2$ 
{start to increase, eventually deviating {significantly 
from their values} at the initial RG scale. Thus, quantum} corrections to the four-quark couplings
appear to amplify~$U_{\rm A}(1)$ symmetry breaking when the phase governed by spontaneous
symmetry breaking is approached from above, provided that~$U_{\rm A}(1)$ symmetry breaking is 
explicitly broken at the initial RG scale. 

In accordance with our observation that the scalar-pseudoscalar channel is most dominant at 
small chemical potential (see, e.g., left panel of Fig.~\ref{fig:rgflowmufq}), 
we also note that~$R_2 \gg R_1$ in this part of the phase diagram, 
see left panel of Fig.~\ref{fig:rgflowmusr}.
For~$\mu\gtrsim \mu_{\chi}$, {the CSC channel then dominates the dynamics 
and~thus $R_1$ and $R_2$ are of the same order of magnitude as both depend on the 
CSC coupling, see} the right panels of Figs.~\ref{fig:rgflowmufq} 
and~\ref{fig:rgflowmusr}. Thus, our results suggest that the dynamically generated violation of
the~$U_{\rm A}(1)$ symmetry is driven by the dynamics of the pions at small 
chemical potential whereas it is driven by the
dynamics of diquark degrees of freedom associated with the CSC 
channel at large chemical potential.

Let us finally compare the phase diagram obtained from our {\it Fierz}-complete study
employing $U_{\rm A}(1)$-symmetry violating initial conditions 
with the one obtained from a manifestly~$U_{\rm A}(1)$-symmetric
{\it Fierz}-complete study. The latter has been calculated by tuning the 
couplings~$\bar{\lambda} _{\text{($\sigma $-$\pi $)}}$ and~$\bar{\lambda}_\SpPm$ at the initial 
RG scale such
that the sum rule~\eqref{eq:SUA2} is fulfilled and the same value for 
the symmetry scale~$k_0$ as in the case 
with $U_{\rm A}(1)$-symmetry violating initial conditions is obtained in the vacuum limit.
If we choose the initial conditions in this way, i.e. such that they respect the~$U_{\rm A}(1)$ symmetry,
then this symmetry remains intact in the RG flow for all values of the RG scale, at least 
for those values of the temperature and the quark chemical potential for which the four-quark 
couplings remain finite on all scales~$k\leq\Lambda$. For values of the temperature 
and the quark chemical potential at which the four-quark couplings diverge at a finite scale~$k_{\text{cr}}(T,\mu)$,
the sum rules~\eqref{eq:SUA1} and~\eqref{eq:SUA2} are only fulfilled for~$k> k_{\text{cr}}(T,\mu)$. Below the 
symmetry breaking scale, the $U_{\rm A}(1)$ symmetry may potentially be broken spontaneously, e.g. alongside
with the chiral symmetry. However, this cannot be resolved in our present study.

As already discussed above, a quantitative comparison of the results from our~$U_{\rm A}(1)$-symmetric calculation 
with the ones from our explicitly~$U_{\rm A}(1)$-violating calculation is generically difficult and
has to be taken with some care. Leaving this concern aside for a moment, we observe that the phase transition lines 
from both studies agree almost identically on the scale of the plot. 
Only for larger values of the 
quark chemical potential, we find that the two phase transition lines start to deviate from each other.
In particular, we observe that, along the phase transition line, a 
chemical potential~$\mu_{\chi}/k_0$ associated with a change in the hierarchy 
of the four-quark couplings exists 
in both cases. Even the corresponding values of~$\mu_{\chi}/k_0$  
agree almost identically. Even more, for~$\mu < \mu_{\chi}$, we find that the 
scalar-pseudoscalar channel dominates the dynamics of the theory in both cases, see 
solid lines in Fig.~\ref{fig:pdsym}. 
For~$\mu > \mu_{\chi}$, 
the dynamics is then dominated by the CSC channel in case of $U_{\rm A}(1)$-symmetry violating
initial conditions, see blue dashed line in Fig.~\ref{fig:pdsym}. 
In case of our~$U_{\rm A}(1)$-symmetric study, 
however, we have a dominance of 
the $\VpAParAdj$-channel in this regime as depicted by the red dashed-dotted line 
in Fig.~\ref{fig:pdsym}; see Eq.~\eqref{Eq:VpAParAdj} for the definition of this channel. The condensate associated with
this channel also breaks the color symmetry of our theory. In mean-field studies (see, e.g., Ref.~\cite{Buballa:2003qv} for a review), the 
appearance of a corresponding condensate has also been discussed. However, its generation has been found to be induced
by a simultaneous formation of a color-symmetry breaking diquark condensate. 
In accordance with this, we observe
that the four most dominant channels along the phase transition line for~$\mu>\mu_{\chi}$ are the 
$\VpAParAdj$-, $(S+P)_{-}^\mathrm{adj}$-, CSC- and~$\VmAPerAdj$-channel 
in our present study. These channels
are all associated with the formation of a color-symmetry breaking condensate. 
{Apart from the $(V\pm A)_{\parallel}$-channels, color-singlet}
channels are found to be subdominant in this 
part of the phase diagram. 
The observed difference in the dominance pattern at large chemical potential 
in the $U_{\rm A}(1)$-symmetric and~$U_{\rm A}(1)$-violating calculation
may point to 
the importance of explicit~$U_{\rm A}(1)$ breaking for the formation of the 
conventional CSC ground state at intermediate and large values of the 
chemical potential as discussed in early {seminal works on 
color superconductivity, see, e.g., Refs.~\cite{Alford:1997zt,Rapp:1997zu,Berges:1998rc,Son:1998uk,%
*Pisarski:1999tv,*Pisarski:1999bf,*Schafer:1999jg,*Brown:1999aq,*Hong:1999fh,*Evans:1999at}.} 

By simply looking at the shape of the phase boundary, one may be tempted to conclude
that $U_{\rm A}(1)$ breaking does not strongly affect the position of the phase transition line.
However, this may be a too bold statement at this point as the same value of~$k_0$ in the two studies
potentially corresponds to different values of the low-energy observables and therefore renders a 
direct quantitative comparison difficult, see our discussion above. 
In any case, the apparent insensitivity 
of the phase transition line under a ``transition" from~$U_{\rm A}(1)$-symmetry violating 
initial conditions to~$U_{\rm A}(1)$-symmetric initial conditions (while keeping the vacuum symmetry
breaking scale fixed) is still an interesting observation. 
At least at small values of the chemical potential,
the latter can in principle be understood from an 
analysis of the large-$\Nc$ limit which we shall consider next.

\subsection{Large-$\Nc$ limit}\label{sec:ln}

In order to better understand the phase structure at small chemical potential, we now analyze our RG 
flow equations in the large-$\Nc$ limit, i.e. we only take into account the leading order of the right-hand sides
of our flow equations in an expansion in powers of~$\Nc$. For the scalar-pseudoscalar coupling, 
for example, we then obtain
 \begin{widetext}
 \begingroup
\allowdisplaybreaks[3]
\begin{align*}
 &\partial_t  \lambda _{\text{($\sigma $-$\pi $)}} = 2 \lambda _{\text{($\sigma $-$\pi $)}}+32\Nc 
 v_4 \Big(-4 \lambda _{\text{($\sigma $-$\pi $)}}^2- 8 \lambda _{\text{($\sigma
    $-$\pi $)}} \lambda _{(S+P)_-}- 8 \lambda _{(S+P)_-}^2- 2 \lambda
    _{\text{($\sigma $-$\pi $)}} \lambda _{(S+P)_-^{\text{adj}}} \\
   &\qquad\qquad\qquad\qquad\qquad\qquad\quad  - 4 \lambda _{(S+P)_-} \lambda
    _{(S+P)_-^{\text{adj}}}+ \lambda _{\text{($\sigma $-$\pi $)}} \lambda _{( V+A )_{\parallel
    }^{\text{adj}}}+ 2 \lambda _{\text{($\sigma $-$\pi $)}} \lambda _{\text{csc}} \\
   & \qquad\qquad\qquad\qquad\qquad\qquad\quad\quad + 4 \lambda
    _{(S+P)_-} \lambda _{\text{csc}}+ 2 \lambda _{(S+P)_-^{\text{adj}}} \lambda _{\text{csc}}\Big)
    l_{\text{$\parallel $+}}^{\text{(F)}}\left(\tau ,0,-i \tilde{\mu}_{\tau}\right) \\
    & \qquad\qquad\qquad\qquad +16\Nc v_4 \Big(- 8 \lambda _{\text{($\sigma $-$\pi $)}}^2- 16 \lambda _{\text{($\sigma
    $-$\pi $)}} \lambda _{(S+P)_-}- 16 \lambda _{(S+P)_-}^2- 4 \lambda
    _{\text{($\sigma $-$\pi $)}} \lambda _{(S+P)_-^{\text{adj}}} \\ 
   &\qquad\qquad\qquad\qquad\qquad\qquad\quad  - 8 \lambda _{(S+P)_-} \lambda
    _{(S+P)_-^{\text{adj}}}- \frac{4}{3} \lambda _{(S+P)_-^{\text{adj}}}^2+ 2 \lambda
    _{\text{($\sigma $-$\pi $)}} \lambda _{( V+A )_{\parallel }^{\text{adj}}} - \frac{1}{3}\lambda _{( V+A
    )_{\parallel }^{\text{adj}}}^2 \nn \\ 
    & \qquad\qquad\qquad\qquad\qquad\qquad\quad\quad   +4 \lambda _{\text{($\sigma $-$\pi $)}} \lambda
    _{\text{csc}}+ 8 \lambda _{(S+P)_-} \lambda _{\text{csc}}+ \frac{4}{3} \lambda _{(S+P)_-^{\text{adj}}}
    \lambda _{\text{csc}}- \frac{4}{3} \lambda _{\text{csc}}^2\Big) l_{\text{$\bot
    $+}}^{\text{(F)}}\left(\tau ,0,-i \tilde{\mu}_{\tau}\right)\,.
    \numberthis\label{eq:largeN1}
 \end{align*}
\endgroup
 \end{widetext}
For the remaining nine four-quark couplings, we find that the right-hand sides of their flow equations
do not contain terms quadratic in the scalar-pseudoscalar coupling~$\lambda _{\text{($\sigma $-$\pi $)}}$ 
but {at most terms} linear in~$\lambda _{\text{($\sigma $-$\pi $)}}$ in the large-$\Nc$ limit.
At first glance, this may not appear noteworthy. However, by setting all four-quark couplings but the 
scalar-pseudoscalar coupling to zero on the right-hand sides of the flow equations, we therefore observe
that only the right-hand side of the flow equation of the scalar-pseudoscalar coupling remains finite. Indeed, from
Eq.~\eqref{eq:largeN1}, we deduce that
\be
\partial_t  \lambda _{\text{($\sigma $-$\pi $)}} &=& 
2 \lambda _{\text{($\sigma $-$\pi $)}} - 128\Nc 
 v_4  \lambda _{\text{($\sigma $-$\pi $)}}^2  \Big( l_{\text{$\parallel $+}}^{\text{(F)}}\left(\tau ,0,-i \tilde{\mu}_{\tau}\right) 
 \nn\\
&&  \qquad\qquad\qquad\qquad +\, l_{\text{$\bot
    $+}}^{\text{(F)}}\left(\tau ,0,-i \tilde{\mu}_{\tau}\right) \Big)\,.
\ee
Note that this flow equation is identical {to Eq.~\eqref{eq:1cftmu} in the large-$\Nc$ limit.}

The right-hand sides of the flow equations of the remaining 
nine couplings are identical to zero when we set all four-quark couplings but
the scalar-pseudoscalar coupling to zero in the large-$\Nc$ limit. 
Thus, we have found a non-trivial fixed point of the RG flow 
at
\be
 \lambda _{\text{($\sigma $-$\pi $)}}^{\ast}=\frac{2\pi^2}{\Nc}\quad\text{and}\quad
  \lambda _{j}^{\ast}=0\,, 
  \label{eq:njlfp}
\ee
which ``sits" on the pure scalar-pseudoscalar axis of our ten-dimensional space spanned by the four-quark couplings.
Here, we {have $j\in {\mathcal B}$ but $j\neq \text{($\sigma $-$\pi $)}$ 
and~${\mathcal B}$ denotes} the set of indices associated with our {\it Fierz}-complete basis of four-quark interactions. 

The fixed point~\eqref{eq:njlfp} has only one IR repulsive direction, namely the one
associated with the scalar-pseudoscalar axis. The remaining nine directions are all IR attractive. This observation already 
suggests that the scalar-pseudoscalar channel dominates the low-energy dynamics, provided that we initiate the RG 
flow sufficiently close to this fixed point.\footnote{We do not aim at a precise determination of the size of the 
associated domain of attraction.}
We add that, the dynamics of our {\it Fierz}-complete system is governed by~$2^{10}=1024$ fixed points. 
Depending on the temperature and the quark chemical potential, some of these fixed points appear in complex-conjugated
pairs as we shall discuss in Subsec.~\ref{sec:2c}.
\begin{figure}[t]
\begin{center}
  \includegraphics[width=1\linewidth]{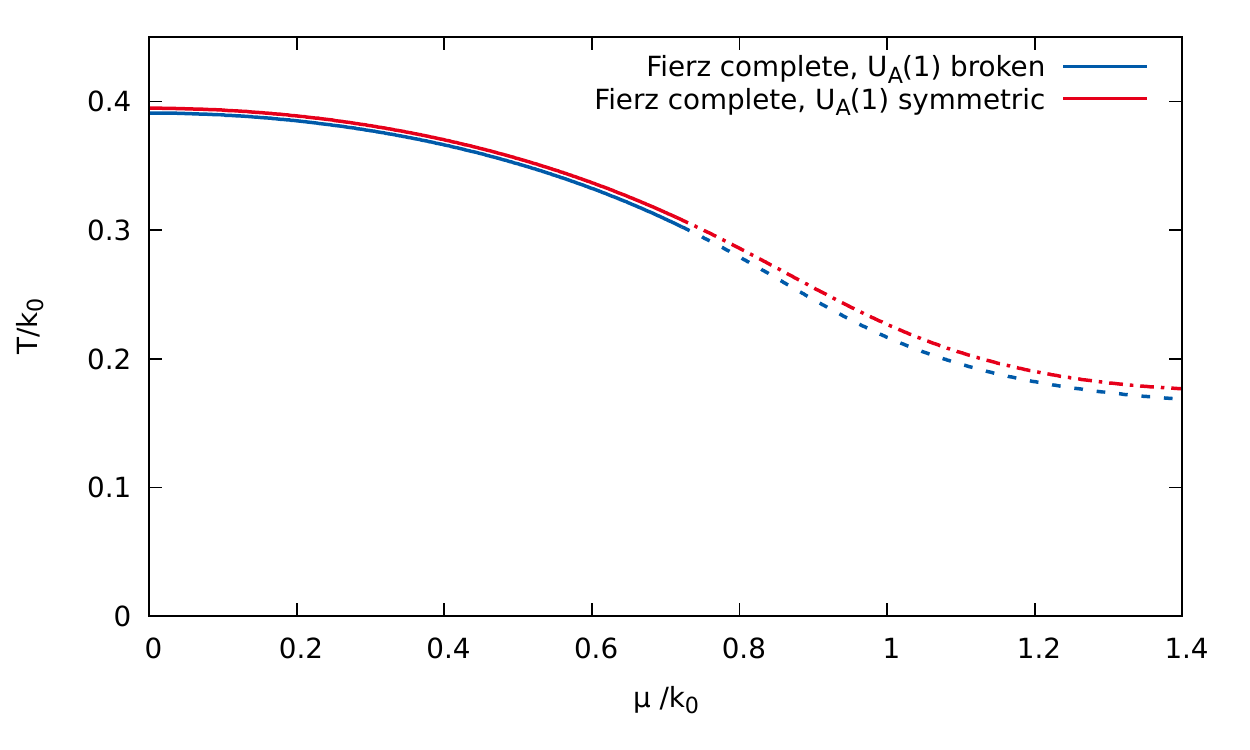}
\end{center}
\caption{(color online) 
 Phase boundary associated with the spontaneous breakdown of at least one of the fundamental symmetries 
of our NJL-type model as obtained from a 
manifestly~$U_{\rm A}(1)$-symmetric {\it Fierz}-complete study of the 
ansatz~\eqref{eq:GLO} (red lines) and from a {\it Fierz}-complete study with 
broken~$U_{\rm A}(1)$ {symmetry (blue lines)}, 
see main text for details.}
\label{fig:pdsym}
\end{figure}

We emphasize that, for any finite value of~$\Nc$, we do not find an interacting fixed point on the 
pure scalar-pseudoscalar axis.\footnote{The scalar-pseudoscalar axis may be viewed as the axis associated with conventional NJL model 
studies taking into account only this channel.} In fact,
not only the flow equation of the scalar-pseudoscalar coupling contains terms 
proportional to the square of the scalar-pseudoscalar coupling but they also appear in the flow 
equations of other four-quark couplings. These terms 
now dynamically generate interactions 
in channels other than the scalar-pseudoscalar channel, 
pushing the fixed point~\eqref{eq:njlfp} away from the scalar-pseudoscalar axis. 
We add that the very same behavior 
has also been observed in the vacuum limit of 
the~$U_{\rm A}(1)$-symmetric NJL model in the large-$\Nc$ limit~\cite{Braun:2011pp} 
and the three-dimensional {\it Thirring} model in the 
large-$\Nf$ limit~\cite{Gies:2010st}.

The existence 
of the fixed point~\eqref{eq:njlfp} and its properties provides us with an explanation of the phase structure 
at small quark chemical potential. First of all, from the standpoint of 
model studies, the existence of this fixed point in the large-$\Nc$ limit
implies that the system always remains on the 
scalar-pseudoscalar axis, provided that we only choose a finite initial value for the scalar-pseudoscalar
coupling and set all the other couplings to zero. Thus, the other 
channels do not contribute at all. Given the scale-fixing procedure underlying our
calculations, it then follows that the phase boundary found in the scalar-pseudoscalar one-channel
approximation agrees identically with the one from the {\it Fierz}-complete 
study in the large-$\Nc$ limit.\footnote{Note that, strictly speaking, invariance under 
{\it Fierz} transformations is violated in the large-$\Nc$ limit.}

Beyond the large-$\Nc$ limit, the fixed point~\eqref{eq:njlfp} is pushed away from the scalar-pseudoscalar axis and now
all four-quark interactions are generated dynamically even if only the scalar-pseudoscalar coupling
is chosen to be finite at the initial RG scale, see, e.g., Fig.~\ref{fig:rgflowvac} and 
also our discussion in Subsec.~\ref{sec:2c} below. 
However, the observed agreement of the results 
for the phase boundary from the one-channel and the {\it Fierz}-complete study
{suggests that this fixed} point still controls the dynamics of the theory at small quark chemical potential, see Fig.~\ref{fig:pd}.
Even non-universal quantities such as the curvature of the phase boundary at~$\mu=0$ appear to be independent
of the inclusion of the dynamics described by the channels other than the scalar-pseudoscalar channel. Only for 
large values of the quark chemical potential, $\mu > \mu_{\chi}$, the influence of the other channels becomes
significant. Recall that, in the mean-field approximation,~$\mu_{\chi}$ is of the order of the vacuum constituent quark mass.

A word of caution needs to be added to this intriguing observation: If we choose initial conditions such that not only the 
scalar-pseudoscalar coupling is finite at the initial RG scale
but also other four-quark couplings, then the RG flow 
{may be potentially controlled} by a different 
interacting fixed point, even at small chemical potential. As a consequence, the phase boundary in this regime may become more sensitive to the 
dynamics described by the full set of four-quark interactions. For example, one may choose a $U_{\rm A}(1)$-symmetric starting point
of the RG flow by tuning the couplings~$\bar{\lambda} _{\text{($\sigma $-$\pi $)}}$ and~$\bar{\lambda}_\SpPm$ such
that the sum rule~\eqref{eq:SUA2} is fulfilled. However, even in this case, we observe that, at small~$\mu$, the phase boundary obtained 
from a {\it Fierz}-complete $U_{\rm A}(1)$-symmetric study agrees very well with the one from our one-channel 
approximation as well as with the one from our {\it Fierz}-complete study taking $U_{\rm A}(1)$-symmetry breaking into account, 
see, e.g., Fig.~\ref{fig:pdsym}.

\subsection{Symmetry breaking mechanisms}\label{sec:2c}

Let us finally analyze the mechanisms underlying the phase structure at large chemical potential where corrections 
beyond the large-$\Nc$ approximation become important. Looking at the modulus of the four-quark couplings 
depicted in Fig.~\ref{fig:rgflowmufq}, we observe that the scalar-pseudoscalar coupling and the CSC coupling 
are the two most dominant couplings in the range of quark chemical potentials studied in this work, at least close to 
and above the phase transition line. For an analysis of the symmetry breaking mechanisms, it therefore appears reasonable
to consider an approximation which only includes the scalar-pseudoscalar coupling and the CSC coupling. The remaining 
eight couplings and their flows are set to zero. The flow equations of such a two-channel approximation then {read
\begin{widetext}
{\allowdisplaybreaks
\be
\partial_t  \lambda _{\text{($\sigma $-$\pi $)}} &=& 2 \lambda _{\text{($\sigma $-$\pi $)}} + 64 v_4 \left(-(2\Nc + 1)\lambda
   _{\text{($\sigma $-$\pi $)}}^2 + (\Nc+1)\lambda _{\text{($\sigma $-$\pi $)}}  \lambda _{\text{csc}}\right)
   l_{\text{$\parallel $+}}^{\text{(F)}}\left(\tau ,0,-i \mu _{\tau }\right)\nn\\
   &&\qquad +64 v_4
   \Big( -(2 \Nc+1) \lambda _{\text{($\sigma
   $-$\pi $)}}^2 
   +\frac{1}{3} (3\Nc - 1)\lambda _{\text{($\sigma $-$\pi $)}} \lambda _{\text{csc}} 
   -\frac{1}{3} (\Nc - 2) \lambda _{\text{csc}}^2\Big) l_{\text{$\bot
   $+}}^{\text{(F)}}\left(\tau ,0,-i \mu _{\tau }\right)\,,
   \label{eq:2casp} \\
\partial_t  \lambda _{\text{csc}} &=& 2 \lambda _{\text{csc}} + 64 v_4 \left(- \lambda _{\text{($\sigma $-$\pi $)}}^2 + (\Nc -2)
   \lambda _{\text{csc}}^2\right) l_{\text{$\parallel $+}}^{\text{(F)}}\left(\tau ,0,-i \mu
   _{\tau }\right)\nn\\
   &&\hspace*{1.5cm}+64 v_4 \left(-\lambda _{\text{($\sigma $-$\pi $)}}^2-2 \lambda
   _{\text{($\sigma $-$\pi $)}} \lambda _{\text{csc}}+4 \lambda _{\text{csc}}^2\right)
   l_{\parallel \pm }^{\text{(F)}}\left(\tau ,0,-i \mu _{\tau }\right)\nn\\
   &&\hspace*{1.5cm}\hspace*{1.5cm}+64 \lambda
   _{\text{($\sigma $-$\pi $)}}^2 v_4 l_{\text{$\bot $+}}^{\text{(F)}}\left(\tau ,0,-i
   \mu _{\tau }\right)\nn\\
   &&\hspace*{1.5cm}\hspace*{1.5cm}\hspace*{1.5cm}+64 v_4 \left( \lambda _{\text{($\sigma $-$\pi $)}}^2-2
   \lambda _{\text{($\sigma $-$\pi $)}} \lambda _{\text{csc}}+4 \lambda _{\text{csc}}^2\right)
   l_{\bot \pm }^{\text{(F)}}\left(\tau ,0,-i \mu _{\tau }\right)\,.
   \label{eq:2cacsc}
\ee
}
\end{widetext}
The} initial conditions are chosen as in our {\it Fierz}-complete study, i.e. we set the CSC coupling 
to zero at the initial RG scale and only tune the scalar-pseudoscalar coupling such that the value for 
the symmetry breaking scale in the vacuum limit is identical to its value in the {\it Fierz}-complete 
study,~$k_0=k_{\text{cr}}(T=0,\mu=0)$,
which, in turn, is identical to the value of the critical scale in our mean-field approximation.
From {the set of the two} flow equations~\eqref{eq:2casp} and~\eqref{eq:2cacsc}, 
we immediately deduce that the CSC coupling is dynamically generated in the RG flow 
although we set it to zero at the initial RG scale.

An asset of our two-channel approximation is that it allows for a comparatively simple but still 
detailed analysis of the RG flow of our system and its fixed-point structure. 
Of course, such an analysis is also possible 
for more than two couplings but it then clearly becomes more involved. In any case, in our two-channel approximation,
we have four fixed points~${\mathcal F}_j=(\lambda _{\text{($\sigma $-$\pi $)},j}^{\ast},\lambda _{\text{csc},j}^{\ast})$ 
in total. At~$T=0$,~$\mu=0$ and~$\Nc=3$, their coordinates {are
\be
&&{\mathcal F}_1\big|_{\Nc=3}=(0,0)\,,\\
&&{\mathcal F}_2\big|_{\Nc=3} \approx (5.165, -1.088)\,, \label{eq:F2}\\
&&{\mathcal F}_3\big|_{\Nc=3}\approx (1.262 - {\rm i}1.567,-8.728 - {\rm i}0.841)\,,\\
&&{\mathcal F}_4\big|_{\Nc=3}\approx (1.262 + {\rm i}1.567,-8.728 + {\rm i}0.841)\,, 
\ee
where~${\mathcal F}_1$ is} the $\Nc$-independent Gau\ss ian fixed-point with two IR attractive directions.
Apparently, ${\mathcal F}_3$ and~${\mathcal F}_4$ form a pair of complex conjugate fixed points. 
The coordinates of the non-Gau\ss ian fixed points up to order~$1/\Nc^2$ in a large $\Nc$-expansion {read
{\allowdisplaybreaks
\be
{\mathcal F}_2(\Nc)&=&\left(\frac{2\pi^2}{\Nc} - \frac{3\pi^2}{2\Nc^2},-\frac{\pi^2}{\Nc^2}\right)\,,\nn\\
{\mathcal F}_3(\Nc)&=& \bigg(-\frac{(3+{\rm i}\sqrt{23})\pi^2}{\Nc} \!+\! \frac{({235\sqrt{10}} + {\rm i}{6 \sqrt{{11481}}} )\pi^2}{4\sqrt{10} \Nc^2}, \nn\\ 
&& \qquad\qquad\qquad -\frac{16\pi^2}{\Nc}\!+\!\frac{(393 - {\rm i}13 \sqrt{23})\pi^2}{2\Nc^2}\bigg)\,,\nn\\
{\mathcal F}_4(\Nc)&=&\bigg(-\frac{(3 - {\rm i}\sqrt{23})\pi^2}{\Nc}\! +\! \frac{({235\sqrt{10}} + {\rm i}{6 \sqrt{{11481}}} )\pi^2}{4\sqrt{10} \Nc^2}, \nn\\ 
&& \qquad\qquad\qquad -\frac{16\pi^2}{\Nc}\!+\!\frac{(393 + {\rm i}13 \sqrt{23})\pi^2}{2\Nc^2}\bigg)\,.\nn
\ee
These} expansions have been} {extracted from the full analytic $\Nc$-dependent expressions 
for the coordinates of the fixed points. 
We observe} {that the suitably $\Nc$-rescaled fixed point~$\Nc\cdot{\mathcal F}_2$ 
is} shifted onto the scalar-pseudoscalar axis for~$\Nc\to\infty$. Moreover, we find that this fixed point has 
one IR repulsive and one IR attractive direction. Thus, 
this fixed point corresponds to the fixed point~\eqref{eq:njlfp} in 
the full {\it Fierz}-complete set of RG flow equations. 
\begin{figure}[t]
\begin{center}
  \includegraphics[width=0.95\linewidth]{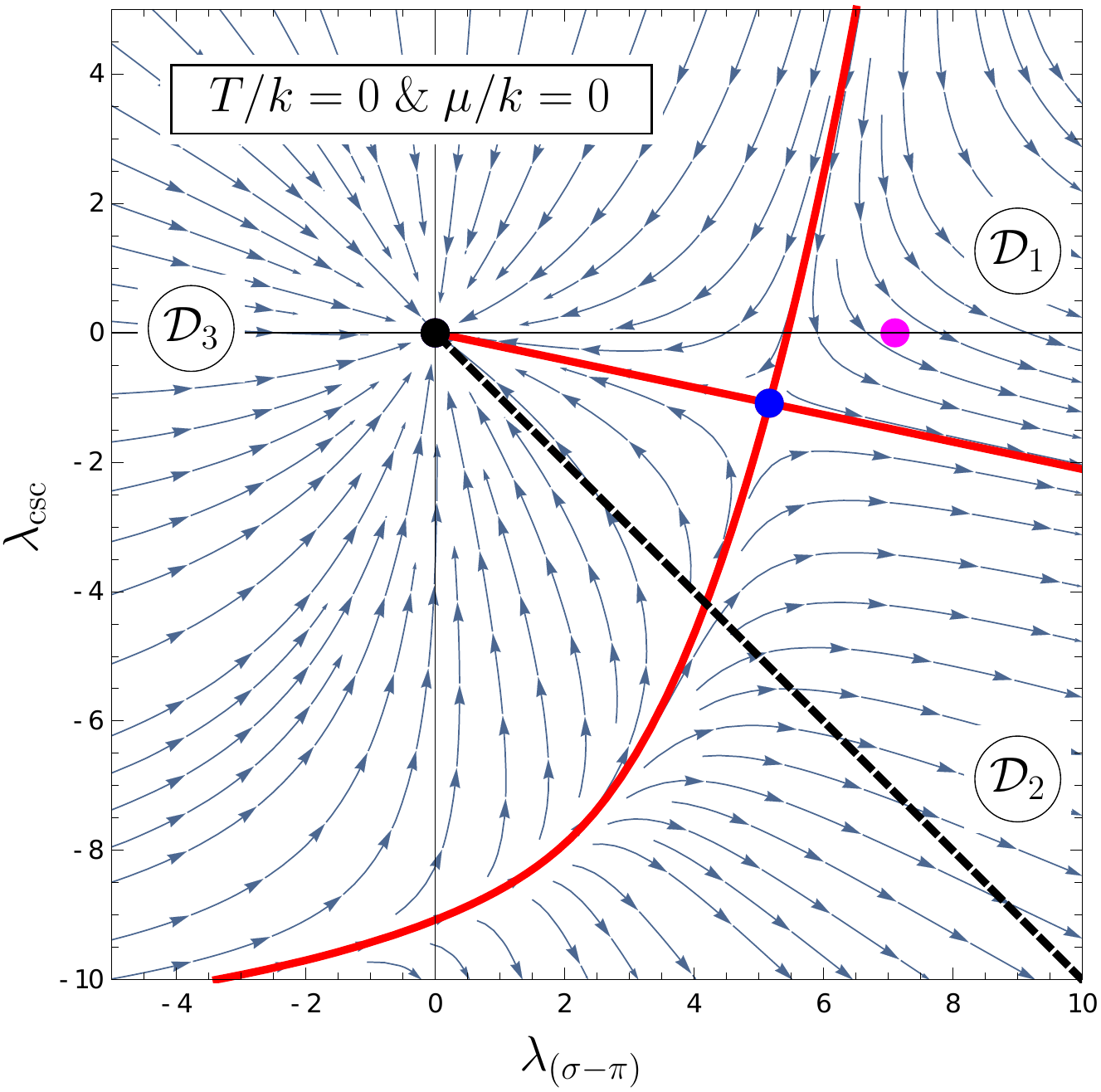}
\end{center}
\caption{(Color online) RG flow of the two-channel approximation at zero temperature and chemical potential in the plane 
spanned by the scalar-pseudoscalar coupling  and the CSC coupling. The black dot represents the Gau\ss ian fixed point 
whereas the blue dot represents the real-valued non-Gau\ss ian fixed-point, see Eq.~\eqref{eq:F2}. The pink dot depicts our choice 
for the initial condition. The RG trajectory starting at this point describes 
four-quark couplings diverging at a finite scale~$k_0=k_{\text{cr}}$, while approaching a separatrix (red solid line) as indicated by the arrows. 
The dominance of the scalar-pseudoscalar channel is illustrated by the position of the 
corresponding separatrix relative to the bisectrix (dashed back line). 
The different domains separated by the separatrices (red solid lines) are labelled~${\mathcal D}_1$,~${\mathcal D}_2$, and~${\mathcal D}_3$.}
\label{fig:rgflow00}
\end{figure}

In the following, we shall not consider the large-$\Nc$ limit any further. In order to have spontaneous 
symmetry breaking in the IR limit, we then choose the initial {condition of the scalar-pseudoscalar coupling to be} 
greater than~$\lambda _{\text{($\sigma $-$\pi $)},2}^{\ast}$ but still set the initial value of the CSC coupling to zero, see our 
discussion above. As a consequence, we also find for this two-channel approximation that the low-energy dynamics
is {dominated by the scalar-pseudoscalar channel. The} RG flow of this two-channel approximation is depicted in 
Fig.~\ref{fig:rgflow00}.
\begin{figure}[t]
\begin{center}
  \includegraphics[width=0.95\linewidth]{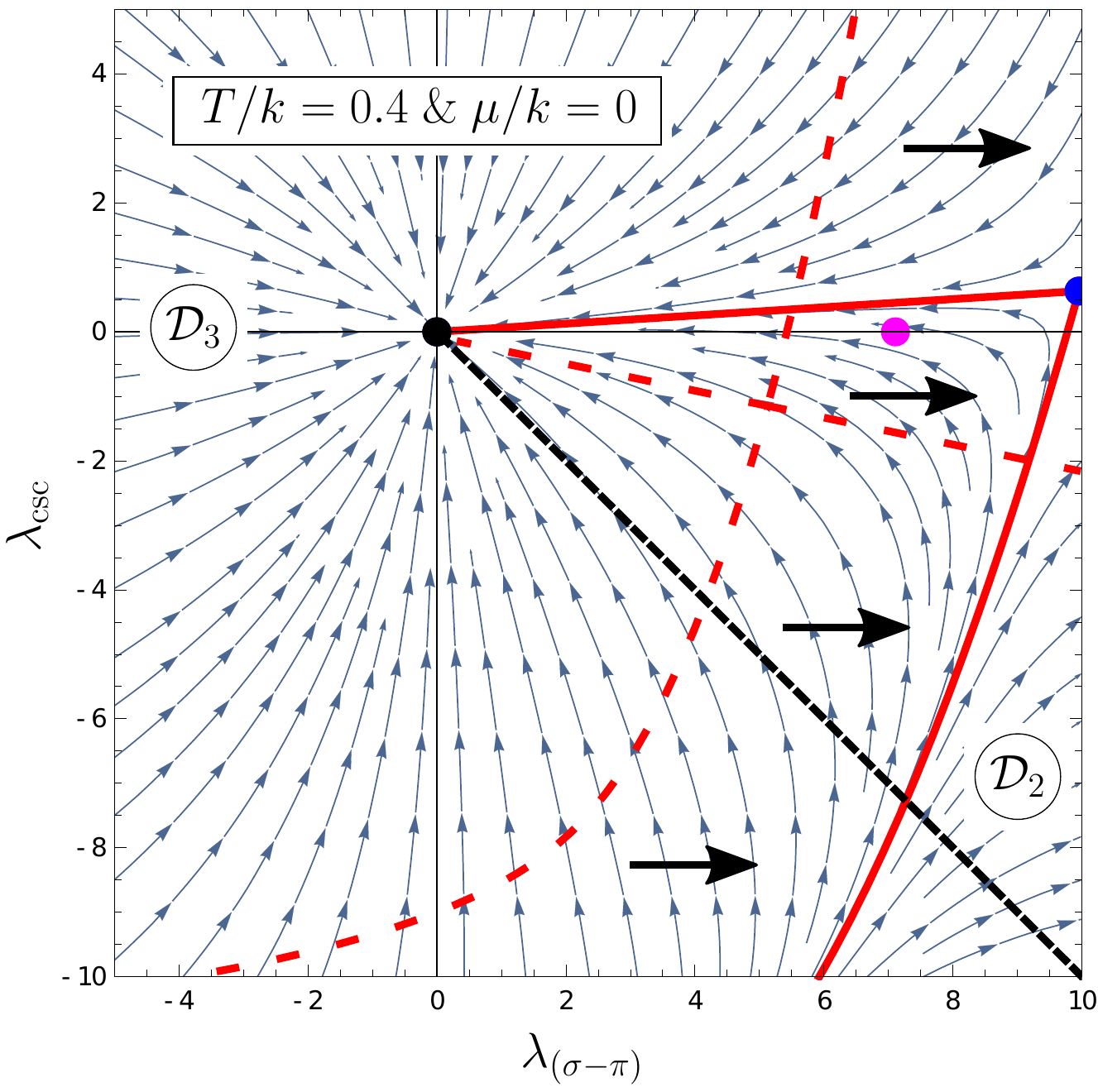}
\end{center}
\caption{(Color online) RG flow of the two-channel approximation at~$T/k=0.4$ and~$\mu=0$ in the plane 
spanned by the scalar-pseudoscalar coupling  and the CSC coupling. The black dot represents the Gau\ss ian fixed point. 
The blue dot represents the real-valued non-Gau\ss ian fixed-point. The pink dot depicts our choice 
for the initial condition. The dashed black line is the bisectrix of the bottom right quadrant. 
The different domains separated by the separatrices (red solid lines) 
are labelled~${\mathcal D}_2$, and~${\mathcal D}_3$, see also Fig.~\ref{fig:rgflow00}. ${\mathcal D}_1$ is not shown.
The black arrows indicate the shift of the real-valued non-Gau\ss ian fixed point together with the domains~${\mathcal D}_1$ 
and~${\mathcal D}_2$ when~$T/k$ is increased, see main text for details. The dashed red lines 
depict the position of the separatrices in the vacuum limit, see also Fig.~\ref{fig:rgflow00}.}
\label{fig:rgflowt}
\end{figure}

Next, let us discuss symmetry restoration at finite temperature and quark chemical potential with the aid of 
our two-channel approximation. The fixed points now become pseudo fixed-points due to the presence 
of a dimensionful external parameter, namely the temperature.\footnote{The same holds true in case 
of a finite chemical potential.} As a consequence, the 
position of the non-Gau\ss ian fixed points is shifted as a function of the dimensionless temperature~$T/k$ 
and therefore also the positions of the separatrices connecting 
the fixed points are shifted. This is illustrated in Fig.~\ref{fig:rgflowt} for 
the RG flow in the plane spanned by the scalar-pseudoscalar and the CSC coupling at~$T/k=0.4$ and~$\mu=0$.
While the fixed points~${\mathcal F}_3$ and~${\mathcal F}_4$ remain complex-valued when~$T/k$ is increased, 
the behavior of the real-valued (pseudo) non-Gau\ss ian fixed-point suggests that, for initial 
conditions chosen to be fixed in the domain~${\mathcal D}_1$ (see, e.g., pink dot in Fig.~\ref{fig:rgflowt}), 
a critical temperature~$T_{\text{cr}}$ exists above which the four-quark couplings do not
diverge anymore at a finite RG scale~$k_{\text{cr}}$ but remain finite on all scales and approach zero in the IR limit,~$k\to 0$.
In other words, there is no (spontaneous) symmetry breaking above the critical temperature. 
At least at high temperature, such a behavior is indeed expected since the quarks become 
effectively stiff degrees of freedom due to their thermal Matsubara mass~$\sim T$.
This {mechanism has already been} discussed in detail in Refs.~\cite{Braun:2011pp,Braun:2017srn} {and underlies
symmetry restoration when the temperature is increased.}
\begin{figure*}[t]
\begin{center}
  \includegraphics[width=0.47\linewidth]{figT0mu0.pdf}
 \includegraphics[width=0.47\linewidth]{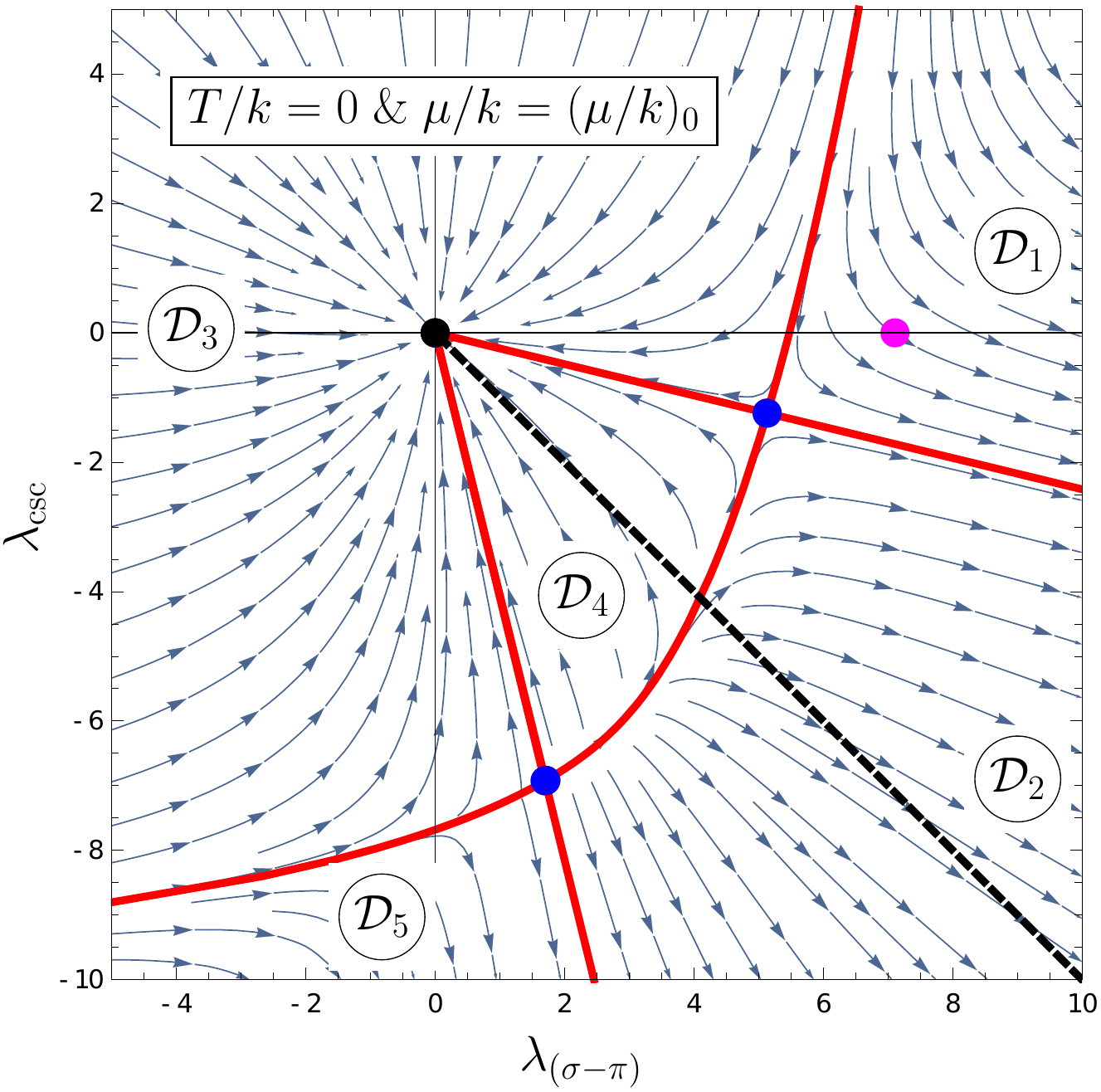}
 \includegraphics[width=0.47\linewidth]{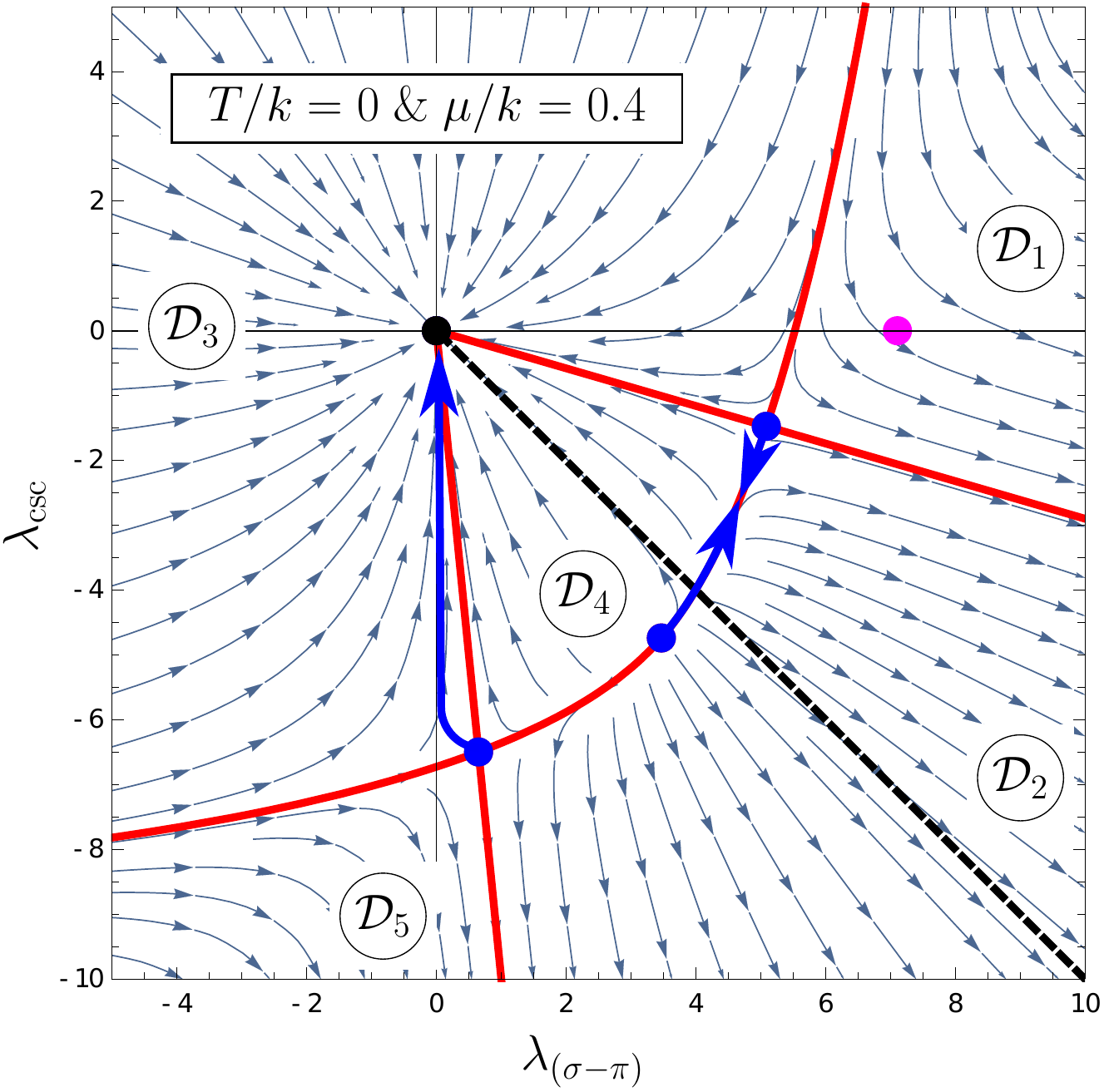}
  \includegraphics[width=0.47\linewidth]{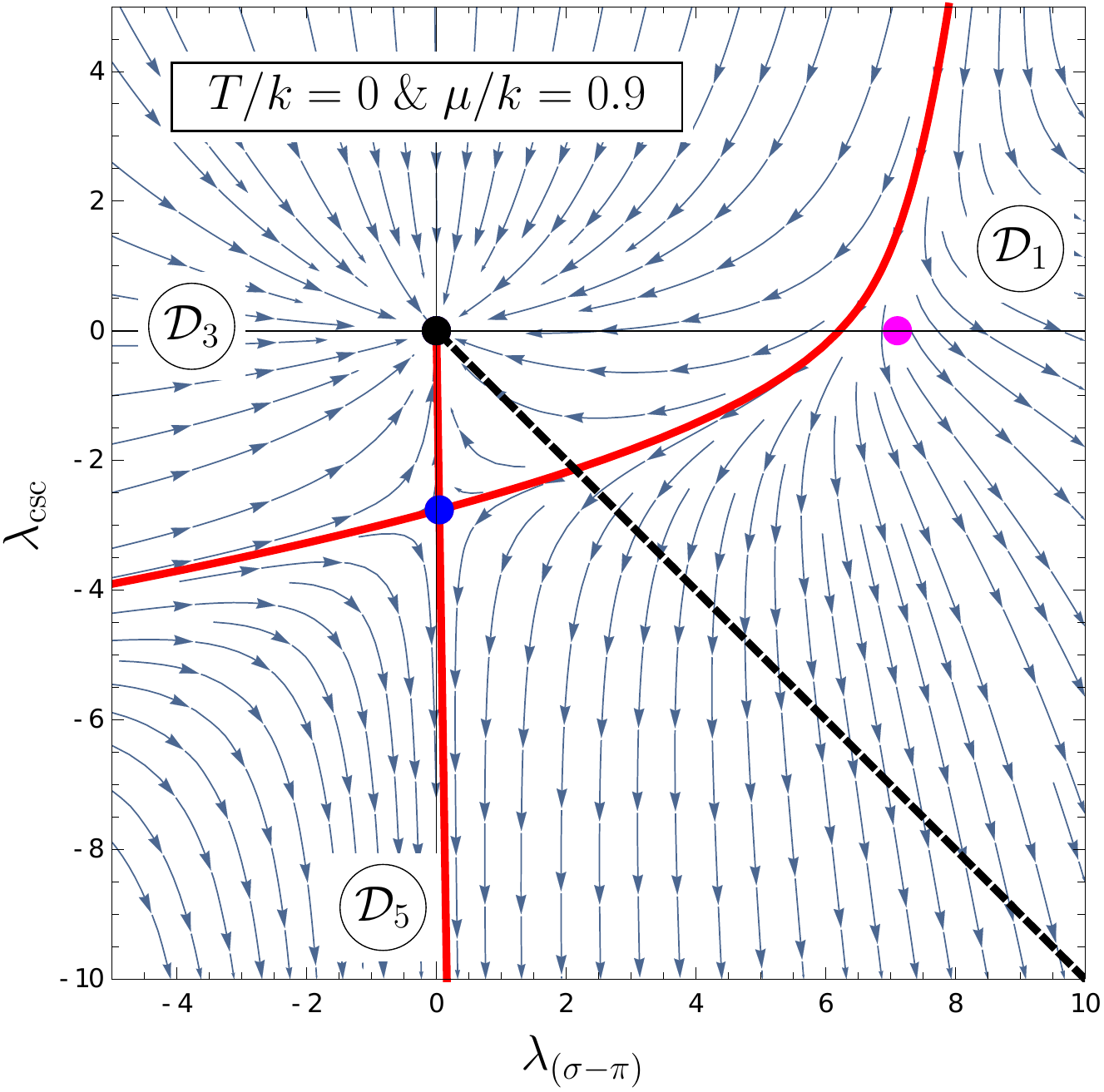}
\end{center}
\caption{(Color online) RG flow of the two-channel approximation at~$T=0$ for 
different values of~$\mu/k$:~$\mu/k=0$ (top left panel; same as Fig.~\ref{fig:rgflow00}),~$\mu/k=(\mu/k)_0\approx 0.298$ (top right panel),
$\mu/k=0.4$ (bottom left panel), and~$\mu/k = 0.9$ (bottom right panel)
in the plane 
spanned by the scalar-pseudoscalar coupling  and the CSC coupling. The black dot represents the Gau\ss ian fixed point. 
Blue dots represent real-valued non-Gau\ss ian fixed points. The pink dot depicts our choice 
for the initial condition. The dashed black line is the bisectrix of the bottom right quadrant. 
Different domains~${\mathcal D}_i$ are separated by separatrices (red solid lines).
The blue arrows in the bottom left panel indicate the shift of the real-valued non-Gau\ss ian fixed points 
when~$\mu/k$ is increased, see main text for details.}
\label{fig:rgflowmu}
\end{figure*}

Before we continue with a discussion of the mechanisms underlying symmetry breaking at zero temperature 
and finite quark chemical potential, we would
like to comment on the curvature of the finite-temperature phase boundary at small~$\mu$. We observe
that the curvature extracted from our two-channel approximation agrees almost identically with the curvatures found
in the one-channel approximation as well as in the {\it Fierz}-complete study. 
The agreement of the latter two can be understood in terms of the fixed-point structure as discussed above.
Of course, the agreement of the curvatures obtained from the two-channel approximation and the {\it Fierz}-complete study can also 
be understood from their fixed-point structure. However, we would like to recall that the flow equations of the 
two-channel approximation suffer from the {\it Fierz} ambiguity. Our two-channel approximation 
has been extracted from the {\it Fierz}-complete set of equations by only taking into 
account the scalar-pseudoscalar coupling and the CSC coupling. The remaining 
eight couplings and their flows have been set to zero. This is well justified for the purpose of analyzing
the mechanisms underlying the structure of the phase diagram found in the
{\it Fierz}-complete study. In practice, however, the flow equations for a two-channel approximation 
are not obtained from the full {\it Fierz}-complete set but by rather only taking into account the 
scalar-pseudoscalar channel and the CSC channel in our ansatz for the effective action~\eqref{eq:GLO}. 
Owing to the freedom of performing {\it Fierz} transformations, the 
set of flow equations for these two couplings resulting from such a {\it Fierz}-incomplete ansatz 
will differ from the one used in our present analysis.
For example, terms associated with a {\it Feynman} diagram of the type 
depicted in the right panel of Fig.~\ref{fig:fd} may be found to contribute to the flow of the scalar-pseudoscalar coupling.
As we have seen in our discussion of the large-$\Nc$ limit, such contributions are parametrically suppressed by factors of~$1/\Nc$ compared 
to those associated with {\it Feynman} diagrams of the type 
shown in the left panel of Fig.~\ref{fig:fd}. Thus, these contributions drop out for~$\Nc\to\infty$. For finite~$\Nc$, 
however, they may still alter the curvature {significantly, see Ref.~\cite{Braun:2017srn}.}

Let us now turn to the discussion of the dense regime of the phase diagram.
At zero temperature, we do not observe symmetry restoration in our {\it Fierz}-complete study
when the quark chemical potential is increased, see Fig.~\ref{fig:pd}. As can also be seen in Fig.~\ref{fig:pd},
the same behavior is found in our present two-channel approximation. 
Even though we do not observe symmetry restoration at zero temperature 
when the chemical potential~$\mu$ is increased, we find that the ``hierarchy" of the channels changes as 
a function of~$\mu$, i.e. the CSC channel becomes the most dominant channel for sufficiently large values 
of the chemical potential,~$\mu\gtrsim \mu_{\chi}$. Note that a dominance of the CSC channel is associated with 
a divergence of the RG flow into the direction defined by the CSC coupling. Given our choice for the initial 
condition {(see the pink dot in Figs.~\ref{fig:rgflow00},~\ref{fig:rgflowt} and~\ref{fig:rgflowmu}), such} a dominance is not immediately apparent. In fact, even if we chose the initial  
condition to be located in the domain~${\mathcal D}_2$, we would still observe a dominance  of the scalar-pseudoscalar channel 
at low energies. Thus, a dominance of the CSC channel is prohibited by the vacuum fixed point structure.
This can be traced back to the fact that the fixed point~${\mathcal F}_2$ has one IR attractive and one IR repulsive 
direction.\footnote{Recall that the fixed point~${\mathcal F}_2$ corresponds to the fixed point~\eqref{eq:njlfp} in the {\it Fierz}-complete study.}
Increasing the quark chemical potential starting from the vacuum limit, we find that the 
fixed point structure together with the position of the {separatrices remain unchanged}
up to a ``critical value"~$(\mu/k)_0$ of the dimensionless chemical 
potential.\footnote{We observe slight changes of the fixed point structure and the associated 
positions of the separatrices for~$\mu/k \lesssim (\mu/k)_0$ 
which arise due to a mild violation of the {\it Silver}-{\it Blaze} property by our covariant regularization scheme, 
see Ref.~\cite{Braun:2017srn} for a detailed discussion of this aspect.}
At~$\mu/k = (\mu/k)_0$, we then observe that two new real-valued fixed points emerge in the plane spanned 
by the scalar-pseudoscalar coupling and the CSC coupling, which ``sit" on top of each other, 
see top right panel of Fig.~\ref{fig:rgflowmu}. These two ``new" fixed points are nothing but the fixed points~${\mathcal F}_3$
and~${\mathcal F}_4$ which become real-valued at~$\mu/k = (\mu/k)_0$. As a consequence of this 
``creation" of two real-valued fixed points, new separatrices emerge in the plane spanned by the two four-quark couplings which 
divide the parameter space into five domains~${\mathcal D}_i$, see, e.g., top right panel of Fig.~\ref{fig:rgflowmu}.
Still, an RG trajectory associated with a dominance of the CSC coupling cannot be established for initial conditions located 
in the domain~${\mathcal D}_1$. Increasing~$\mu/k$ further, 
the two new real-valued non-Gau\ss ian (pseudo) fixed points are shifted in different directions as indicated by the 
blue arrows in the bottom left panel of Fig.~\ref{fig:rgflowmu}. One of these two fixed points has one attractive and one repulsive 
direction and is shifted towards the Gau\ss ian fixed point. The other one is shifted towards
the fixed point~${\mathcal F}_2$. At sufficiently large~$\mu/k > (\mu/k)_0$, the latter two then annihilate 
each other in the sense that they become complex-valued fixed points. This annihilation also removes the separatrix separating
the domains~${\mathcal D}_1$ and~${\mathcal D}_2$. As a consequence, any initial condition of the RG flow 
located in the domain~${\mathcal D}_1$ now yields an RG trajectory eventually pointing into the direction associated 
with the CSC coupling with the two couplings diverging at a finite critical scale~$k_{\text{cr}}$. In other words,
for sufficiently large~$\mu/k$, the low-energy physics is 
potentially 
dominated by the dynamics associated with the CSC channel.\footnote{Note that, as we solve the RG flow 
from $k=\Lambda$ to $k\to 0$, the dimensionless chemical potential~$\mu/k$ changes from~$\mu/\Lambda \gtrsim 0$ to~$\mu/k \to \infty$. In terms of the 
RG ``time"~$t = \ln (k/\Lambda)$, however, 
the four-quark couplings may already diverge at a finite value of~$k$ before the RG flow
fully changes its direction at a certain value of~$\mu/k$. 
Large values of~$\mu/k$ may therefore not be reached in the RG flow and the 
scalar-pseudoscalar channel may still dominate the dynamics at sufficiently small values of~$\mu$.} 
The remaining real-valued fixed point at large~$\mu/k$ is eventually shifted towards the Gau\ss ian fixed point 
for~$\mu/k \to \infty$. As has been shown in Ref.~\cite{Braun:2017srn}, {the merging of the latter two
fixed points is associated
with the {\it Cooper} instability. Indeed, this behavior leaves its imprint 
in the~$\mu$-dependence of the symmetry breaking scale, 
exhibiting} the typical  
BCS-type exponential scaling behavior.

\section{Conclusions}\label{sec:conc}

In this work we have used RG flow equations of four-quark couplings to analyze the phase structure of an NJL-type model with two quark flavors 
coming in~$\Nc$ colors at leading order of the derivative expansion 
of the effective action. With our study, we aimed at an understanding on how {\it Fierz}-incomplete approximations
affect the predictive power of this class of models which still underlies to a large extent our understanding of the 
dynamics of QCD at high density. 
Our present leading-order approximation of the effective action already includes corrections beyond the 
often employed mean-field approximation. Note that 
such corrections are ultimately required to preserve the invariance of the results {under {\it Fierz} transformations~\cite{Jaeckel:2002rm}.}

Our results suggest that {\it Fierz}-incompleteness strongly affects the phase structure. For example, the phase transition temperature 
at large chemical potential almost increases by a factor of two compared to a study which only includes the conventional scalar-pseudoscalar
interaction channel together with a channel associated with the formation of a colorsuperconducting ground state. Although we do not have 
direct access to the gap within our present study, the observed shift of the phase boundary may suggest that the use of {\it Fierz}-incomplete approximations 
also affects the magnitude of the gap in the high-density regime. This is in accord with mean-field studies of 
this regime (see, e.g., Refs.~\cite{Klevansky:1992qe,Buballa:2003qv,Fukushima:2011jc,Andersen:2014xxa} for reviews).
However, we rush to add that the strength of the effect 
is expected to depend on the details of the chosen basis of four-quark interactions and
the actual choice for the initial conditions of the RG flow equations, i.e. the choice for the parameters 
appearing in the classical action. 

Despite the fact that our present study relies on a {\it Fierz}-complete approximation, it is clear that our results are mostly qualitative. 
In fact, our present study based on the analysis of RG flow equations of four-quark interactions at leading order of the derivative expansion
is limited with respect to
a determination of the properties of the actual ground state in the phase governed by spontaneous symmetry breaking.
In order to gain at least some insight into the structure of the ground state, we have analyzed the ``hierarchy" of the four-quark
interactions (in terms of their strength) as a function of the temperature and the quark chemical potential, following 
the analysis in Ref.~\cite{Braun:2017srn}. A dominance of a given channel may then be considered as an indication for 
the formation of a corresponding condensate. Of course, such an analysis has to be taken with some care as a dominance of a given
channel may not necessarily entail condensation in this channel. Moreover, more than one condensate may be formed, e.g., at large 
chemical potential. Still, it allows us to gain some insight into the dynamics underlying the phase structure. Interestingly, in
our present {\it Fierz}-complete study, we observe that the dynamics close to the phase boundary at small quark chemical potential
is clearly dominated by the scalar-pseudoscalar interaction channel whereas the channel associated with the formation of 
the most conventional color superconducting condensate dominates the dynamics at large chemical potential. In the latter regime,
the scalar-pseudoscalar channel is found to be only subdominant. Even more, the channels associated with the formation of color-symmetry breaking condensates 
are most dominant in this regime.

In order to understand better the dynamics underlying the phase structure, we have analyzed our {\it Fierz}-complete study in several ways. For example, 
we {have monitored the} strength of~$U_{\rm A}(1)$ symmetry breaking and even studied a~$U_{\rm A}(1)$-symmetric variation of our model 
which indicated that the ``hierarchy" of the channels changes 
at large chemical potential in this case. Moreover, we considered our 
flow equation in the large-$\Nc$ limit which revealed the existence of a fixed point which controls the dynamics at least for small 
values of the chemical potential, provided the initial conditions have been chosen to be located in a specific domain in the space spanned
by our set of four-quark couplings. At large chemical potential, the leading order of the large-$\Nc$ expansion cannot 
be used to explain the phase structure since channels subleading in this expansion become important. 
With the aid of a suitably 
chosen two-channel 
approximation, however, we have found that the phase structure and the dominance of the color superconducting channel 
at large chemical potential is a consequence of 
an intriguing creation and annihilation of pairs of (pseudo) fixed points.

Finally, we emphasize again that, at the present order of the derivative expansion, our analysis is still qualitative regarding the determination of the
actual properties of the ground state. In order to {unambiguously determine the ground state properties, a}
{calculation of the {full at least ten-dimensional order-parameter}
potential would in principle be required, representing an ambitious continuation of, e.g., recent beyond-mean field 
calculations of the order-parameter potential with a scalar-pseudoscalar and a diquark 
channel~\cite{Khan:2015puu,Strodthoff:2011tz,*Kamikado:2012bt,*Strodthoff:2013cua} as well as with 
a scalar-pseudoscalar and a vector channel~\cite{Rennecke:2015eba}.
Nevertheless, our present analysis already provides new} insights into the phase structure and the ground-state properties 
of NJL-type models at finite temperature and density, and may therefore also be helpful for future studies of bulk quantities, such as the equation 
of state at high density.

{\it Acknowledgments.--~} The authors would like to thank H.~Gies and J.~M.~Pawlowski
for useful discussions and comments on the manuscript. 
As members of the {\it fQCD collaboration}~\cite{fQCD}, the authors also would like to 
thank the other members of this collaboration for discussions.
J.B. acknowledges support by HIC for FAIR within the LOEWE program of the State of Hesse. 
This work is supported by the DFG through grant SFB 1245.


\appendix

\section{RG formalism}\label{app:RG}
For the computation of the RG flow equations 
of the various couplings and renormalization factors, 
we have employed the {\it Wetterich} equation~\cite{Wetterich:1992yh} which represents an RG equation for the
(scale-dependent) quantum effective {action~$\Gamma_k$:
\be
\!\!\!\!\!\!\del_t \Gamma_k[\Phi] &=& -\frac{1}{2} \text{Tr} \left \lbrace [ \Gamma^{(1,1)}_k [\Phi] + R_k^{\psi} ] ^{-1}\cdot (\del_t R_k^{\psi}) \right \rbrace\,.
\label{eq:WetterichEquation}
\ee
Here,~$\Gamma^{(1,1)}_k$} denotes the second functional derivative of the (scale-dependent) quantum effective action~$\Gamma_k$
with respect to the fermion fields summarized in the field {vector $\Phi^T(q) = \left(\psi^T (q),\bar{\psi}(-q)\right)$.}
For an explicit calculation of RG flow equations, we have to specify the regulator function~$R_k^{\psi}$ 
which encodes the regularization scheme. In this work, we have employed a four-dimensional so-called 
{\it Fermi}-surface-adapted regulator function as introduced in Ref.~\cite{Braun:2017srn}: 
$R_k^{\psi}=-({p}\fslash+{\rm i}\gamma_0\mu)\,r_{\psi}$.
Here, the regulator shape function~$r_{\psi}$ is defined as
\be
r_{\psi}=\frac{1}{\sqrt{1-{\rm e}^{-\bar{\omega}_{+}\bar{\omega}_{-}}}}-1\,,
\ee
where $\bar{\omega}_{\pm}=\omega_{\pm}/k$ {and~$\omega_{\pm}^2=p_0^2 + (|\vec{p}^{\,}|\pm \mu)^2$, 
see Ref.~\cite{Braun:2017srn} for a detailed discussion of the properties of this regulator.}

\section{RG flow equations}\label{app:fcset}
{For an introduction to the derivation of RG flow equations of four-fermion interactions, 
we refer the reader to Ref.~\cite{Braun:2011pp}. A 
general expression for the flow equations of pointlike four-fermion interactions
of a general class of relativistic fermionic theories with a continuous chiral $U_{\rm L}(\Nf)\otimes U_{\rm R}(\Nf)$ 
symmetry in a 
{\it Fierz}-complete setting can be found in Ref.~\cite{Gehring:2015vja}.
For the derivation of the flow equations 
of our present model 
with a~$SU(\Nc)\otimes SU_\text{L}(2)\otimes SU_\text{R}(2)\otimes U_\text{V}(1)$ symmetry,
we} have made use of existing software packages~\cite{Huber:2011qr,Cyrol:2016zqb}. 
In the following, we list the set of flow equations underlying our 
{\it Fierz}-complete study for general $\Nc$ and $\Nf=2$:
\begin{widetext}
{
\allowdisplaybreaks
\be
\partial_t  \lambda _{\text{($\sigma $-$\pi $)}} &=& 2 \lambda _{\text{($\sigma $-$\pi $)}} +64 v_4 \Big(-\lambda _{\text{($\sigma $-$\pi $)}}^2-4 \lambda _{\text{($\sigma $-$\pi $)}} \lambda _{(S+P)_-}-4
   \lambda _{(S+P)_-}^2+\lambda _{\text{($\sigma $-$\pi $)}} \lambda _{( V+A )_{\parallel }}+\lambda
   _{\text{($\sigma $-$\pi $)}} \lambda _{( V-A )_{\parallel }}\br
   +3 \lambda _{\text{($\sigma $-$\pi $)}} \lambda _{( V+A )_{\bot
   }}-\lambda _{( V+A )_{\parallel }^{\text{adj}}} \lambda _{( V+A )_{\bot }}+\lambda _{\text{($\sigma $-$\pi $)}} \lambda _{(
   V-A )_{\bot }}+2 \lambda _{\text{($\sigma $-$\pi $)}} \lambda _{( V-A )_{\bot }^{\text{adj}}}-\frac{1}{N_c^2}\lambda
   _{(S+P)_-^{\text{adj}}}^2\br
   +\frac{2}{N_c}\lambda _{\text{($\sigma $-$\pi $)}} \lambda
   _{(S+P)_-^{\text{adj}}}+\frac{4}{N_c}\lambda _{(S+P)_-} \lambda _{(S+P)_-^{\text{adj}}}+\frac{1}{N_c}\lambda
   _{(S+P)_-^{\text{adj}}}^2-\frac{1}{2 N_c}\lambda _{\text{($\sigma $-$\pi $)}} \lambda _{( V+A )_{\parallel
   }^{\text{adj}}}\br
   -\frac{1}{2 N_c}\lambda _{\text{($\sigma $-$\pi $)}} \lambda _{( V-A )_{\bot }^{\text{adj}}}-2 N_c
   \lambda _{\text{($\sigma $-$\pi $)}}^2 -4 N_c \lambda _{\text{($\sigma $-$\pi $)}} \lambda _{(S+P)_-} -4 N_c
   \lambda _{(S+P)_-}^2 -N_c\lambda _{\text{($\sigma $-$\pi $)}} \lambda _{(S+P)_-^{\text{adj}}} \br
   -2 N_c \lambda
   _{(S+P)_-} \lambda _{(S+P)_-^{\text{adj}}} +\frac{N_c}{2} \lambda _{\text{($\sigma $-$\pi $)}} \lambda _{( V+A )_{\parallel
   }^{\text{adj}}} +\lambda _{\text{($\sigma $-$\pi $)}} \lambda _{\text{csc}}-\lambda _{(S+P)_-^{\text{adj}}} \lambda
   _{\text{csc}}+N_c \lambda _{\text{($\sigma $-$\pi $)}}  \lambda _{\text{csc}}\br
   +2 N_c \lambda _{(S+P)_-}  \lambda
   _{\text{csc}}+N_c\lambda _{(S+P)_-^{\text{adj}}}  \lambda _{\text{csc}}\Big) l_{\text{$\parallel
   $+}}^{\text{(F)}}\left(\tau ,0,-i \tilde{\mu} _{\tau }\right)\br
   +64 v_4 \Big(-\lambda _{\text{($\sigma $-$\pi $)}} \lambda _{( V+A )_{\parallel }}+\lambda _{\text{($\sigma $-$\pi $)}} \lambda _{(
   V+A )_{\bot }}
   +\lambda _{( V+A )_{\parallel }^{\text{adj}}} \lambda _{( V+A )_{\bot }}
   +\frac{1}{2 N_c}\lambda _{\text{($\sigma $-$\pi
   $)}} \lambda _{( V+A )_{\parallel }^{\text{adj}}}\Big) l_{\parallel \pm }^{\text{(F)}}\left(\tau ,0,-i \tilde{\mu}
   _{\tau }\right)\br
   +64 v_4 \Big(-\lambda _{\text{($\sigma $-$\pi $)}}^2-4 \lambda _{\text{($\sigma $-$\pi $)}} \lambda _{(S+P)_-}-4
   \lambda _{(S+P)_-}^2-\frac{2}{3} \lambda _{\text{($\sigma $-$\pi $)}} \lambda
   _{(S+P)_-^{\text{adj}}}-\frac{4}{3} \lambda _{(S+P)_-} \lambda _{(S+P)_-^{\text{adj}}}\br
   +\lambda _{\text{($\sigma $-$\pi $)}}
   \lambda _{( V+A )_{\parallel }}+\frac{1}{3} \lambda _{\text{($\sigma $-$\pi $)}} \lambda _{( V-A )_{\parallel }}-\frac{1}{3}
   \lambda _{( V+A )_{\parallel }} \lambda _{( V+A )_{\parallel }^{\text{adj}}}+3 \lambda _{\text{($\sigma $-$\pi $)}} \lambda
   _{( V+A )_{\bot }}-\frac{2}{3} \lambda _{( V+A )_{\parallel }^{\text{adj}}} \lambda _{( V+A )_{\bot }}\br 
   +\frac{1}{3} \lambda
   _{\text{($\sigma $-$\pi $)}} \lambda _{( V-A )_{\bot }}+\frac{2}{3} \lambda _{\text{($\sigma $-$\pi $)}} \lambda _{( V-A
   )_{\bot }^{\text{adj}}}-\frac{1}{N_c^2}\lambda _{(S+P)_-^{\text{adj}}}^2+\frac{2}{N_c}\lambda _{\text{($\sigma
   $-$\pi $)}} \lambda _{(S+P)_-^{\text{adj}}}+\frac{4}{N_c} \lambda _{(S+P)_-} \lambda _{(S+P)_-^{\text{adj}}}\br
   +\frac{5
   }{3 N_c}\lambda _{(S+P)_-^{\text{adj}}}^2-\frac{1}{2 N_c}\lambda _{\text{($\sigma $-$\pi $)}} \lambda _{( V+A
   )_{\parallel }^{\text{adj}}}+\frac{1}{6
   N_c}\lambda _{( V+A )_{\parallel }^{\text{adj}}}^2-\frac{1}{6 N_c}\lambda _{\text{($\sigma $-$\pi $)}} \lambda _{( V-A )_{\bot }^{\text{adj}}}-2 N_c \lambda
   _{\text{($\sigma $-$\pi $)}}^2 \br
   -4 N_c \lambda _{\text{($\sigma $-$\pi $)}} \lambda _{(S+P)_-} -4 N_c\lambda
   _{(S+P)_-}^2 -N_c\lambda _{\text{($\sigma $-$\pi $)}} \lambda _{(S+P)_-^{\text{adj}}} -2 N_c \lambda _{(S+P)_-}
   \lambda _{(S+P)_-^{\text{adj}}} -\frac{N_c}{3} \lambda _{(S+P)_-^{\text{adj}}}^2 \br
   +\frac{N_c}{2} \lambda
   _{\text{($\sigma $-$\pi $)}} \lambda _{( V+A )_{\parallel }^{\text{adj}}} -\frac{N_c}{12} \lambda _{( V+A
   )_{\parallel }^{\text{adj}}}^2 -\frac{1}{3}\lambda _{\text{($\sigma $-$\pi $)}} \lambda _{\text{csc}}-\frac{4}{3}
   \lambda _{(S+P)_-} \lambda _{\text{csc}}-\frac{1}{3} \lambda _{(S+P)_-^{\text{adj}}} \lambda _{\text{csc}}\br
   +\frac{2}{3 N_c}\lambda
   _{(S+P)_-^{\text{adj}}} \lambda _{\text{csc}}+N_c \lambda _{\text{($\sigma $-$\pi $)}}  \lambda _{\text{csc}}+2 N_c
   \lambda _{(S+P)_-}  \lambda _{\text{csc}}
   +\frac{N_c}{3} \lambda _{(S+P)_-^{\text{adj}}}  \lambda _{\text{csc}}+\frac{2
   }{3}\lambda _{\text{csc}}^2\br
   -\frac{N_c}{3}  \lambda _{\text{csc}}^2\Big) l_{\text{$\bot $+}}^{\text{(F)}}\left(\tau ,0,-i
   \tilde{\mu} _{\tau }\right)\br
   +64 v_4 \Big(\frac{1}{3} \lambda _{\text{($\sigma $-$\pi $)}} \lambda _{( V+A )_{\parallel }}+\frac{1}{3} \lambda _{( V+A
   )_{\parallel }} \lambda _{( V+A )_{\parallel }^{\text{adj}}}-\frac{5}{3} \lambda _{\text{($\sigma $-$\pi $)}} \lambda _{(
   V+A )_{\bot }}-\frac{2}{3} \lambda _{( V+A )_{\parallel }^{\text{adj}}} \lambda _{( V+A )_{\bot }}\br
   -\frac{1}{6 N_c}\lambda
   _{\text{($\sigma $-$\pi $)}} \lambda _{( V+A )_{\parallel }^{\text{adj}}}-\frac{1}{6 N_c}\lambda _{( V+A )_{\parallel
   }^{\text{adj}}}^2\Big) l_{\bot \pm }^{\text{(F)}}\left(\tau ,0,-i \tilde{\mu} _{\tau }\right)\,,\nn
\ee

\be
\partial_t  \lambda _{\text{csc}} &=& 2 \lambda _{\text{csc}} + 64 v_4 \Big(-\lambda _{\text{($\sigma $-$\pi $)}}^2+2 \lambda _{\text{($\sigma $-$\pi $)}} \lambda _{( V+A
   )_{\parallel }^{\text{adj}}}-\lambda _{( V+A )_{\parallel }^{\text{adj}}}^2+3 \lambda _{( V-A )_{\bot }}
   \lambda _{( V-A )_{\bot }^{\text{adj}}}-\frac{3}{2 N_c}\lambda _{( V-A )_{\bot }^{\text{adj}}}^2\br
   +\frac{3 N_c}{4}
   \lambda _{( V-A )_{\bot }^{\text{adj}}}^2 +2 \lambda _{( V-A )_{\parallel }} \lambda
   _{\text{csc}}-\frac{3}{2} \lambda _{( V-A )_{\bot }^{\text{adj}}} \lambda _{\text{csc}}
   +\frac{3 N_c}{2} \lambda _{( V-A )_{\bot
   }^{\text{adj}}}  \lambda _{\text{csc}}-2 \lambda _{\text{csc}}^2\br
   +N_c \lambda _{\text{csc}}^2\Big) l_{\text{$\parallel
   $+}}^{\text{(F)}}\left(\tau ,0,-i \tilde{\mu} _{\tau }\right)\br
   + 64 v_4 \Big(-\lambda _{\text{($\sigma $-$\pi $)}}^2-4 \lambda _{\text{($\sigma $-$\pi $)}} \lambda _{(S+P)_-}-4
   \lambda _{(S+P)_-}^2-4 \lambda _{\text{($\sigma $-$\pi $)}} \lambda _{(S+P)_-^{\text{adj}}}-8 \lambda
   _{(S+P)_-} \lambda _{(S+P)_-^{\text{adj}}}-\lambda _{(S+P)_-^{\text{adj}}}^2\br
   -3 \lambda _{( V-A )_{\bot }}
   \lambda _{( V-A )_{\bot }^{\text{adj}}}-\frac{1}{N_c^2}\lambda _{(S+P)_-^{\text{adj}}}^2+\frac{2}{N_c}\lambda
   _{\text{($\sigma $-$\pi $)}} \lambda _{(S+P)_-^{\text{adj}}}+\frac{4}{N_c} \lambda _{(S+P)_-} \lambda
   _{(S+P)_-^{\text{adj}}}+\frac{4}{N_c} \lambda _{(S+P)_-^{\text{adj}}}^2+\br
   \frac{3}{2 N_c} \lambda _{( V-A
   )_{\bot }^{\text{adj}}}^2-2 \lambda _{\text{($\sigma $-$\pi $)}} \lambda _{\text{csc}}-4 \lambda _{(S+P)_-}
   \lambda _{\text{csc}}+2 \lambda _{(S+P)_-^{\text{adj}}} \lambda _{\text{csc}}-\lambda _{( V-A )_{\parallel }} \lambda
   _{\text{csc}}-3 \lambda _{( V-A )_{\bot }} \lambda _{\text{csc}}\br
   +\frac{3}{2} \lambda _{( V-A )_{\bot }^{\text{adj}}} \lambda
   _{\text{csc}}+\frac{2}{N_c} \lambda _{(S+P)_-^{\text{adj}}} \lambda _{\text{csc}}+\frac{3}{2 N_c} \lambda _{( V-A )_{\bot
   }^{\text{adj}}} \lambda _{\text{csc}}+4 \lambda _{\text{csc}}^2\Big) l_{\parallel \pm }^{\text{(F)}}\left(\tau
   ,0,-i \tilde{\mu} _{\tau }\right)\br
   +64 v_4 \Big(\lambda _{\text{($\sigma $-$\pi $)}}^2-2 \lambda _{\text{($\sigma $-$\pi $)}} \lambda _{( V+A
   )_{\parallel }^{\text{adj}}}+\lambda _{( V+A )_{\parallel }^{\text{adj}}}^2+\lambda _{( V-A )_{\parallel }}
   \lambda _{( V-A )_{\bot }^{\text{adj}}}-2 \lambda _{( V-A )_{\bot }} \lambda _{( V-A )_{\bot
   }^{\text{adj}}}+\frac{1}{N_c}\lambda _{( V-A )_{\bot }^{\text{adj}}}^2\br
   -\frac{N_c}{2} \lambda _{( V-A
   )_{\bot }^{\text{adj}}}^2 +2 \lambda _{( V-A )_{\bot }} \lambda _{\text{csc}}+\frac{1}{2} \lambda _{( V-A
   )_{\bot }^{\text{adj}}} \lambda _{\text{csc}}
   -\frac{1}{N_c}\lambda _{( V-A )_{\bot }^{\text{adj}}} \lambda
   _{\text{csc}}
   -\frac{N_c}{2} \lambda _{( V-A )_{\bot }^{\text{adj}}}  \lambda _{\text{csc}}\Big) l_{\text{$\bot
   $+}}^{\text{(F)}}\left(\tau ,0,-i \tilde{\mu} _{\tau }\right)\br
   +64 v_4 \Big(\lambda _{\text{($\sigma $-$\pi $)}}^2+4 \lambda _{\text{($\sigma $-$\pi $)}} \lambda _{(S+P)_-}+4
   \lambda _{(S+P)_-}^2+\lambda _{(S+P)_-^{\text{adj}}}^2-\lambda _{( V-A )_{\parallel }} \lambda
   _{( V-A )_{\bot }^{\text{adj}}}-2 \lambda _{( V-A )_{\bot }} \lambda _{( V-A )_{\bot }^{\text{adj}}}\br
   +\frac{1}{N_c^2}\lambda
   _{(S+P)_-^{\text{adj}}}^2-\frac{2}{N_c}\lambda _{\text{($\sigma $-$\pi $)}} \lambda
   _{(S+P)_-^{\text{adj}}}-\frac{4}{N_c} \lambda _{(S+P)_-} \lambda _{(S+P)_-^{\text{adj}}}+\frac{1}{N_c}\lambda _{( V-A
   )_{\bot }^{\text{adj}}}^2-2 \lambda _{\text{($\sigma $-$\pi $)}} \lambda _{\text{csc}}\br
   -4 \lambda _{(S+P)_-}\lambda _{\text{csc}}+2 \lambda _{(S+P)_-^{\text{adj}}} \lambda _{\text{csc}}-\lambda _{( V-A )_{\parallel }} \lambda
   _{\text{csc}}-3 \lambda _{( V-A )_{\bot }} \lambda _{\text{csc}}+\frac{3}{2} \lambda _{( V-A )_{\bot }^{\text{adj}}} \lambda
   _{\text{csc}}\br
   +\frac{2}{N_c} \lambda _{(S+P)_-^{\text{adj}}} \lambda _{\text{csc}}+\frac{3}{2 N_c} \lambda _{( V-A )_{\bot
   }^{\text{adj}}} \lambda _{\text{csc}}+4 \lambda _{\text{csc}}^2\Big) l_{\bot \pm }^{\text{(F)}}\left(\tau ,0,-i
   \tilde{\mu} _{\tau }\right)\,,\nn
\ee

\be
\partial_t \lambda_\SpPmAdj &=& 2 \lambda_\SpPmAdj +64 v_4 \Big(\lambda _{\text{($\sigma $-$\pi $)}}^2+2 \lambda _{(S+P)_-^{\text{adj}}} \lambda _{( V+A
   )_{\parallel }}-\frac{3}{2} \lambda _{\text{($\sigma $-$\pi $)}} \lambda _{( V+A )_{\parallel }^{\text{adj}}}+\lambda
   _{(S+P)_-} \lambda _{( V+A )_{\parallel }^{\text{adj}}}\br
   +\lambda _{( V+A )_{\parallel }^{\text{adj}}}^2+2
   \lambda _{\text{($\sigma $-$\pi $)}} \lambda _{( V+A )_{\bot }}+4 \lambda _{(S+P)_-} \lambda _{( V+A )_{\bot }}+2 \lambda
   _{(S+P)_-^{\text{adj}}} \lambda _{( V+A )_{\bot }}-3 \lambda _{( V-A )_{\bot }} \lambda _{( V-A )_{\bot
   }^{\text{adj}}}\br
   -\frac{3}{2 N_c} \lambda _{(S+P)_-^{\text{adj}}} \lambda _{( V+A )_{\parallel }^{\text{adj}}}-\frac{2}{N_c} \lambda
   _{(S+P)_-^{\text{adj}}} \lambda _{( V+A )_{\bot }}+\frac{3}{2
   N_c} \lambda _{( V-A )_{\bot }^{\text{adj}}}^2+\frac{N_c}{2} \lambda _{(S+P)_-^{\text{adj}}} \lambda _{( V+A )_{\parallel }^{\text{adj}}} \br
   -\frac{3 N_c}{4} \lambda
   _{( V-A )_{\bot }^{\text{adj}}}^2 +2 \lambda _{( V+A )_{\parallel }} \lambda _{\text{csc}}-2 \lambda _{( V-A
   )_{\parallel }} \lambda _{\text{csc}}-\frac{1}{2} \lambda _{( V+A )_{\parallel }^{\text{adj}}} \lambda
   _{\text{csc}}+\frac{3}{2} \lambda _{( V-A )_{\bot }^{\text{adj}}} \lambda _{\text{csc}}\br
   -\frac{1}{N_c}\lambda _{( V+A )_{\parallel
   }^{\text{adj}}} \lambda _{\text{csc}}+\frac{N_c}{2} \lambda _{( V+A )_{\parallel }^{\text{adj}}}  \lambda
   _{\text{csc}}-\frac{3N_c}{2} \lambda _{( V-A )_{\bot }^{\text{adj}}}  \lambda _{\text{csc}}+2 \lambda _{\text{csc}}^2-N_c
   \lambda _{\text{csc}}^2\Big) l_{\text{$\parallel $+}}^{\text{(F)}}\left(\tau ,0,-i \tilde{\mu} _{\tau }\right)\br
   +64 v_4\Big(\lambda _{\text{($\sigma $-$\pi $)}}^2+4 \lambda _{\text{($\sigma $-$\pi $)}} \lambda _{(S+P)_-}+4
   \lambda _{(S+P)_-}^2+4 \lambda _{\text{($\sigma $-$\pi $)}} \lambda _{(S+P)_-^{\text{adj}}}+8 \lambda
   _{(S+P)_-} \lambda _{(S+P)_-^{\text{adj}}}\br
   +\lambda _{(S+P)_-^{\text{adj}}}^2-\lambda _{(S+P)_-^{\text{adj}}}
   \lambda _{( V-A )_{\parallel }}+2 \lambda _{\text{($\sigma $-$\pi $)}} \lambda _{( V-A )_{\bot }}+4 \lambda _{(S+P)_-}
   \lambda _{( V-A )_{\bot }}-\lambda _{(S+P)_-^{\text{adj}}} \lambda _{( V-A )_{\bot }}\br
   -\frac{1}{2} \lambda _{\text{($\sigma
   $-$\pi $)}} \lambda _{( V-A )_{\bot }^{\text{adj}}}-\lambda _{(S+P)_-} \lambda _{( V-A )_{\bot }^{\text{adj}}}+\lambda
   _{(S+P)_-^{\text{adj}}} \lambda _{( V-A )_{\bot }^{\text{adj}}}+3 \lambda _{( V-A )_{\bot }} \lambda _{( V-A )_{\bot
   }^{\text{adj}}}+\frac{1}{N_c^2}\lambda _{(S+P)_-^{\text{adj}}}^2\br
   +\frac{1}{N_c^2}\lambda _{(S+P)_-^{\text{adj}}} \lambda
   _{( V-A )_{\bot }^{\text{adj}}}-\frac{2}{N_c} \lambda _{\text{($\sigma $-$\pi $)}} \lambda
   _{(S+P)_-^{\text{adj}}}-\frac{4}{N_c} \lambda _{(S+P)_-} \lambda _{(S+P)_-^{\text{adj}}}-\frac{4}{N_c} \lambda
   _{(S+P)_-^{\text{adj}}}^2\br
   -\frac{2}{N_c} \lambda _{(S+P)_-^{\text{adj}}} \lambda _{( V-A )_{\bot
   }}-\frac{1}{N_c}\lambda _{\text{($\sigma $-$\pi $)}} \lambda _{( V-A )_{\bot }^{\text{adj}}}-\frac{2}{N_c} \lambda _{(S+P)_-}
   \lambda _{( V-A )_{\bot }^{\text{adj}}}
   +\frac{1}{N_c}\lambda _{(S+P)_-^{\text{adj}}} \lambda _{( V-A )_{\bot
   }^{\text{adj}}}\br
   -\frac{3}{2 N_c} \lambda _{( V-A )_{\bot }^{\text{adj}}}^2\Big) l_{\parallel \pm
   }^{\text{(F)}}\left(\tau ,0,-i \tilde{\mu} _{\tau }\right)\br
   +64 v_4 \Big(-\lambda _{\text{($\sigma $-$\pi $)}}^2+\frac{2}{3} \lambda _{\text{($\sigma $-$\pi $)}} \lambda
   _{( V+A )_{\parallel }}+\frac{4}{3} \lambda _{(S+P)_-} \lambda _{( V+A )_{\parallel }}+\frac{2}{3} \lambda
   _{(S+P)_-^{\text{adj}}} \lambda _{( V+A )_{\parallel }}+\frac{11}{6} \lambda _{\text{($\sigma $-$\pi $)}} \lambda _{( V+A
   )_{\parallel }^{\text{adj}}}\br
   -\frac{1}{3} \lambda _{(S+P)_-} \lambda _{( V+A )_{\parallel }^{\text{adj}}}-\lambda _{(
   V+A )_{\parallel }^{\text{adj}}}^2+\frac{4}{3} \lambda _{\text{($\sigma $-$\pi $)}} \lambda _{( V+A )_{\bot
   }}+\frac{8}{3} \lambda _{(S+P)_-} \lambda _{( V+A )_{\bot }}+\frac{10}{3} \lambda _{(S+P)_-^{\text{adj}}} \lambda _{( V+A
   )_{\bot }}\br
   -\lambda _{( V-A )_{\parallel }} \lambda _{( V-A )_{\bot }^{\text{adj}}}+2 \lambda _{( V-A )_{\bot }} \lambda _{(
   V-A )_{\bot }^{\text{adj}}}+\frac{1}{3
   N_c^2}\lambda _{(S+P)_-^{\text{adj}}} \lambda _{( V+A )_{\parallel }^{\text{adj}}}-\frac{2}{3 N_c} \lambda _{(S+P)_-^{\text{adj}}} \lambda _{( V+A )_{\parallel }}\br
   -\frac{1}{3 N_c}\lambda _{\text{($\sigma $-$\pi
   $)}} \lambda _{( V+A )_{\parallel }^{\text{adj}}}-\frac{2}{3 N_c} \lambda _{(S+P)_-} \lambda _{( V+A )_{\parallel
   }^{\text{adj}}}-\frac{1}{6 N_c}\lambda _{(S+P)_-^{\text{adj}}} \lambda _{( V+A )_{\parallel }^{\text{adj}}}-\frac{4}{3 N_c}
   \lambda _{(S+P)_-^{\text{adj}}} \lambda _{( V+A )_{\bot }}\br
   -\frac{1}{N_c}\lambda _{( V-A )_{\bot
   }^{\text{adj}}}^2-\frac{N_c}{6} \lambda _{(S+P)_-^{\text{adj}}} \lambda _{( V+A )_{\parallel }^{\text{adj}}}
   +\frac{N_c}{2} \lambda _{( V-A )_{\bot }^{\text{adj}}}^2 +\frac{1}{6} \lambda _{( V+A )_{\parallel
   }^{\text{adj}}} \lambda _{\text{csc}}+2 \lambda _{( V+A )_{\bot }} \lambda _{\text{csc}}\br
   -2 \lambda _{( V-A )_{\bot }}
   \lambda _{\text{csc}}-\frac{1}{2} \lambda _{( V-A )_{\bot }^{\text{adj}}} \lambda _{\text{csc}}+\frac{1}{N_c}\lambda _{( V-A
   )_{\bot }^{\text{adj}}} \lambda _{\text{csc}}
   -\frac{N_c}{6} \lambda _{( V+A )_{\parallel }^{\text{adj}}}  \lambda
   _{\text{csc}}\br
   +\frac{N_c}{2} \lambda _{( V-A )_{\bot }^{\text{adj}}}  \lambda _{\text{csc}}\Big) l_{\text{$\bot
   $+}}^{\text{(F)}}\left(\tau ,0,-i \tilde{\mu} _{\tau }\right)\br
   +64 v_4 \Big(-\lambda _{\text{($\sigma $-$\pi $)}}^2-4 \lambda _{\text{($\sigma $-$\pi $)}} \lambda _{(S+P)_-}-4
   \lambda _{(S+P)_-}^2-\lambda _{(S+P)_-^{\text{adj}}}^2+\frac{2}{3} \lambda _{\text{($\sigma
   $-$\pi $)}} \lambda _{( V-A )_{\parallel }}+\frac{4}{3} \lambda _{(S+P)_-} \lambda _{( V-A )_{\parallel }}\br
   -\frac{1}{3}
   \lambda _{(S+P)_-^{\text{adj}}} \lambda _{( V-A )_{\parallel }}+\frac{4}{3} \lambda _{\text{($\sigma $-$\pi $)}} \lambda _{(
   V-A )_{\bot }}+\frac{8}{3} \lambda _{(S+P)_-} \lambda _{( V-A )_{\bot }}-\frac{5}{3} \lambda _{(S+P)_-^{\text{adj}}} \lambda
   _{( V-A )_{\bot }}\br
   -\frac{5}{6} \lambda _{\text{($\sigma $-$\pi $)}} \lambda _{( V-A )_{\bot }^{\text{adj}}}-\frac{5}{3}
   \lambda _{(S+P)_-} \lambda _{( V-A )_{\bot }^{\text{adj}}}+\frac{2}{3} \lambda _{(S+P)_-^{\text{adj}}} \lambda _{( V-A
   )_{\bot }^{\text{adj}}}+\lambda _{( V-A )_{\parallel }} \lambda _{( V-A )_{\bot }^{\text{adj}}}\br
   +2 \lambda _{( V-A )_{\bot }}
   \lambda _{( V-A )_{\bot }^{\text{adj}}}-\frac{1}{N_c^2}\lambda _{(S+P)_-^{\text{adj}}}^2+\frac{2}{3 N_c^2} \lambda
   _{(S+P)_-^{\text{adj}}} \lambda _{( V-A )_{\bot}^{\text{adj}}}+\frac{2}{N_c} \lambda _{\text{($\sigma $-$\pi $)}}
   \lambda _{(S+P)_-^{\text{adj}}}\br
   +\frac{4}{N_c} \lambda _{(S+P)_-} \lambda _{(S+P)_-^{\text{adj}}}-\frac{2}{3 N_c} \lambda
   _{(S+P)_-^{\text{adj}}} \lambda _{( V-A )_{\parallel }}-\frac{4}{3 N_c} \lambda _{(S+P)_-^{\text{adj}}} \lambda _{( V-A
   )_{\bot }}-\frac{2}{3 N_c} \lambda _{\text{($\sigma $-$\pi $)}} \lambda _{( V-A )_{\bot }^{\text{adj}}}\br
   -\frac{4}{3 N_c}
   \lambda _{(S+P)_-} \lambda _{( V-A )_{\bot }^{\text{adj}}}+\frac{5}{3 N_c} \lambda _{(S+P)_-^{\text{adj}}} \lambda _{( V-A
   )_{\bot }^{\text{adj}}}-\frac{1}{N_c}\lambda _{( V-A )_{\bot }^{\text{adj}}}^2\Big) l_{\bot \pm
   }^{\text{(F)}}\left(\tau ,0,-i \tilde{\mu} _{\tau }\right)\,,\nn
\ee

\be
\partial_t \lambda_\SpPm &=& 2 \lambda_\SpPm + 64 v_4 \Big(-\frac{1}{2} \lambda _{\text{($\sigma $-$\pi $)}}^2+\lambda _{\text{($\sigma $-$\pi $)}} \lambda
   _{(S+P)_-}+2 \lambda _{(S+P)_-}^2+\frac{1}{2} \lambda _{\text{($\sigma $-$\pi $)}} \lambda _{( V+A
   )_{\parallel }}+2 \lambda _{(S+P)_-} \lambda _{( V+A )_{\parallel }}\br
   -\frac{1}{2} \lambda _{\text{($\sigma $-$\pi $)}}
   \lambda _{( V-A )_{\parallel }}+\lambda _{\text{($\sigma $-$\pi $)}} \lambda _{( V+A )_{\parallel
   }^{\text{adj}}}+\frac{1}{4} \lambda _{(S+P)_-^{\text{adj}}} \lambda _{( V+A )_{\parallel }^{\text{adj}}}-\frac{1}{2}
   \lambda _{( V+A )_{\parallel }^{\text{adj}}}^2-\frac{1}{2} \lambda _{\text{($\sigma $-$\pi $)}} \lambda _{(
   V+A )_{\bot }}\br
   +2 \lambda _{(S+P)_-} \lambda _{( V+A )_{\bot }}+\lambda _{(S+P)_-^{\text{adj}}} \lambda _{( V+A )_{\bot
   }}+\frac{1}{2} \lambda _{( V+A )_{\parallel }^{\text{adj}}} \lambda _{( V+A )_{\bot }}-\frac{1}{2} \lambda _{\text{($\sigma
   $-$\pi $)}} \lambda _{( V-A )_{\bot }}\br
   -\lambda _{\text{($\sigma $-$\pi $)}} \lambda _{( V-A )_{\bot
   }^{\text{adj}}}+\frac{3}{2} \lambda _{( V-A )_{\bot }} \lambda _{( V-A )_{\bot }^{\text{adj}}}-\frac{3}{8} \lambda _{(
   V-A )_{\bot }^{\text{adj}}}^2+\frac{1}{2 N_c^2}\lambda _{(S+P)_-^{\text{adj}}}^2-\frac{1}{4 N_c^2}\lambda
   _{(S+P)_-^{\text{adj}}} \lambda _{( V+A )_{\parallel }^{\text{adj}}}\br
   -\frac{1}{N_c^2}\lambda _{(S+P)_-^{\text{adj}}} \lambda
   _{( V+A )_{\bot }}+\frac{3}{4 N_c^2} \lambda _{( V-A )_{\bot }^{\text{adj}}}^2+\frac{1}{2 N_c}\lambda
   _{\text{($\sigma $-$\pi $)}}^2-\frac{1}{2 N_c}\lambda _{\text{($\sigma $-$\pi $)}} \lambda
   _{(S+P)_-^{\text{adj}}}-\frac{2}{N_c} \lambda _{(S+P)_-} \lambda _{(S+P)_-^{\text{adj}}}\br
   -\frac{1}{2 N_c}\lambda
   _{(S+P)_-^{\text{adj}}}^2-\frac{1}{N_c}\lambda _{\text{($\sigma $-$\pi $)}} \lambda _{( V+A )_{\parallel
   }^{\text{adj}}}-\frac{1}{2 N_c}\lambda _{(S+P)_-} \lambda _{( V+A )_{\parallel }^{\text{adj}}}+\frac{1}{2 N_c}\lambda _{(
   V+A )_{\parallel }^{\text{adj}}}^2\br
   +\frac{1}{N_c}\lambda _{\text{($\sigma $-$\pi $)}} \lambda _{( V+A )_{\bot
   }}+\frac{2}{N_c} \lambda _{(S+P)_-} \lambda _{( V+A )_{\bot }}+\frac{1}{4 N_c}\lambda _{\text{($\sigma $-$\pi $)}} \lambda _{(
   V-A )_{\bot }^{\text{adj}}}-\frac{3}{2
   N_c} \lambda _{( V-A )_{\bot }} \lambda _{( V-A )_{\bot }^{\text{adj}}}\br
   -\frac{3}{4 N_c} \lambda _{( V-A )_{\bot }^{\text{adj}}}^2+2 N_c \lambda _{(S+P)_-}^2
   +N_c\lambda _{(S+P)_-} \lambda _{(S+P)_-^{\text{adj}}} +\frac{N_c}{2} \lambda _{(S+P)_-} \lambda _{( V+A )_{\parallel
   }^{\text{adj}}} +\frac{3N_c}{8} \lambda _{( V-A )_{\bot }^{\text{adj}}}^2 \br
   -\lambda _{\text{($\sigma $-$\pi
   $)}} \lambda _{\text{csc}}+\frac{1}{2} \lambda _{(S+P)_-^{\text{adj}}} \lambda _{\text{csc}}-\lambda _{( V+A )_{\parallel }}
   \lambda _{\text{csc}}+\lambda _{( V-A )_{\parallel }} \lambda _{\text{csc}}+\frac{1}{2} \lambda _{( V+A )_{\parallel
   }^{\text{adj}}} \lambda _{\text{csc}}-\frac{3}{2} \lambda _{( V-A )_{\bot }^{\text{adj}}} \lambda
   _{\text{csc}}\br
   -\frac{1}{2 N_c^2}\lambda _{( V+A )_{\parallel }^{\text{adj}}} \lambda _{\text{csc}}+\frac{1}{N_c}\lambda _{( V+A
   )_{\parallel }} \lambda _{\text{csc}}-\frac{1}{N_c}\lambda _{( V-A )_{\parallel }} \lambda _{\text{csc}}+\frac{1}{4 N_c}\lambda
   _{( V+A )_{\parallel }^{\text{adj}}} \lambda _{\text{csc}}+\frac{3}{4 N_c} \lambda _{( V-A )_{\bot }^{\text{adj}}} \lambda
   _{\text{csc}}\br
   -N_c\lambda _{(S+P)_-}  \lambda _{\text{csc}}-\frac{N_c}{2} \lambda _{(S+P)_-^{\text{adj}}}  \lambda
   _{\text{csc}}-\frac{N_c}{4} \lambda _{( V+A )_{\parallel }^{\text{adj}}}  \lambda _{\text{csc}}+\frac{3N_c}{4} \lambda _{( V-A
   )_{\bot }^{\text{adj}}}  \lambda _{\text{csc}}
   -\frac{3}{2} \lambda _{\text{csc}}^2+\frac{1}{N_c}\lambda
   _{\text{csc}}^2\br
   +\frac{N_c}{2}  \lambda _{\text{csc}}^2\Big) l_{\text{$\parallel $+}}^{\text{(F)}}\left(\tau ,0,-i
   \tilde{\mu} _{\tau }\right)\br
   +64 v_4\Big(-\frac{1}{2} \lambda _{\text{($\sigma $-$\pi $)}}^2-2 \lambda _{\text{($\sigma $-$\pi $)}} \lambda
   _{(S+P)_-}-2 \lambda _{(S+P)_-}^2-2 \lambda _{\text{($\sigma $-$\pi $)}} \lambda _{(S+P)_-^{\text{adj}}}-4
   \lambda _{(S+P)_-} \lambda _{(S+P)_-^{\text{adj}}}\br
   -\frac{1}{2} \lambda _{(S+P)_-^{\text{adj}}}^2+\frac{1}{2}
   \lambda _{\text{($\sigma $-$\pi $)}} \lambda _{( V+A )_{\parallel }}-\frac{1}{2} \lambda _{\text{($\sigma $-$\pi $)}}
   \lambda _{( V-A )_{\parallel }}-\lambda _{(S+P)_-} \lambda _{( V-A )_{\parallel }}-\frac{1}{2} \lambda _{\text{($\sigma
   $-$\pi $)}} \lambda _{( V+A )_{\bot }}\br
   -\frac{1}{2} \lambda _{( V+A )_{\parallel }^{\text{adj}}} \lambda _{( V+A )_{\bot
   }}-\frac{1}{2} \lambda _{\text{($\sigma $-$\pi $)}} \lambda _{( V-A )_{\bot }}-\lambda _{(S+P)_-} \lambda _{( V-A )_{\bot
   }}+\lambda _{(S+P)_-^{\text{adj}}} \lambda _{( V-A )_{\bot }}\br
   +\frac{1}{2} \lambda _{\text{($\sigma $-$\pi $)}} \lambda _{(
   V-A )_{\bot }^{\text{adj}}}+\lambda _{(S+P)_-} \lambda _{( V-A )_{\bot }^{\text{adj}}}-\frac{1}{4} \lambda
   _{(S+P)_-^{\text{adj}}} \lambda _{( V-A )_{\bot }^{\text{adj}}}-\frac{3}{2} \lambda _{( V-A )_{\bot }} \lambda _{( V-A
   )_{\bot }^{\text{adj}}}\br
   +\frac{1}{2 N_c^3}\lambda _{(S+P)_-^{\text{adj}}}^2+\frac{1}{2 N_c^3}\lambda
   _{(S+P)_-^{\text{adj}}} \lambda _{( V-A )_{\bot }^{\text{adj}}}-\frac{1}{N_c^2}\lambda _{\text{($\sigma $-$\pi $)}} \lambda
   _{(S+P)_-^{\text{adj}}}-\frac{2}{N_c^2} \lambda _{(S+P)_-} \lambda _{(S+P)_-^{\text{adj}}}\br
   -\frac{5}{2 N_c^2} \lambda
   _{(S+P)_-^{\text{adj}}}^2-\frac{1}{N_c^2}\lambda _{(S+P)_-^{\text{adj}}} \lambda _{( V-A )_{\bot
   }}-\frac{1}{2 N_c^2}\lambda _{\text{($\sigma $-$\pi $)}} \lambda _{( V-A )_{\bot }^{\text{adj}}}-\frac{1}{N_c^2}\lambda
   _{(S+P)_-} \lambda _{( V-A )_{\bot }^{\text{adj}}}\br
   +\frac{1}{4 N_c^2}\lambda _{(S+P)_-^{\text{adj}}} \lambda _{( V-A )_{\bot
   }^{\text{adj}}}-\frac{3}{4 N_c^2} \lambda _{( V-A )_{\bot }^{\text{adj}}}^2+\frac{1}{2 N_c}\lambda
   _{\text{($\sigma $-$\pi $)}}^2+\frac{2}{N_c} \lambda _{\text{($\sigma $-$\pi $)}} \lambda
   _{(S+P)_-}+\frac{2}{N_c} \lambda _{(S+P)_-}^2\br
   +\frac{3}{N_c} \lambda _{\text{($\sigma $-$\pi $)}} \lambda
   _{(S+P)_-^{\text{adj}}}+\frac{6}{N_c} \lambda _{(S+P)_-} \lambda _{(S+P)_-^{\text{adj}}}+\frac{5}{2 N_c} \lambda
   _{(S+P)_-^{\text{adj}}}^2-\frac{1}{4 N_c}\lambda _{\text{($\sigma $-$\pi $)}} \lambda _{( V+A )_{\parallel
   }^{\text{adj}}}\br
   +\frac{1}{N_c}\lambda _{\text{($\sigma $-$\pi $)}} \lambda _{( V-A )_{\bot }}+\frac{2}{N_c} \lambda
   _{(S+P)_-} \lambda _{( V-A )_{\bot }}-\frac{1}{2
   N_c}\lambda _{(S+P)_-^{\text{adj}}} \lambda _{( V-A )_{\bot }^{\text{adj}}}
   +\frac{3}{2 N_c} \lambda _{( V-A )_{\bot }} \lambda _{( V-A )_{\bot }^{\text{adj}}}\br
   +\frac{3}{4 N_c} \lambda _{( V-A
   )_{\bot }^{\text{adj}}}^2\Big) l_{\parallel \pm }^{\text{(F)}}\left(\tau ,0,-i \tilde{\mu} _{\tau }\right)\br
   +64 v_4 \Big(\frac{1}{2}\lambda _{\text{($\sigma $-$\pi $)}}^2+\lambda _{\text{($\sigma $-$\pi $)}} \lambda
   _{(S+P)_-}+2 \lambda _{(S+P)_-}^2+\frac{1}{3} \lambda _{\text{($\sigma $-$\pi $)}} \lambda
   _{(S+P)_-^{\text{adj}}}+\frac{2}{3} \lambda _{(S+P)_-} \lambda _{(S+P)_-^{\text{adj}}}\br
   -\frac{1}{6} \lambda _{\text{($\sigma
   $-$\pi $)}} \lambda _{( V+A )_{\parallel }}+\frac{2}{3} \lambda _{(S+P)_-} \lambda _{( V+A )_{\parallel }}+\frac{1}{3}
   \lambda _{(S+P)_-^{\text{adj}}} \lambda _{( V+A )_{\parallel }}-\frac{1}{6} \lambda _{\text{($\sigma $-$\pi $)}} \lambda _{(
   V-A )_{\parallel }}-\frac{5}{6} \lambda _{\text{($\sigma $-$\pi $)}} \lambda _{( V+A )_{\parallel
   }^{\text{adj}}}\br
   +\frac{1}{3} \lambda _{(S+P)_-} \lambda _{( V+A )_{\parallel }^{\text{adj}}}-\frac{1}{12} \lambda
   _{(S+P)_-^{\text{adj}}} \lambda _{( V+A )_{\parallel }^{\text{adj}}}+\frac{1}{6} \lambda _{( V+A )_{\parallel }} \lambda _{(
   V+A )_{\parallel }^{\text{adj}}}+\frac{1}{2} \lambda _{( V+A )_{\parallel }^{\text{adj}}}^2\br
   +\frac{1}{6}
   \lambda _{\text{($\sigma $-$\pi $)}} \lambda _{( V+A )_{\bot }}+\frac{10}{3} \lambda _{(S+P)_-} \lambda _{( V+A )_{\bot
   }}+\frac{2}{3} \lambda _{(S+P)_-^{\text{adj}}} \lambda _{( V+A )_{\bot }}+\frac{1}{3} \lambda _{( V+A )_{\parallel
   }^{\text{adj}}} \lambda _{( V+A )_{\bot }}\br
   -\frac{1}{6} \lambda _{\text{($\sigma $-$\pi $)}} \lambda _{( V-A )_{\bot
   }}-\frac{1}{3} \lambda _{\text{($\sigma $-$\pi $)}} \lambda _{( V-A )_{\bot }^{\text{adj}}}+\frac{1}{2} \lambda _{( V-A
   )_{\parallel }} \lambda _{( V-A )_{\bot }^{\text{adj}}}-\lambda _{( V-A )_{\bot }} \lambda _{( V-A )_{\bot
   }^{\text{adj}}}+\frac{1}{4} \lambda _{( V-A )_{\bot }^{\text{adj}}}^2\br
   +\frac{1}{6 N_c^3}\lambda _{(S+P)_-^{\text{adj}}}
   \lambda _{( V+A )_{\parallel }^{\text{adj}}}+\frac{1}{2
   N_c^2}\lambda _{(S+P)_-^{\text{adj}}}^2-\frac{1}{3 N_c^2}\lambda _{(S+P)_-^{\text{adj}}} \lambda _{( V+A )_{\parallel }}-\frac{1}{6 N_c^2}\lambda _{\text{($\sigma $-$\pi
   $)}} \lambda _{( V+A )_{\parallel }^{\text{adj}}}\br
   -\frac{1}{3 N_c^2}\lambda _{(S+P)_-} \lambda _{( V+A )_{\parallel
   }^{\text{adj}}}+\frac{1}{12
   N_c^2}\lambda _{(S+P)_-^{\text{adj}}} \lambda _{( V+A )_{\parallel }^{\text{adj}}}-\frac{2}{3 N_c^2} \lambda _{(S+P)_-^{\text{adj}}} \lambda _{( V+A )_{\bot }}-\frac{1}{2 N_c^2}\lambda _{( V-A )_{\bot
   }^{\text{adj}}}^2\br
   -\frac{1}{2 N_c}\lambda _{\text{($\sigma $-$\pi $)}}^2-\frac{1}{2 N_c}\lambda
   _{\text{($\sigma $-$\pi $)}} \lambda _{(S+P)_-^{\text{adj}}}-\frac{2}{N_c} \lambda _{(S+P)_-} \lambda
   _{(S+P)_-^{\text{adj}}}-\frac{5}{6 N_c} \lambda _{(S+P)_-^{\text{adj}}}^2+\frac{1}{3 N_c}\lambda
   _{\text{($\sigma $-$\pi $)}} \lambda _{( V+A )_{\parallel }}\br
   +\frac{2}{3 N_c} \lambda _{(S+P)_-} \lambda _{( V+A )_{\parallel
   }}+\frac{1}{N_c}\lambda _{\text{($\sigma $-$\pi $)}} \lambda _{( V+A )_{\parallel }^{\text{adj}}}-\frac{1}{2 N_c}\lambda
   _{(S+P)_-} \lambda _{( V+A )_{\parallel }^{\text{adj}}}-\frac{1}{3 N_c}\lambda _{(S+P)_-^{\text{adj}}} \lambda _{( V+A
   )_{\parallel }^{\text{adj}}}\br
   -\frac{7}{12 N_c} \lambda _{( V+A )_{\parallel }^{\text{adj}}}^2+\frac{2}{3 N_c}
   \lambda _{\text{($\sigma $-$\pi $)}} \lambda _{( V+A )_{\bot }}+\frac{4}{3 N_c} \lambda _{(S+P)_-} \lambda _{( V+A )_{\bot
   }}+\frac{1}{12 N_c}\lambda _{\text{($\sigma $-$\pi $)}} \lambda _{( V-A )_{\bot }^{\text{adj}}}\br
   -\frac{1}{2 N_c}\lambda _{( V-A
   )_{\parallel }} \lambda _{( V-A )_{\bot }^{\text{adj}}}+\frac{1}{N_c}\lambda _{( V-A )_{\bot }} \lambda _{( V-A )_{\bot
   }^{\text{adj}}}+\frac{1}{2 N_c}\lambda _{( V-A )_{\bot }^{\text{adj}}}^2+2N_c \lambda
   _{(S+P)_-}^2 \br
   +N_c\lambda _{(S+P)_-} \lambda _{(S+P)_-^{\text{adj}}} +\frac{N_c}{6} \lambda
   _{(S+P)_-^{\text{adj}}}^2 +\frac{N_c}{2} \lambda _{(S+P)_-} \lambda _{( V+A )_{\parallel }^{\text{adj}}}
   +\frac{N_c}{6} \lambda _{(S+P)_-^{\text{adj}}} \lambda _{( V+A )_{\parallel }^{\text{adj}}} \br
   +\frac{N_c}{24} \lambda
   _{( V+A )_{\parallel }^{\text{adj}}}^2 -\frac{N_c}{4} \lambda _{( V-A )_{\bot }^{\text{adj}}}^2
   -\frac{1}{3}\lambda _{\text{($\sigma $-$\pi $)}} \lambda _{\text{csc}}+\frac{2}{3} \lambda _{(S+P)_-} \lambda
   _{\text{csc}}+\frac{1}{6} \lambda _{(S+P)_-^{\text{adj}}} \lambda _{\text{csc}}-\lambda _{( V+A )_{\bot }} \lambda
   _{\text{csc}}\br
   +\lambda _{( V-A )_{\bot }} \lambda _{\text{csc}}+\frac{1}{2} \lambda _{( V-A )_{\bot }^{\text{adj}}} \lambda
   _{\text{csc}}+\frac{1}{2 N_c^2}\lambda _{( V-A )_{\bot }^{\text{adj}}} \lambda _{\text{csc}}-\frac{1}{3 N_c}\lambda
   _{(S+P)_-^{\text{adj}}} \lambda _{\text{csc}}+\frac{1}{12 N_c}\lambda _{( V+A )_{\parallel }^{\text{adj}}} \lambda
   _{\text{csc}}\br
   +\frac{1}{N_c}\lambda _{( V+A )_{\bot }} \lambda _{\text{csc}}-\frac{1}{N_c}\lambda _{( V-A )_{\bot }} \lambda
   _{\text{csc}}-\frac{3}{4 N_c} \lambda _{( V-A )_{\bot }^{\text{adj}}} \lambda _{\text{csc}}-N_c\lambda _{(S+P)_-} 
   \lambda _{\text{csc}}-\frac{N_c}{6} \lambda _{(S+P)_-^{\text{adj}}}  \lambda _{\text{csc}}\br
   -\frac{N_c}{12} \lambda _{( V+A
   )_{\parallel }^{\text{adj}}}  \lambda _{\text{csc}}-\frac{N_c}{4} \lambda _{( V-A )_{\bot }^{\text{adj}}}  \lambda
   _{\text{csc}}-\frac{1}{3}\lambda _{\text{csc}}^2+\frac{N_c}{6}  \lambda _{\text{csc}}^2\Big) l_{\text{$\bot
   $+}}^{\text{(F)}}\left(\tau ,0,-i \tilde{\mu} _{\tau }\right)\br
   +64 v_4 \Big(\frac{1}{2}\lambda _{\text{($\sigma $-$\pi $)}}^2+2 \lambda _{\text{($\sigma $-$\pi $)}} \lambda
   _{(S+P)_-}+2 \lambda _{(S+P)_-}^2+\frac{1}{2} \lambda _{(S+P)_-^{\text{adj}}}^2-\frac{1}{6}
   \lambda _{\text{($\sigma $-$\pi $)}} \lambda _{( V+A )_{\parallel }}-\frac{1}{6} \lambda _{\text{($\sigma $-$\pi $)}} \lambda
   _{( V-A )_{\parallel }}\br
   -\frac{1}{3} \lambda _{(S+P)_-} \lambda _{( V-A )_{\parallel }}+\frac{1}{3} \lambda
   _{(S+P)_-^{\text{adj}}} \lambda _{( V-A )_{\parallel }}-\frac{1}{6} \lambda _{( V+A )_{\parallel }} \lambda _{( V+A
   )_{\parallel }^{\text{adj}}}+\frac{5}{6} \lambda _{\text{($\sigma $-$\pi $)}} \lambda _{( V+A )_{\bot }}\br
   +\frac{1}{3} \lambda
   _{( V+A )_{\parallel }^{\text{adj}}} \lambda _{( V+A )_{\bot }}-\frac{5}{6} \lambda _{\text{($\sigma $-$\pi $)}} \lambda _{(
   V-A )_{\bot }}-\frac{5}{3} \lambda _{(S+P)_-} \lambda _{( V-A )_{\bot }}+\frac{2}{3} \lambda _{(S+P)_-^{\text{adj}}} \lambda
   _{( V-A )_{\bot }}\br
   +\frac{1}{3} \lambda _{\text{($\sigma $-$\pi $)}} \lambda _{( V-A )_{\bot }^{\text{adj}}}+\frac{2}{3}
   \lambda _{(S+P)_-} \lambda _{( V-A )_{\bot }^{\text{adj}}}-\frac{5}{12} \lambda _{(S+P)_-^{\text{adj}}} \lambda _{( V-A
   )_{\bot }^{\text{adj}}}-\frac{1}{2} \lambda _{( V-A )_{\parallel }} \lambda _{( V-A )_{\bot }^{\text{adj}}}\br
   -\lambda _{( V-A
   )_{\bot }} \lambda _{( V-A )_{\bot }^{\text{adj}}}-\frac{1}{2
   N_c^3}\lambda _{(S+P)_-^{\text{adj}}}^2+\frac{1}{3 N_c^3}\lambda _{(S+P)_-^{\text{adj}}} \lambda _{( V-A )_{\bot }^{\text{adj}}}+\frac{1}{N_c^2}\lambda _{\text{($\sigma
   $-$\pi $)}} \lambda _{(S+P)_-^{\text{adj}}}\br
   +\frac{2}{N_c^2} \lambda _{(S+P)_-} \lambda
   _{(S+P)_-^{\text{adj}}}+\frac{1}{2 N_c^2}\lambda _{(S+P)_-^{\text{adj}}}^2-\frac{1}{3 N_c^2}\lambda
   _{(S+P)_-^{\text{adj}}} \lambda _{( V-A )_{\parallel }}-\frac{2}{3 N_c^2} \lambda _{(S+P)_-^{\text{adj}}} \lambda _{( V-A
   )_{\bot }}\br
   -\frac{1}{3 N_c^2}\lambda _{\text{($\sigma $-$\pi $)}} \lambda _{( V-A )_{\bot }^{\text{adj}}}-\frac{2}{3 N_c^2}
   \lambda _{(S+P)_-} \lambda _{( V-A )_{\bot }^{\text{adj}}}+\frac{5}{12 N_c^2} \lambda _{(S+P)_-^{\text{adj}}} \lambda _{( V-A
   )_{\bot }^{\text{adj}}}-\frac{1}{2
   N_c^2}\lambda _{( V-A )_{\bot }^{\text{adj}}}^2\br
   -\frac{1}{2 N_c}\lambda _{\text{($\sigma $-$\pi $)}}^2-\frac{2}{N_c} \lambda _{\text{($\sigma $-$\pi $)}}
   \lambda _{(S+P)_-}-\frac{2}{N_c} \lambda _{(S+P)_-}^2-\frac{1}{N_c}\lambda _{\text{($\sigma $-$\pi $)}} \lambda
   _{(S+P)_-^{\text{adj}}}-\frac{2}{N_c} \lambda _{(S+P)_-} \lambda _{(S+P)_-^{\text{adj}}}\br
   -\frac{1}{2 N_c}\lambda
   _{(S+P)_-^{\text{adj}}}^2+\frac{1}{3
   N_c}\lambda _{\text{($\sigma $-$\pi $)}} \lambda _{( V-A )_{\parallel }}+\frac{2}{3 N_c} \lambda _{(S+P)_-} \lambda _{( V-A )_{\parallel }}+\frac{1}{12 N_c}\lambda _{\text{($\sigma $-$\pi $)}} \lambda
   _{( V+A )_{\parallel }^{\text{adj}}}\br
   +\frac{1}{12
   N_c}\lambda _{( V+A )_{\parallel }^{\text{adj}}}^2+\frac{2}{3 N_c} \lambda _{\text{($\sigma $-$\pi $)}} \lambda _{( V-A )_{\bot }}+\frac{4}{3 N_c} \lambda _{(S+P)_-} \lambda _{(
   V-A )_{\bot }}-\frac{1}{3 N_c}\lambda _{(S+P)_-^{\text{adj}}} \lambda _{( V-A )_{\bot }^{\text{adj}}}\br
   +\frac{1}{2 N_c}\lambda
   _{( V-A )_{\parallel }} \lambda _{( V-A )_{\bot }^{\text{adj}}}+\frac{1}{N_c}\lambda _{( V-A )_{\bot }} \lambda _{( V-A
   )_{\bot }^{\text{adj}}}+\frac{1}{2 N_c}\lambda _{( V-A )_{\bot }^{\text{adj}}}^2\Big) l_{\bot \pm
   }^{\text{(F)}}\left(\tau ,0,-i \tilde{\mu} _{\tau }\right)\,,\nn
\ee
\be
\partial_t \lambda_\VpAPar &=& 2 \lambda_\VpAPar + 64 v_4 \Big(\frac{1}{2}{\lambda _{\text{($\sigma $-$\pi $)}}^2}+2 \lambda _{\text{($\sigma $-$\pi $)}} \lambda
   _{(S+P)_-}+2 \lambda _{(S+P)_-}^2+\frac{1}{2} \lambda _{\text{($\sigma $-$\pi $)}} \lambda
   _{(S+P)_-^{\text{adj}}}+\lambda _{(S+P)_-} \lambda _{(S+P)_-^{\text{adj}}}\br
   +\lambda
   _{(S+P)_-^{\text{adj}}}^2+\frac{1}{2} \lambda _{( V+A )_{\parallel }}^2+2 \lambda _{\text{($\sigma
   $-$\pi $)}} \lambda _{( V-A )_{\parallel }}+\lambda _{( V+A )_{\parallel }} \lambda _{( V-A )_{\parallel }}+\frac{1}{8}
   \lambda _{( V+A )_{\parallel }^{\text{adj}}}^2+\frac{3}{2} \lambda _{( V+A )_{\bot }}^2\br
   -3
   \lambda _{( V+A )_{\parallel }} \lambda _{( V-A )_{\bot }}+\frac{3}{4} \lambda _{\text{($\sigma $-$\pi $)}} \lambda _{( V-A
   )_{\bot }^{\text{adj}}}-\frac{1}{2 N_c^2}\lambda _{(S+P)_-^{\text{adj}}}^2-\frac{1}{8 N_c^2}\lambda _{( V+A
   )_{\parallel }^{\text{adj}}}^2-\frac{1}{N_c^2}\lambda _{\text{($\sigma $-$\pi $)}} \lambda _{( V-A )_{\bot
   }^{\text{adj}}}\br
   -\frac{1}{2 N_c}\lambda _{(S+P)_-^{\text{adj}}}^2-\frac{1}{2 N_c}\lambda _{( V+A )_{\parallel
   }^{\text{adj}}} \lambda _{( V+A )_{\bot }}+\frac{2}{N_c} \lambda _{\text{($\sigma $-$\pi $)}} \lambda _{( V-A )_{\bot
   }}+\frac{1}{N_c}\lambda _{\text{($\sigma $-$\pi $)}} \lambda _{( V-A )_{\bot }^{\text{adj}}}\br
   +\frac{3}{2 N_c} \lambda _{( V+A
   )_{\parallel }} \lambda _{( V-A )_{\bot }^{\text{adj}}}-4N_c \lambda _{( V+A )_{\parallel }} \lambda _{( V-A
   )_{\parallel }} -\frac{3N_c}{2} \lambda _{( V+A )_{\parallel }} \lambda _{( V-A )_{\bot }^{\text{adj}}} -\lambda
   _{\text{($\sigma $-$\pi $)}} \lambda _{\text{csc}}-\lambda _{(S+P)_-} \lambda _{\text{csc}}\br
   +\frac{3}{2} \lambda
   _{(S+P)_-^{\text{adj}}} \lambda _{\text{csc}}-\lambda _{( V+A )_{\parallel }} \lambda _{\text{csc}}-\frac{1}{N_c^2}\lambda
   _{(S+P)_-^{\text{adj}}} \lambda _{\text{csc}}+\frac{2}{N_c} \lambda _{\text{($\sigma $-$\pi $)}} \lambda
   _{\text{csc}}+\frac{2}{N_c} \lambda _{(S+P)_-} \lambda _{\text{csc}}\br
   -\frac{1}{2 N_c}\lambda _{(S+P)_-^{\text{adj}}} \lambda
   _{\text{csc}}+N_c\lambda _{( V+A )_{\parallel }}  \lambda _{\text{csc}}+\lambda _{\text{csc}}^2-\frac{1}{N_c}\lambda
   _{\text{csc}}^2\Big) l_{\text{$\parallel $+}}^{\text{(F)}}\left(\tau ,0,-i \tilde{\mu} _{\tau }\right)\br
   +64 v_4 \Big(-\frac{1}{2} \lambda _{\text{($\sigma $-$\pi $)}}^2-\frac{1}{2} \lambda _{( V+A )_{\parallel
   }}^2+\frac{1}{4} \lambda _{\text{($\sigma $-$\pi $)}} \lambda _{( V+A )_{\parallel }^{\text{adj}}}-\frac{1}{8}
   \lambda _{( V+A )_{\parallel }^{\text{adj}}}^2-\frac{3}{2} \lambda _{( V+A )_{\bot
   }}^2\br
   +\frac{1}{8 N_c^2}\lambda _{( V+A )_{\parallel }^{\text{adj}}}^2+\frac{2}{N_c} \lambda _{\text{($\sigma
   $-$\pi $)}} \lambda _{( V+A )_{\bot }}+\frac{1}{2
   N_c}\lambda _{( V+A )_{\parallel }^{\text{adj}}} \lambda _{( V+A )_{\bot }}\Big) l_{\parallel \pm }^{\text{(F)}}\left(\tau ,0,-i \tilde{\mu} _{\tau }\right)\br
   +64 v_4 \Big(\frac{1}{2} \lambda _{\text{($\sigma $-$\pi $)}} \lambda _{(S+P)_-^{\text{adj}}}+\lambda _{(S+P)_-} \lambda
   _{(S+P)_-^{\text{adj}}}+\frac{1}{3} \lambda _{(S+P)_-^{\text{adj}}}^2-2 \lambda _{\text{($\sigma $-$\pi $)}}
   \lambda _{( V-A )_{\parallel }}-\lambda _{( V+A )_{\parallel }} \lambda _{( V-A )_{\parallel }}\br
   -\frac{1}{24} \lambda
   _{( V+A )_{\parallel }^{\text{adj}}}^2+\lambda _{( V+A )_{\parallel }} \lambda _{( V+A )_{\bot }}+\lambda _{(
   V+A )_{\bot }}^2+3 \lambda _{( V+A )_{\parallel }} \lambda _{( V-A )_{\bot }}-\frac{3}{4} \lambda _{\text{($\sigma
   $-$\pi $)}} \lambda _{( V-A )_{\bot }^{\text{adj}}}\br
   +\frac{1}{3
   N_c^2}\lambda _{(S+P)_-^{\text{adj}}}^2+\frac{1}{12 N_c^2}\lambda _{( V+A )_{\parallel }^{\text{adj}}}^2+\frac{2}{3 N_c^2} \lambda _{\text{($\sigma $-$\pi
   $)}} \lambda _{( V-A )_{\bot }^{\text{adj}}}-\frac{1}{3 N_c}\lambda _{\text{($\sigma $-$\pi $)}} \lambda
   _{(S+P)_-^{\text{adj}}}\br
   -\frac{2}{3 N_c} \lambda _{(S+P)_-} \lambda _{(S+P)_-^{\text{adj}}}-\frac{1}{2 N_c}\lambda
   _{(S+P)_-^{\text{adj}}}^2+\frac{2}{3
   N_c} \lambda _{\text{($\sigma $-$\pi $)}} \lambda _{( V-A )_{\parallel }}-\frac{1}{6 N_c}\lambda _{( V+A )_{\parallel }} \lambda _{( V+A )_{\parallel }^{\text{adj}}}\br
   -\frac{1}{3 N_c}\lambda _{( V+A
   )_{\parallel }^{\text{adj}}} \lambda _{( V+A )_{\bot }}-\frac{4}{3 N_c} \lambda _{\text{($\sigma $-$\pi $)}} \lambda _{( V-A
   )_{\bot }}+\frac{1}{3 N_c}\lambda _{\text{($\sigma $-$\pi $)}} \lambda _{( V-A )_{\bot }^{\text{adj}}}-\frac{3}{2 N_c} \lambda
   _{( V+A )_{\parallel }} \lambda _{( V-A )_{\bot }^{\text{adj}}}\br
   +4N_c \lambda _{( V+A )_{\parallel }} \lambda _{( V-A
   )_{\parallel }} +\frac{3N_c}{2} \lambda _{( V+A )_{\parallel }} \lambda _{( V-A )_{\bot }^{\text{adj}}} +\lambda
   _{\text{($\sigma $-$\pi $)}} \lambda _{\text{csc}}+\lambda _{(S+P)_-} \lambda _{\text{csc}}+\frac{1}{6} \lambda
   _{(S+P)_-^{\text{adj}}} \lambda _{\text{csc}}\br
   +\lambda _{( V+A )_{\parallel }} \lambda _{\text{csc}}+\frac{1}{3 N_c^2}\lambda
   _{(S+P)_-^{\text{adj}}} \lambda _{\text{csc}}-\frac{2}{3
   N_c} \lambda _{\text{($\sigma $-$\pi $)}} \lambda _{\text{csc}}-\frac{2}{3 N_c} \lambda _{(S+P)_-} \lambda _{\text{csc}}-\frac{1}{6
   N_c}\lambda _{(S+P)_-^{\text{adj}}} \lambda _{\text{csc}}\br
   -N_c\lambda _{( V+A )_{\parallel }}  \lambda _{\text{csc}}-\frac{1}{6}\lambda _{\text{csc}}^2+\frac{1}{3 N_c}\lambda
   _{\text{csc}}^2\Big) l_{\text{$\bot $+}}^{\text{(F)}}\left(\tau ,0,-i \tilde{\mu} _{\tau }\right)\br
   +64v_4 \Big(\frac{1}{4} \lambda _{\text{($\sigma $-$\pi $)}} \lambda _{( V+A )_{\parallel }^{\text{adj}}}-\lambda _{( V+A
   )_{\parallel }} \lambda _{( V+A )_{\bot }}+\lambda _{( V+A )_{\bot }}^2-\frac{1}{3 N_c^2}\lambda _{\text{($\sigma $-$\pi
   $)}} \lambda _{( V+A )_{\parallel }^{\text{adj}}}-\frac{1}{12 N_c^2}\lambda _{( V+A )_{\parallel
   }^{\text{adj}}}^2\br
   +\frac{2}{3
   N_c} \lambda _{\text{($\sigma $-$\pi $)}} \lambda _{( V+A )_{\parallel }}+\frac{1}{6 N_c}\lambda _{( V+A )_{\parallel }} \lambda _{( V+A )_{\parallel }^{\text{adj}}}
   -\frac{4}{3 N_c} \lambda
   _{\text{($\sigma $-$\pi $)}} \lambda _{( V+A )_{\bot }}\br
   -\frac{1}{3 N_c}\lambda _{( V+A )_{\parallel }^{\text{adj}}} \lambda
   _{( V+A )_{\bot }}\Big) l_{\bot \pm }^{\text{(F)}}\left(\tau ,0,-i \tilde{\mu} _{\tau }\right)\,,\nn
\ee

\be
\partial_t  \lambda _\VpAPer &=& 2 \lambda _\VpAPer 
+64 v_4 \Big(-\frac{1}{2 N_c} \lambda _{\SpPmAdj}^2 +\frac{1}{2} \lambda _{\SpPmAdj}^2+\frac{1}{2} \lambda_{\Csc} \lambda _{\SpPmAdj}+\frac{1}{2} \lambda _{\SigmaPion} \lambda _{\SpPmAdj} \nn \\
&& +\lambda _{\SpPm} \lambda_{\SpPmAdj} -\frac{1}{2 N_c} \lambda _{\Csc} \lambda _{\SpPmAdj} + \lambda _{\VpAPer}^2+\lambda _{\Csc} \lambda_{\SigmaPion} + \lambda _{\Csc} \lambda _{\SpPm} \nn \\
&& -2 \lambda _{\SigmaPion} \lambda _{\VmAPer} -\frac{1}{4} \lambda_{\SigmaPion} \lambda _{\VmAPerAdj} + \lambda _{\Csc} \lambda _{\VpAPer} + \lambda _{\VmAPar} \lambda _{\VpAPer}+\lambda_{\VmAPer} \lambda _{\VpAPer} \nn \\
&& +\lambda _{\VpAPar} \lambda _{\VpAPer} - N_c \lambda _{\Csc} \lambda _{\VpAPer} + 4 N_c \lambda_{\VmAPer} \lambda _{\VpAPer} +\frac{N_c}{2} \lambda _{\VmAPerAdj} \lambda _{\VpAPer} \nn \\
&& +\frac{1}{N_c} \lambda _{\SigmaPion}\lambda _{\VmAPerAdj} -\frac{1}{2 N_c}\lambda _{\VmAPerAdj} \lambda _{\VpAPer} - \frac{1}{2 N_c} \lambda _{\VpAParAdj} \lambda_{\VpAPer} \Big) \lFParallelP \nn \\
&&+ 64 v_4 \Big(\lambda _{\VpAPer}^2-\lambda _{\VpAPar} \lambda _{\VpAPer}+\frac{1}{2 N_c}\lambda_{\VpAParAdj} \lambda _{\VpAPer}+\frac{1}{4} \lambda _{\SigmaPion} \lambda _{\VpAParAdj}\Big) \lFParallelPM \nn \\
&& + 64 v_4 \Big(\frac{1}{6} \lambda _{\Csc}^2+\frac{1}{3} \lambda _{\SigmaPion} \lambda _{\Csc}+\frac{1}{3} \lambda _{\SpPm} \lambda _{\Csc}+\frac{1}{2} \lambda _{\SpPmAdj} \lambda _{\Csc}+\frac{1}{3} \lambda _{\VpAPer} \lambda_{\Csc} \nn \\
&&-\frac{N_c}{3} \lambda _{\VpAPer} \lambda _{\Csc}-\frac{1}{6N_c} \lambda _{\SpPmAdj} \lambda _{\Csc} + \frac{1}{6} \lambda _{\SigmaPion}^2+\frac{2}{3} \lambda _{\SpPm}^2+\frac{1}{2} \lambda _{\SpPmAdj}^2+\frac{1}{6} \lambda _{\VpAPar}^2 \nn \\
&& +\frac{7}{6} \lambda _{\VpAPer}^2 + \frac{2}{3} \lambda _{\SigmaPion} \lambda _{\SpPm} + \frac{1}{2} \lambda _{\SigmaPion} \lambda _{\SpPmAdj} + \lambda _{\SpPm} \lambda _{\SpPmAdj} -\frac{2}{3} \lambda _{\SigmaPion} \lambda _{\VmAPer} \nn \\
&& - \frac{1}{12} \lambda _{\SigmaPion} \lambda _{\VmAPerAdj} + \frac{1}{3} \lambda _{\VmAPar} \lambda_{\VpAPer} + \frac{1}{3} \lambda _{\VmAPer} \lambda _{\VpAPer} + \frac{2}{3} \lambda _{\VpAPar} \lambda_{\VpAPer}\nn \\
&& + \frac{4}{3} N_c \lambda _{\VmAPer} \lambda _{\VpAPer} +\frac{N_c}{6} \lambda _{\VmAPerAdj} \lambda _{\VpAPer} -\frac{1}{2 N_c} \lambda _{\SpPmAdj}^2 - \frac{1}{3 N_c}\lambda _{\SigmaPion} \lambda _{\SpPmAdj} \nn \\
&& - \frac{2 }{3 N_c} \lambda_{\SpPm} \lambda _{\SpPmAdj} + \frac{1}{3 N_c} \lambda _{\SigmaPion} \lambda _{\VmAPerAdj} - \frac{1}{6 N_c} \lambda_{\VpAPar} \lambda _{\VpAParAdj} -\frac{1}{6 N_c} \lambda _{\VmAPerAdj} \lambda _{\VpAPer}\nn \\
&& - \frac{1}{3 N_c} \lambda_{\VpAParAdj} \lambda _{\VpAPer} + \frac{1}{6 N_c^2} \lambda _{\SpPmAdj}^2 + \frac{1}{24 N_c^2} \lambda _{\VpAParAdj}^2 \Big) \lFOrthogonalP \nn \\
&&+ 64 v_4 \Big(-\frac{1}{6} \lambda _{\SigmaPion}^2-\frac{1}{12} \lambda _{\VpAParAdj} \lambda_{\SigmaPion}-\frac{1}{6} \lambda _{\VpAPar}^2-\frac{1}{24} \lambda _{\VpAParAdj}^2-\frac{7}{6} \lambda_{\VpAPer}^2+\frac{2}{3} \lambda _{\VpAPar} \lambda _{\VpAPer} \nn \\
&&+\frac{1}{6 N_c} \lambda _{\VpAPar} \lambda _{\VpAParAdj} -\frac{1}{3 N_c} \lambda _{\VpAParAdj} \lambda _{\VpAPer} -\frac{1}{24 N_c^2} \lambda _{\VpAParAdj}^2 \Big) \lFOrthogonalPM \,,\nn
\ee
\vspace*{-0.5cm}
\be
\partial_t  \lambda _\VmAPar &=& 2 \lambda _\VmAPar 
+ 64 v_4 \Big(\frac{1}{2} N_c \lambda _{\Csc}^2-\frac{1}{2} \lambda _{\Csc}^2-\lambda _{\VmAPar} \lambda_{\Csc}-\frac{3}{2} \lambda _{\VmAPerAdj} \lambda _{\Csc}+ N_c \lambda _{\VmAPar} \lambda _{\Csc} \nn \\
&& + \frac{3}{2} N_c \lambda_{\VmAPerAdj} \lambda _{\Csc} -\frac{1}{2} \lambda _{\SigmaPion}^2 +\frac{3}{2} \lambda _{\VmAPar}^2+\frac{3}{2} \lambda _{\VmAPer}^2-\frac{1}{2} \lambda _{\VpAParAdj}^2-3 \lambda _{\VmAPar} \lambda _{\VmAPer}\nn \\
&& +\frac{3}{2} \lambda_{\VmAPer} \lambda _{\VmAPerAdj}+2 \lambda _{\SigmaPion} \lambda _{\VpAPar}  +\lambda _{\SigmaPion} \lambda_{\VpAParAdj}-2 N_c \lambda _{\VmAPar}^2 +\frac{3}{8} N_c \lambda _{\VmAPerAdj}^2 \nn \\
&& -2 N_c \lambda _{\VpAPar}^2  - \frac{3}{2} N_c \lambda _{\VmAPar} \lambda _{\VmAPerAdj} - \frac{3}{4 N_c} \lambda _{\VmAPerAdj}^2  + \frac{1}{2 N_c} \lambda_{\VpAParAdj}^2 \nn \\
&& + \frac{3}{2 N_c} \lambda _{\VmAPar} \lambda _{\VmAPerAdj}  - \frac{3}{2 N_c} \lambda _{\VmAPer} \lambda_{\VmAPerAdj} - \frac{1}{N_c} \lambda _{\SigmaPion} \lambda _{\VpAParAdj} \nn \\
&& +\frac{3}{8 N_c^2} \lambda _{\VmAPerAdj}^2 \Big) \lFParallelP \nn \\
&& + 64 v_4 \Big( -\frac{3}{2} \lambda _{\SigmaPion}^2 - 6 \lambda _{\SpPm} \lambda _{\SigmaPion}-3 \lambda _{\SpPmAdj} \lambda _{\SigmaPion} + \frac{3 }{N_c} \lambda _{\SpPmAdj} \lambda _{\SigmaPion} - 6 \lambda_{\SpPm}^2 \nn \\
&& -\frac{3}{2} \lambda _{\SpPmAdj}^2  -\frac{1}{2} \lambda _{\VmAPar}^2 - \frac{3}{2} \lambda_{\VmAPer}^2-\frac{3}{8} \lambda _{\VmAPerAdj}^2-6 \lambda _{\SpPm} \lambda _{\SpPmAdj} \nn \\
&& -\frac{3}{2} \lambda _{\VmAPer} \lambda _{\VmAPerAdj}  +\frac{3 }{N_c} \lambda _{\SpPmAdj}^2 + \frac{3}{4 N_c} \lambda _{\VmAPerAdj}^2+\frac{6}{N_c} \lambda_{\SpPm} \lambda _{\SpPmAdj} \nn \\
&&+ \frac{3}{2 N_c} \lambda _{\VmAPer} \lambda _{\VmAPerAdj}-\frac{3}{2 N_c^2} \lambda_{\SpPmAdj}^2 -\frac{3}{8 N_c^2} \lambda _{\VmAPerAdj}^2 \Big) \lFParallelPM \nn \\ 
&& + 64 v_4 \Big(\frac{1}{2} \lambda _{\SigmaPion}^2-2 \lambda _{\VpAPar} \lambda _{\SigmaPion}-\lambda _{\VpAParAdj} \lambda_{\SigmaPion}+\frac{1}{N_c} \lambda _{\VpAParAdj} \lambda _{\SigmaPion} -\lambda _{\VmAPar}^2-\lambda_{\VmAPer}^2 \nn \\
&&+\frac{1}{2} \lambda _{\VpAParAdj}^2+\lambda _{\Csc} \lambda _{\VmAPar}+4 \lambda _{\VmAPar} \lambda_{\VmAPer}+\lambda _{\Csc} \lambda _{\VmAPerAdj}+\frac{1}{2} \lambda _{\VmAPar} \lambda _{\VmAPerAdj} \nn \\
&&-\lambda_{\VmAPer} \lambda _{\VmAPerAdj}+2 N_c \lambda _{\VmAPar}^2 -\frac{N_c}{4} \lambda _{\VmAPerAdj}^2 + 2 N_c \lambda_{\VpAPar}^2 - N_c \lambda _{\Csc} \lambda _{\VmAPar} \nn \\
&& - N_c \lambda _{\Csc} \lambda _{\VmAPerAdj}  +\frac{3}{2} N_c \lambda_{\VmAPar} \lambda _{\VmAPerAdj} +\frac{1}{2 N_c} \lambda _{\VmAPerAdj}^2 -\frac{1}{2 N_c} \lambda _{\VpAParAdj}^2 \nn \\
&& - \frac{2}{N_c} \lambda _{\VmAPar} \lambda _{\VmAPerAdj}  + \frac{1}{N_c} \lambda _{\VmAPer} \lambda_{\VmAPerAdj}  -\frac{1}{4 N_c^2} \lambda _{\VmAPerAdj}^2 \Big) \lFOrthogonalP \nn \\
&& + 64 v_4 \Big(\frac{1}{2} \lambda_{\SigmaPion}^2+2 \lambda _{\SpPm} \lambda _{\SigmaPion}+\lambda _{\SpPmAdj} \lambda _{\SigmaPion}-\frac{1}{N_c} \lambda_{\SpPmAdj} \lambda _{\SigmaPion} +2 \lambda _{\SpPm}^2+\frac{1}{2} \lambda _{\SpPmAdj}^2 \nn \\
&& -\lambda_{\VmAPer}^2-\frac{1}{4} \lambda _{\VmAPerAdj}^2+2 \lambda _{\SpPm} \lambda _{\SpPmAdj}-\lambda _{\VmAPar} \lambda_{\VmAPer}-\frac{1}{2} \lambda _{\VmAPar} \lambda _{\VmAPerAdj} \nn \\
&& -\lambda _{\VmAPer} \lambda _{\VmAPerAdj}-\frac{1}{N_c} \lambda_{\SpPmAdj}^2 + \frac{1}{2 N_c} \lambda _{\VmAPerAdj}^2 -\frac{2 }{N_c} \lambda _{\SpPm} \lambda_{\SpPmAdj} \nn \\
&& + \frac{1}{2 N_c} \lambda _{\VmAPar} \lambda _{\VmAPerAdj} + \frac{1}{N_c} \lambda _{\VmAPer} \lambda_{\VmAPerAdj} +\frac{1}{2 N_c^2} \lambda _{\SpPmAdj}^2 -\frac{1}{4 N_c^2} \lambda _{\VmAPerAdj}^2 \Big) \lFOrthogonalPM \,,\nn
\ee
\be
\partial_t  \lambda _\VmAPer &=& 2 \lambda _\VmAPer 
+ 64 v_4 \Big(\frac{N_c}{2}  \lambda _{\Csc}^2+\frac{1}{N_c}\lambda _{\Csc}^2-\frac{3}{2} \lambda _{\Csc}^2+\lambda _{\VmAPer} \lambda_{\Csc}-\frac{3}{2} \lambda _{\VmAPerAdj} \lambda _{\Csc}-N_c \lambda _{\VmAPer}  \lambda _{\Csc} \nn \\
&& +\frac{N_c}{2} \lambda_{\VmAPerAdj} \lambda _{\Csc}  + \frac{1}{N_c} \lambda _{\VmAPerAdj} \lambda _{\Csc} -\frac{1}{2} \lambda_{\SigmaPion}^2-\frac{5}{8} \lambda _{\VmAPerAdj}^2-\frac{1}{2} \lambda _{\VpAParAdj}^2+2 \lambda _{\VmAPar} \lambda_{\VmAPer} \nn \\
&& +\frac{3}{2} \lambda _{\VmAPer} \lambda _{\VmAPerAdj} +\lambda _{\SigmaPion} \lambda _{\VpAParAdj}-2 \lambda_{\SigmaPion} \lambda _{\VpAPer}+2 N_c \lambda _{\VmAPer}^2 +\frac{3}{8} N_c \lambda _{\VmAPerAdj}^2 \nn \\
&& + 2 N_c \lambda_{\VpAPer}^2  + \frac{N_c}{2} \lambda _{\VmAPer} \lambda _{\VmAPerAdj} + \frac{1}{N_c} \lambda _{\SigmaPion}^2 - \frac{3}{4 N_c}  \lambda _{\VmAPerAdj}^2 +\frac{1}{2 N_c} \lambda _{\VpAParAdj}^2 \nn \\
&& - \frac{2}{N_c}  \lambda _{\VmAPer} \lambda_{\VmAPerAdj} - \frac{1}{N_c} \lambda _{\SigmaPion} \lambda _{\VpAParAdj} + \frac{1}{N_c^2} \lambda_{\VmAPerAdj}^2 \Big) \lFParallelP \nn \\
&& + 64 v_4 \Big(\frac{1}{N_c} \lambda _{\SigmaPion}^2 -\frac{1}{2} \lambda_{\SigmaPion}^2-2 \lambda _{\SpPm} \lambda _{\SigmaPion}-\lambda _{\SpPmAdj} \lambda _{\SigmaPion}+\frac{4}{N_c} \lambda_{\SpPm} \lambda _{\SigmaPion} \nn \\
&& +\frac{3}{N_c} \lambda _{\SpPmAdj} \lambda _{\SigmaPion}  - \frac{2}{N_c^2} \lambda_{\SpPmAdj} \lambda _{\SigmaPion} - 2 \lambda _{\SpPm}^2-\frac{1}{2} \lambda _{\SpPmAdj}^2-\lambda_{\VmAPer}^2-\frac{1}{4} \lambda _{\VmAPerAdj}^2 \nn \\
&& -2 \lambda _{\SpPm} \lambda _{\SpPmAdj}  -\lambda _{\VmAPar} \lambda_{\VmAPer}-\frac{3}{2} \lambda _{\VmAPer} \lambda _{\VmAPerAdj}+\frac{4}{N_c} \lambda _{\SpPm}^2 + \frac{2}{N_c} \lambda_{\SpPmAdj}^2 \nn \\
&& +\frac{3}{4 N_c} \lambda _{\VmAPerAdj}^2  + \frac{6}{N_c} \lambda _{\SpPm} \lambda _{\SpPmAdj} + \frac{3}{2 N_c} \lambda _{\VmAPer} \lambda _{\VmAPerAdj} - \frac{5}{2 N_c^2} \lambda _{\SpPmAdj}^2 \nn \\
&&  - \frac{1}{2 N_c^2}\lambda_{\VmAPerAdj}^2 - \frac{4}{N_c^2} \lambda _{\SpPm} \lambda _{\SpPmAdj} + \frac{1}{N_c^3} \lambda_{\SpPmAdj}^2 \Big) \lFParallelPM \nn \\
&&+ 64 v_4 \Big(-\frac{1}{3 N_c}\lambda _{\Csc}^2+\frac{1}{3} \lambda _{\Csc}^2+\frac{1}{3} \lambda _{\VmAPer} \lambda_{\Csc}+\lambda _{\VmAPerAdj} \lambda _{\Csc}-\frac{N_c}{3} \lambda _{\VmAPer}  \lambda _{\Csc} \nn \\
&& -\frac{N_c}{3} \lambda_{\VmAPerAdj} \lambda _{\Csc} - \frac{2 }{3 N_c}\lambda_{\VmAPerAdj} \lambda _{\Csc} + \frac{1}{2} \lambda_{\SigmaPion}^2 + \frac{1}{6} \lambda _{\VmAPar}^2+\frac{3}{2} \lambda _{\VmAPer}^2+\frac{13}{24} \lambda_{\VmAPerAdj}^2 \nn \\
&& +\frac{1}{2} \lambda _{\VpAParAdj}^2  -\frac{1}{3} \lambda _{\VmAPar} \lambda _{\VmAPer}+\frac{1}{2}\lambda _{\VmAPar} \lambda _{\VmAPerAdj}-\lambda _{\VmAPer} \lambda _{\VmAPerAdj}-\lambda _{\SigmaPion} \lambda_{\VpAParAdj} \nn \\ 
&& -\frac{2}{3} \lambda _{\SigmaPion} \lambda _{\VpAPer} +\frac{2}{3} N_c \lambda _{\VmAPer}^2 -\frac{N_c}{4}\lambda _{\VmAPerAdj}^2 +\frac{2}{3} N_c \lambda _{\VpAPer}^2 +\frac{N_c}{6} \lambda _{\VmAPer} \lambda _{\VmAPerAdj} \nn \\
&&  -\frac{1}{3 N_c} \lambda _{\SigmaPion}^2 +\frac{1}{2 N_c} \lambda _{\VmAPerAdj}^2 - \frac{1}{2 N_c} \lambda _{\VpAParAdj}^2 - \frac{1}{2 N_c} \lambda _{\VmAPar} \lambda _{\VmAPerAdj} \nn \\
&& + \frac{5}{6 N_c}  \lambda _{\VmAPer} \lambda _{\VmAPerAdj} + \frac{1}{N_c}\lambda _{\SigmaPion} \lambda _{\VpAParAdj} - \frac{19}{24 N_c^2}  \lambda _{\VmAPerAdj}^2 \Big) \lFOrthogonalP \nn \\ 
&& + 64 v_4 \Big(-\frac{1}{3 N_c} \lambda _{\SigmaPion}^2 + \frac{1}{6} \lambda _{\SigmaPion}^2+\frac{2}{3} \lambda_{\SpPm} \lambda _{\SigmaPion}+\frac{1}{3} \lambda _{\SpPmAdj} \lambda _{\SigmaPion} - \frac{4}{3 N_c}  \lambda _{\SpPm} \lambda_{\SigmaPion} \nn \\
&& - \frac{1}{N_c} \lambda _{\SpPmAdj} \lambda _{\SigmaPion} + \frac{2}{3 N_c^2}  \lambda _{\SpPmAdj} \lambda_{\SigmaPion} + \frac{2}{3} \lambda _{\SpPm}^2 + \frac{1}{6} \lambda _{\SpPmAdj}^2-\frac{1}{6} \lambda_{\VmAPar}^2-\frac{7}{6} \lambda _{\VmAPer}^2 \nn \\
&& - \frac{7}{24} \lambda _{\VmAPerAdj}^2 + \frac{2}{3} \lambda _{\SpPm} \lambda _{\SpPmAdj} - \frac{2}{3} \lambda _{\VmAPar} \lambda _{\VmAPer} - \frac{1}{2} \lambda _{\VmAPar} \lambda_{\VmAPerAdj} \nn \\
&& - \lambda _{\VmAPer} \lambda _{\VmAPerAdj}  - \frac{4}{3 N_c}\lambda _{\SpPm}^2 - \frac{2}{3 N_c} \lambda_{\SpPmAdj}^2 + \frac{1}{2 N_c} \lambda _{\VmAPerAdj}^2 - \frac{2}{N_c}  \lambda _{\SpPm} \lambda_{\SpPmAdj} \nn \\
&& + \frac{1}{2 N_c} \lambda _{\VmAPar} \lambda _{\VmAPerAdj}  + \frac{1}{N_c} \lambda _{\VmAPer} \lambda_{\VmAPerAdj} +\frac{5}{6 N_c^2}  \lambda _{\SpPmAdj}^2 - \frac{5}{24 N_c^2} \lambda _{\VmAPerAdj}^2 \nn \\
&&  + \frac{4}{3 N_c^2} \lambda_{\SpPm} \lambda _{\SpPmAdj} -\frac{1}{3 N_c^3}\lambda _{\SpPmAdj}^2 \Big) \lFOrthogonalPM \,,\nn
\ee
\be
\partial_t  \lambda _\VpAParAdj &=& 2 \lambda _\VpAParAdj 
+ 64 v_4 \Big(2 N_c \lambda _{\Csc} \lambda _{\SpPmAdj}-\frac{2}{N_c} \lambda _{\Csc} \lambda _{\SpPmAdj} + N_c \lambda _{\Csc}^2-\frac{2}{N_c} \lambda _{\SigmaPion} \lambda _{\VmAPerAdj} \nn \\
&& + N_c \lambda _{\SpPmAdj}^2  - \frac{2}{N_c} \lambda_{\SpPmAdj}^2 + \frac{3}{2 N_c} \lambda _{\VmAPerAdj} \lambda _{\VpAParAdj} + \frac{N_c}{4} \lambda_{\VpAParAdj}^2-\frac{1}{2 N_c} \lambda _{\VpAParAdj}^2 \nn \\
&& + 4 \lambda _{\Csc} \lambda _{\SigmaPion} + 4 \lambda _{\Csc} \lambda _{\SpPm} -2 \lambda _{\Csc} \lambda _{\SpPmAdj}-\lambda _{\Csc} \lambda _{\VpAParAdj}-2 \lambda _{\Csc}^2+2 \lambda _{\SigmaPion} \lambda _{\SpPmAdj} \nn \\
&& +4 \lambda _{\SigmaPion} \lambda _{\VmAPer} +2 \lambda _{\SigmaPion} \lambda_{\VmAPerAdj}  + 4 \lambda _{\SpPm} \lambda _{\SpPmAdj}+\lambda _{\VmAPar} \lambda _{\VpAParAdj} \nn \\
&& -3 \lambda _{\VmAPer} \lambda _{\VpAParAdj} +\lambda _{\VpAPar} \lambda _{\VpAParAdj}  -\lambda _{\VpAParAdj} \lambda_{\VpAPer} \Big) \lFParallelP \nn \\
&& + 64 v_4 \Big(\frac{1}{2 N_c}\lambda _{\VpAParAdj}^2 \!+\! 4 \lambda _{\SigmaPion} \lambda_{\VpAPer}\!-\!\lambda _{\VpAPar} \lambda _{\VpAParAdj}
\!+\!\lambda _{\VpAParAdj} \lambda _{\VpAPer}\Big) \lFParallelPM \nn \\
&& + 64 v_4 \Big(-\frac{2}{3} N_c \lambda _{\Csc} \lambda _{\SpPmAdj}+\frac{2}{3 N_c} \lambda _{\Csc} \lambda _{\SpPmAdj} -\frac{N_c}{3} \lambda _{\Csc}^2+\frac{4}{3 N_c} \lambda _{\SigmaPion} \lambda _{\VmAPerAdj} \nn \\
&& - \frac{N_c}{3} \lambda _{\SpPmAdj}^2+\frac{2}{3 N_c} \lambda _{\SpPmAdj}^2 - \frac{3}{2 N_c} \lambda _{\VmAPerAdj} \lambda _{\VpAParAdj} - \frac{N_c}{12} \lambda _{\VpAParAdj}^2 + \frac{1}{6 N_c}\lambda _{\VpAParAdj}^2 \nn \\
&& - \frac{4}{3} \lambda _{\Csc} \lambda _{\SigmaPion}  -\frac{4}{3} \lambda _{\Csc} \lambda _{\SpPm}+\frac{2}{3} \lambda _{\Csc} \lambda_{\SpPmAdj}+\lambda _{\Csc} \lambda _{\VpAParAdj}+\frac{2}{3} \lambda _{\Csc}^2-\frac{2}{3} \lambda _{\SigmaPion} \lambda _{\SpPmAdj} \nn \\
&& +\frac{4}{3} \lambda _{\SigmaPion} \lambda _{\VmAPar}\! -\!\frac{8}{3} \lambda _{\SigmaPion} \lambda_{\VmAPer}\!+\!\frac{2}{3} \lambda _{\SigmaPion} \lambda _{\VmAPerAdj}\!-\!\frac{4}{3} \lambda _{\SpPm} \lambda_{\SpPmAdj}\!-\!\lambda _{\VmAPar} \lambda _{\VpAParAdj} \nn \\ 
&&  +3 \lambda _{\VmAPer} \lambda _{\VpAParAdj} -\frac{1}{3} \lambda_{\VpAPar} \lambda _{\VpAParAdj}+\frac{1}{3} \lambda _{\VpAParAdj} \lambda _{\VpAPer}\Big) \lFOrthogonalP \nn \\
&& + 64 v_4 \Big(-\frac{2}{3 N_c} \lambda _{\SigmaPion} \lambda _{\VpAParAdj}-\frac{1}{6 N_c}\lambda_{\VpAParAdj}^2 + \frac{4}{3} \lambda _{\SigmaPion} \lambda _{\VpAPar}-\frac{8}{3} \lambda _{\SigmaPion} \lambda_{\VpAPer} \nn \\
&& +\frac{1}{3} \lambda _{\VpAPar} \lambda _{\VpAParAdj}-\frac{5}{3} \lambda _{\VpAParAdj} \lambda_{\VpAPer}\Big) \lFOrthogonalPM \,,\nn
\ee
\be
\partial_t  \lambda _\VmAPerAdj &=& 2 \lambda _\VmAPerAdj 
+ 64 v_4 \Big(-2 N_c \lambda_{\Csc} \lambda _{\VmAPerAdj}-N_c \lambda _{\Csc}^2-\frac{5}{4} N_c \lambda _{\VmAPerAdj}^2+\frac{2}{N_c} \lambda_{\VmAPerAdj}^2\nn \\ 
&& + 3 \lambda _{\Csc} \lambda _{\VmAPerAdj}  +2 \lambda _{\Csc}^2- 2 \lambda _{\SigmaPion} \lambda_{\VpAParAdj} + 2 \lambda _{\SigmaPion}^2+2 \lambda _{\VmAPar} \lambda _{\VmAPerAdj}\nn \\
&& -4 \lambda _{\VmAPer} \lambda_{\VmAPerAdj} +\lambda _{\VmAPerAdj}^2  +\lambda _{\VpAParAdj}^2\Big) \lFParallelP \nn \\ 
&& +  64 v_4 \Big( -\frac{4}{N_c} \lambda _{\SigmaPion} \lambda _{\SpPmAdj} - \frac{8}{N_c} \lambda_{\SpPm} \lambda _{\SpPmAdj} - \frac{4}{N_c} \lambda _{\SpPmAdj}^2 + \frac{2}{N_c^2}  \lambda_{\SpPmAdj}^2 \nn \\
&& - \frac{1}{2 N_c} \lambda _{\VmAPerAdj}^2  + 8 \lambda _{\SigmaPion} \lambda _{\SpPm} + 4 \lambda_{\SigmaPion} \lambda _{\SpPmAdj} + 2 \lambda _{\SigmaPion}^2 + 8 \lambda _{\SpPm} \lambda _{\SpPmAdj}\nn \\ 
&&  + 8 \lambda_{\SpPm}^2 + 2 \lambda _{\SpPmAdj}^2 -\lambda _{\VmAPar} \lambda _{\VmAPerAdj}+\lambda _{\VmAPer} \lambda_{\VmAPerAdj} \Big) \lFParallelPM \nn \\
&& + 64 v_4 \Big(\frac{4}{3} N_c \lambda _{\Csc} \lambda _{\VmAPerAdj}+\frac{N_c}{3} \lambda_{\Csc}^2+\frac{13}{12} N_c \lambda _{\VmAPerAdj}^2-\frac{7}{3 N_c} \lambda _{\VmAPerAdj}^2 - \lambda _{\Csc} \lambda_{\VmAPerAdj} \nn \\
&& -\frac{2}{3} \lambda _{\Csc}^2+2 \lambda _{\SigmaPion} \lambda _{\VpAParAdj} -\frac{2}{3} \lambda_{\SigmaPion}^2-\frac{4}{3} \lambda _{\VmAPar} \lambda _{\VmAPerAdj}+\frac{14}{3} \lambda _{\VmAPer} \lambda_{\VmAPerAdj} \nn \\
&& +\frac{1}{3} \lambda _{\VmAPerAdj}^2  -\lambda _{\VpAParAdj}^2\Big) \lFOrthogonalP \nn \\
&& + 64 v_4 \Big( \frac{4}{3 N_c}\lambda_{\SigmaPion} \lambda _{\SpPmAdj} + \frac{8}{3 N_c} \lambda _{\SpPm} \lambda _{\SpPmAdj} + \frac{4}{3 N_c} \lambda_{\SpPmAdj}^2 - \frac{2}{3 N_c^2} \lambda _{\SpPmAdj}^2 \nn \\
&& + \frac{1}{6 N_c}\lambda _{\VmAPerAdj}^2 - \frac{8}{3} \lambda _{\SigmaPion} \lambda _{\SpPm}-\frac{4}{3} \lambda _{\SigmaPion} \lambda _{\SpPmAdj} - \frac{2}{3} \lambda_{\SigmaPion}^2 - \frac{8}{3} \lambda _{\SpPm} \lambda _{\SpPmAdj}\nn \\
&&  -\frac{8}{3} \lambda _{\SpPm}^2 -\frac{2}{3} \lambda_{\SpPmAdj}^2 +\frac{1}{3} \lambda _{\VmAPar} \lambda _{\VmAPerAdj}-\frac{1}{3} \lambda _{\VmAPer} \lambda_{\VmAPerAdj} \Big) \lFOrthogonalPM 
\,.\nn 
\ee
}
\end{widetext}
\bibliography{qcd}

\end{document}